\documentclass[12pt]{article}

\usepackage{amssymb,amsmath}
\usepackage[dvips]{graphicx}
\usepackage{setspace}

\def\blfootnote{\xdef\@thefnmark{}\@footnotetext} 

\long\def\symbolfootnote[#1]#2{\begingroup%
\def\thefootnote{\fnsymbol{footnote}}\footnote[#1]{#2}\endgroup}

\newcommand{\sla}[1]{{\raisebox{0.8pt}{\makebox[0pt][l]{\hspace{-2.5pt}\bf{
          /}}}{#1}}} 
\newcommand{\de}{\partial}

\newcommand{\slap}{{\raisebox{-0.5pt}{\makebox[0pt][l]{\hspace{-4.4pt}\bf{
          /}}}{p}}} 
\newcommand{\slaA}{{\raisebox{1pt}{\makebox[0pt][l]{\hspace{-2.8pt}\bf{
          /}}}{A}}} 
\newcommand{\sladx}{{\raisebox{0.65pt}{\makebox[0pt][l]{\hspace{-2.4pt}\bf{
          /}}}{\dot{X}_E}}} 

\newcommand{\la}{\langle}
\newcommand{\ra}{\rangle}
\newcommand{\f}[2]{\frac{#1}{#2}}

\newcommand{\outs}[1]{\la {\rm out}\,\, #1 }
\newcommand{\ins}[1]{ #1 \,\, {\rm in} \ra}

\newcommand{\ket}[1]{| #1 \,\ra}

\newcommand{\barq}{\protect{\bar{q}}}
\newcommand{\barQ}{\protect{\bar{Q}}}

\newcommand{\hS}[1]{\hat{S}_{#1}}
\newcommand{\dl}[1]{\delta_{#1}}

\newcommand{\br}{\la 0 |}
\newcommand{\ke}{| 0 \ra}

\newcommand{\Texp}{{\rm Texp}}
\newcommand{\Lim}{{\rm Lim}}

\newcommand{\DX}{[{\cal D}X]}
\newcommand{\DXp}{[{\cal D}X']}

\newcommand{\DXE}{[{\cal D}X_E]}
\newcommand{\DXpE}{[{\cal D}X_E']}
\newcommand{\DZ}{[{\cal D}Z]}
\newcommand{\DZp}{[{\cal D}Z']}
\newcommand{\DP}{[{\cal D}\Pi]}
\newcommand{\DPp}{[{\cal D}\Pi']}
\newcommand{\DPE}{[{\cal D}\Pi_E]}
\newcommand{\DPpE}{[{\cal D}\Pi_E']}
\newcommand{\DR}{[{\cal D}\rho]}
\newcommand{\DRp}{[{\cal D}\rho']}
\newcommand{\wf}{\varphi}
\newcommand{\vr}{\vec{R}}
\newcommand{\vq}{\vec{q}}

\newcommand{\M}{{\cal M}}
\newcommand{\W}{{\cal W}}
\newcommand{\A}{{\cal A}}
\newcommand{\tr}{{\rm tr}}
\newcommand{\m}{\tilde{m}}
\newcommand{\Sp}{{\cal S}}
\newcommand{\Sps}{{\rm S}}

\newcommand{\dlx}[1]{\delta^{(#1)}}
\newcommand{\para}{\parallel}
\newcommand{\diag}{{\rm diag}}

\newcommand{\acosh}{{\rm arccosh}}

\newcommand{\LL}{L_0}
\newcommand{\tLL}{\tilde{L}_0}
\newcommand{\LLe}{L_{E0}}

\renewcommand{\Re}{{\rm Re\,}}
\renewcommand{\Im}{{\rm Im\,}}

\newcommand{\nsection}[1]{\setcounter{equation}{0}\section{#1}
  \vspace{5mm}}
\newcommand{\nonsection}[1]{\setcounter{equation}{0}\section*{#1}
  \vspace{5mm}} 

\newcommand{\mytitle}[1]{
\begin{center}
  \vspace*{1.0cm}
  \begin{doublespace}
    {\Large \bf #1}
  \end{doublespace}
  \vspace*{0.5cm}
  {\large Matteo Giordano
        \symbolfootnote[1]{E-mail: giordano@unizar.es}
      }\\
  \vspace*{0.5cm}{\normalsize
    { 
      Departamento de F\'isica Te\'orica, 
      Universidad de Zaragoza, \\
      Calle Pedro Cerbuna 12,
      E--50009 Zaragoza, Spain
    }
  }\\  \vspace*{0.5cm}
\end{center}
}

\begin{document}

\mytitle{Wilson-loop formalism for Reggeon exchange in soft 
  high-energy scattering}

\begin{abstract}
  We derive a nonperturbative expression for the non-vacuum, 
  $q\barq$-Reggeon-ex\-change contribution to the meson-meson elastic
  scattering amplitude at high energy and low momentum transfer, in
  the framework of QCD. Describing the mesons in terms of 
  colourless $q\barq$ dipoles, the problem is reduced to the
  two-fermion-exchange contribution to the dipole-dipole scattering
  amplitudes, which is expressed as a path integral, over the
  trajectories of the exchanged fermions, of the expectation value of
  a certain Wilson loop. We also show how the resulting expression can
  be reconstructed from a corresponding quantity in the Euclidean
  theory, by means of analytic continuation. Finally, we make contact 
  with previous work on Reggeon exchange in the gauge/gravity duality
  approach.
\end{abstract}

\nsection{Introduction}

The problem of hadronic high-energy scattering at low transferred
momentum, i.e., in the so-called {\it soft} high-energy regime, has
been challenging theoretical physicists for many decades, since well
before the discovery of Quantum Chromodynamics (QCD). Nowadays, it is
generally believed that QCD is the fundamental, microscopic theory
underlying strong interactions, and thus it should provide an
explanation of {soft} high-energy scattering from first
principles. However, {soft} high-energy processes are 
characterised by two different energy scales, the total center-of-mass
energy $\sqrt{s}$, which is a large scale, and the transferred
momentum $\sqrt{|t|}$, which is fixed, and smaller than or of the
order of the typical hadronic scale, $\sqrt{|t|}\lesssim 1 {\rm GeV}
\ll \sqrt{s}$. As a consequence, the study of these processes requires
the investigation of the nonperturbative regime of QCD, which has not
been completely understood yet.

From a phenomenological point of view, {soft} high-energy
hadron-hadron scattering processes can be described, in the language 
of Regge theory, in terms of the exchange of ``families'' of states
between the interacting hadrons. These ``families'' correspond to the
singularities in the complex-angular-momentum plane of the amplitude
in the crossed channel, and their position as a function of the
transferred momentum defines the corresponding ``Regge trajectory''
$\alpha(t)$ (see, e.g., Ref.~\cite{Collins}).  The leading
contribution to elastic scattering amplitudes at high energy comes
from the so-called {\it Pomeron}, which carries the quantum numbers of
the vacuum, while subleading non-vacuum contributions are usually
called {\it Reggeons}, and correspond to various non-vacuum
quantum-number exchanges.  

One of the aims of the theoretical study of {soft} high-energy
reactions in the framework of QCD is an explanation from first
principles of these phenomenological concepts. As regards the Pomeron,
a nonperturbative approach to the problem has been formulated  
long ago~\cite{Nacht}. This approach is based on the description of
the interacting hadrons in terms of partons, which together with the
LSZ reduction formulas~\cite{LSZ1,LSZ2}, and the eikonal approximation
for propagators in an external field, leads to the Wilson-loop
formalism for {soft} high energy 
scattering~\cite{DFK,Nachtr,Meggiolaro96,BN,Meggiolaro00,Dosch,LLCM1}.  
The resulting expressions for the scattering amplitudes have been
investigated by means of various nonperturbative techniques, including
the Stochastic Vacuum Model~\cite{DFK,Nachtr,BN,Dosch,LLCM1,LLCM2},
the Instanton Liquid Model~\cite{ILM, instantons}, the AdS/CFT
correspondence for non-confining~\cite{Jani1,adsbounds,GPS} and
confining~\cite{Jani2,Janik,Kharzeev} backgrounds, and Lattice Gauge
Theory~\cite{lattice,instantons,lat_pomeron}, taking advantage, in
most cases, of the analytic continuation of the amplitudes in
Euclidean
space~\cite{Meggiolaro97,Meggiolaro98,Meggiolaro02,Meggiolaro05,
  crossing,crossing2,EMduality}. These works are concerned with the
leading behaviour of the elastic scattering amplitudes at high energy,
and so non-vacuum Reggeon-exchange contributions are neglected from
the onset. 

To our knowledge, the only attempt at an extension of this approach to
the problem of subleading contributions, i.e., to Reggeon exchange, is
the one discussed in Ref.~\cite{Jani}, and recently reanalysed
in Ref.~\cite{reggeon_duality}.\footnote{Recent works on Reggeon
  exchange, following different approaches, include Ref.~\cite{Yoshi},
  where a unified treatment of the signature-odd partner of the Pomeron,
  the so-called ``Odderon'', and of the signature-odd Reggeons is
  proposed, and Ref.~\cite{Makeenko}, where the Regge behaviour of
  scattering amplitudes in QCD is obtained in an effective string
  approach.} In those works, the Reggeon-exchange amplitude is put
into a relation with the expectation value of certain Euclidean Wilson
loops, describing the exchange of a (Reggeised) quark-antiquark pair
between the interacting hadrons. More precisely, the loop contours are
made up of a fixed part, corresponding to the eikonal trajectories of
the ``spectator'' fermions, and a ``floating'' part, corresponding to
the trajectories of the exchanged fermions. The Reggeon-exchange
scattering amplitude is obtained by summing up the contributions of
these loops, through a path-integration over the trajectories of the
exchanged fermions, and performing an appropriate analytic
continuation to Minkowski space-time. An estimate of the
Reggeon-exchange amplitude is then obtained, by relating the
Wilson-loop expectation value, via gauge/gravity duality, to minimal
surfaces in a curved confining
metric~\cite{Wilson,Wilson2,GO,DGO,Sonn0,Sonn1,Sonn2}, having the loop
contour as boundary, and by evaluating the path integral by means 
of a saddle-point approximation. The resulting amplitude is of
Regge-pole type with a linear Regge trajectory in the massless-quark
case~\cite{Jani}; the inclusion of the effects of a nonzero quark mass
leaves unchanged the linearity and the slope of the trajectory, while
modifying the slope of the amplitude at $t=0$ and its shrinkage with 
energy~\cite{reggeon_duality}. The slope of the Regge trajectory is
equal to the inverse string tension $\alpha'_{\rm eff}=1/(2\pi\sigma)$
appearing in the confining potential: this is a first step into
understanding the relation between the Wilson-loop formalism and the
usual picture of Regge poles in the crossed channel. 

The results of Refs.~\cite{Jani,reggeon_duality} are therefore in
qualitative agreement with the phenomenology. Nevertheless, two points
are left unclear. First of all, although the authors of Ref.~\cite{Jani} 
give reasonable arguments for the validity of the proposed expression
for the Reggeon-exchange amplitude, they do not provide a direct
derivation from first principles. In particular, they do not take into
account the fact that the fermions involved in the Reggeon-exchange
process are partons inside of a hadron. The more detailed discussion
of Ref.~\cite{reggeon_duality} mentions these problems, but does not
provide a direct derivation either. The second issue is the use of
analytic continuation to Euclidean space. The analytic-continuation
relation used in Refs.~\cite{Jani,reggeon_duality} is the one which
has been proved to be correct for Pomeron 
exchange~\cite{Meggiolaro97,Meggiolaro98,Meggiolaro02,Meggiolaro05,
  crossing,crossing2,EMduality}. Although it seems reasonable that it 
should work also in the Reggeon-exchange case, this is not guaranteed
{\it a priori}.

The aim of this paper is precisely to clarify these two points. We
provide a derivation of the Reggeon-exchange amplitude in the high 
energy, low transferred-momentum limit, in the framework of QCD,
considering in particular the elastic scattering of two mesons.   
Using the partonic description of hadrons and the LSZ reduction
approach discussed in~\cite{Nacht}, the meson-meson scattering
amplitude is reconstructed from the scattering amplitude of 
two colourless $q\barq$ dipoles, which in turn is decomposed into a
sum of terms corresponding to elastic and inelastic processes at the
partonic level. While Pomeron exchange corresponds to the
parton-elastic process, Reggeon exchange is identified with the
process in which two valence fermions are exchanged between the
interacting hadrons. Exploiting then the path-integral representation
for fermion propagators~\cite{Fradkin} in an external non-Abelian
gauge field~\cite{Brandt,Brandt1,Polyakov,Korchemsky,Korchemsky2}, we
reduce the corresponding amplitude to a path-integral over the
trajectories of the exchanged fermions of the (properly normalised)
expectation value of a certain Wilson loop. Finally, using the
techniques of~\cite{EMduality}, we show how the amplitude in Minkowski
space can be reconstructed from a Euclidean quantity by means of
analytic continuation, under appropriate analyticity assumptions. 

The plan of the paper is the following. In Section \ref{review} we
review the main assumptions and the techniques used in the
nonperturbative approach to {soft} high-energy scattering, in 
particular in the case of elastic meson-meson scattering. In Section
\ref{Pomeron} we rederive the Pomeron-exchange amplitude using the
path-integral representation for the fermion propagator in an external
non-Abelian gauge field. In Section \ref{amplitude} we apply similar
techniques in order to derive a nonperturbative expression for the
Reggeon-exchange amplitude. In Section \ref{an_cont} we derive the
analytic continuation relations which allow to reconstruct the
physical, Minkowskian amplitude from an appropriate Euclidean
quantity. In Section \ref{saddle} we make contact with the work of
Refs.~\cite{Jani,reggeon_duality}. Finally, in Section \ref{concl} we
discuss our conclusions and show some prospects for the future. 


\nsection{Nonperturbative approach to soft high-energy scattering}
\label{review}

\begin{figure}[t]
  \centering
    \includegraphics[width=0.6\textwidth]{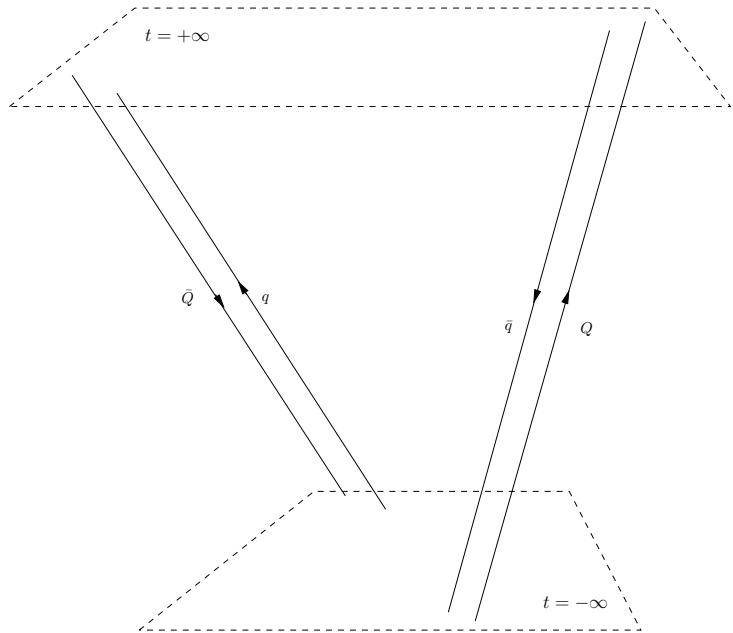}
  \caption{Space-time picture of the
    Pomeron-exchange process.} 
  \label{fig:0}
\end{figure}

In the {soft} high-energy regime, perturbation theory is not
completely reliable because of the presence of two different and
widely separated energy scales, and a genuine nonperturbative approach
is required. Such an approach has been proposed long ago in
Ref.~\cite{Nacht}, based on certain assumptions which can be justified
in the given energy regime. The basic idea is that, by choosing
appropriately the resolution for the hadronic wave functions, the
interacting hadrons can be reliably described in terms of partons over
a small time interval, during which the partonic state of the hadrons
does not change qualitatively, i.e., annihilation and production
processes can be neglected. Moreover, this time interval is chosen to
be also larger than the typical interaction time, so that the partons
can be approximately considered as good {\it in} (resp.~{\it out})
states at the beginning (resp.~at the end) of the small
time-window. One can then reconstruct the hadron-hadron scattering
amplitudes from the scattering amplitudes of partons, which in turn
can be expressed in terms of vacuum expectation values of products of
field operators through the LSZ reduction approach~\cite{LSZ1,LSZ2}.  

The next step is to evaluate the partonic amplitudes. Due to the high
energy of the interacting hadrons, partons which carry a finite
fraction of the longitudinal momentum of the hadrons travel
approximately on straight, almost lightlike trajectories; moreover,
since the transferred momentum is small, these trajectories are left
practically unchanged by the soft diffusion process (see
Fig.~\ref{fig:0}). The dominant contribution to the partonic
scattering amplitudes comes therefore from the elastic component,
which involves the exchange of soft gluons between the interacting
partons. Finally, by making use of an eikonal approximation for the 
parton propagators~\cite{Nacht,Nachtr,Meggiolaro96,Meggiolaro00}, it is
possible to obtain approximate (but nonperturbative) expressions for
the partonic amplitudes, in terms of the correlation function of
lightlike Wilson lines in the appropriate representation, running
along the classical trajectories of the partons. In the language of
Regge theory, the resulting amplitude should describe the exchange of
``Pomerons'' between the interacting particles, i.e., the exchange of
``Reggeons'' carrying the quantum numbers of the vacuum, and in the
following we will refer to it as the Pomeron-exchange amplitude. 

Our purpose in this paper is to investigate a subdominant contribution
to the high-energy scattering amplitude, involving the exchange
between the hadrons of a pair of valence partons, i.e., an inelastic
process at the partonic level. Since, according to the description
given above, partons which carry a finite fraction of the hadronic
momenta travel undisturbed along their classical, straight-line
trajectories, this kind of process can take place only when the
valence partons are ``wee'', i.e., only when they carry a vanishingly
small fraction of the longitudinal momentum. This would agree with
Feynman's description of high-energy processes~\cite{FeynH}, according
to which only ``wee'' partons can be exchanged between the scattering
hadrons. Clearly, the eikonal approximation cannot be used to describe
the propagation of the exchanged fermions, and different techniques
are required. Nevertheless, it remains a viable approximation for the 
``spectator'' partons, carrying a finite fraction of longitudinal
momentum. The scattering amplitude corresponding to this process
describes the exchange of a different ``Reggeon'' between the interacting
particles, which this time carries non-vacuum quantum numbers. In the
following we will refer to it simply as the Reggeon-exchange
amplitude (understanding that the Reggeon we refer to is not the
Pomeron, of course).

In the remaining part of this Section, we discuss in some detail the
decomposition of the hadronic amplitudes in terms of the partonic
ones, and the reduction of the latter by means of the LSZ
formula. Moreover, we introduce the path-integral formalism for the
propagators, which will be used in the following Sections in order to
obtain a representation of the partonic amplitudes in terms of Wilson
loops. For definiteness, we focus on the case of elastic
meson-meson scattering.

\subsection{Elastic meson-meson scattering}

The $S$-matrix element we want to evaluate is ($p_f\equiv p_1'+p_2'$,
$p_i\equiv p_1+p_2$)  
\begin{equation}
  \label{eq:S_meson}
 S_{fi}= \outs{M_1(p_1') M_2(p_2')} | \ins{M_1(p_1) M_2(p_2)}  =
 \delta_{fi} + i(2\pi)^4\delta^{(4)}(p_f-p_i) \A_{fi}\, , 
\end{equation}
where $M_{1,2}$ denote two mesons, which for simplicity are taken with
the following flavour content, $M_1=Q\barq, M_2=q\barQ$, and therefore
with the same mass $m$.   
Here $p^1_1\simeq p_1^0$, $p^1_2\simeq -p_2^0$,
since we are considering highly energetic mesons travelling in the
$x^1$ direction, and moreover $p_{1,2}'\simeq p_{1,2}$. More
precisely,\footnote{When indices are omitted, explicit expressions of
  Minkowskian four-vectors are given in terms of contravariant
  components. We adopt the ``mostly minus'' convention for the metric
  tensor.} 
\begin{equation}
  \label{eq:momenta}
  \begin{aligned}
    p_1 &= m\left(\cosh\f{\chi}{2},\sinh\f{\chi}{2},\vec{0}_\perp\right) \equiv
    mu_1\, ,\\ 
    p_2 &= m\left(\cosh\f{\chi}{2},-\sinh\f{\chi}{2},\vec{0}_\perp\right) \equiv
    mu_2\, , \\
    q &= p_2'-p_2 = p_1-p_1' \simeq (0,0,\vec{q}_\perp)\,,
  \end{aligned}
\end{equation}
where $\chi$ is the hyperbolic angle between the classical
trajectories of the mesons,
\begin{equation}
  \label{eq:hyp_angle}
  \cosh\chi = \f{s}{2m^2}-1\,.
\end{equation}
In the high-energy limit we are interested in, $\chi$ is large and
approximately equal to $\chi \simeq \log(s/m^2)$. In particular,
$(p_i'-p_i)\cdot u_i \simeq 0$, $i=1,2$, where the dot stands for the
Minkowskian scalar product.\footnote{We use timelike momenta,
  appropriate for massive partons, rather than lightlike momenta as in
  the original derivation of~\cite{Nacht}, in order to regularise from
  the onset the problem of infrared divergencies~\cite{Verlinde}.}     

We adopt a simple description of the mesons as superpositions of
colourless $q\barq$ dipoles \cite{DFK,Nachtr,Dosch} (see
also~\cite{Muel,Muel2,Muel3}); 
after the evaluation of the dipole-dipole scattering amplitude, the
mesonic amplitude is reconstructed by folding with the appropriate
wave functions. In a first approximation, we neglect the gluonic
component of the wave functions, and so we do not consider the case in
which gluons carrying a finite fraction of the meson momenta take part
in the process. The approach can however be generalised to take into
account these contributions. Alternatively, the present approach can
be seen as a description of mesons in terms of {\it constituent}
$q\barq$ dipoles, with gluonic and sea-quark contributions included in
the wave functions; of course, the meson wave functions would be
different in the two cases. We then describe the mesons as follows:
\begin{equation}
  \label{eq:mesons}
  \begin{aligned}
    & \ket{M_1(p_1)} = \int d\mu_1\, \ket{d_1(\mu_1)} \,,
    && && \ket{M_2(p_2)} = \int d\mu_2\, \ket{d_2(\mu_2)} \,,
  \end{aligned}
\end{equation}
where we have introduced the dipole states
\begin{equation}
  \label{eq:dipoles}
  \begin{aligned}
&    \ket{d_1(\mu_1)} =  \f{1}{\sqrt{N_c}}
\sum_{i,j}\delta_{i j} \ket{Q_{s_Q i}(p_{Q})\, 
      \barq_{t_\barq j}(p_{\barq})}\, , \quad \mu_1 =
    (p_{Q},s_Q;p_{\barq},t_\barq)\, ,\\
& p_Q = \left(\zeta_1 p_1^0,\zeta_1 p_1^1,
      \f{\vec p_{1\perp}}{2}+\vec{k}_{1\perp}\right)\, 
    , \quad p_\barq = p_1 - p_Q\,,\\
&    \ket{d_2(\mu_2)}= \f{1}{\sqrt{N_c}}
\sum_{i,j}\delta_{i j}\ket{q_{s_q i}(p_{q})\,\barQ_{t_\barQ
    j}(p_{\barQ})}\, , \quad \mu_2 = 
(p_{q},s_q;p_{\barQ},t_\barQ)\, , \\
& p_q = \left(\zeta_2 p_2^0,\zeta_2 p_2^1,
    \f{\vec p_{2\perp}}{2}+\vec{k}_{2\perp}\right)\, , \quad p_\barQ = p_2
  - p_q\, .
  \end{aligned}
\end{equation}
Similar relations hold for the meson states with primed variables. 
Here the quarks $q$ and $Q$ have (Lagrangian) masses $m_q$ and $m_Q$
respectively, the quark and antiquark states are normalised according
to the relativistic normalisation, 
\begin{equation}
  \label{eq:rel_norm}
  \begin{aligned}
    \la X(\vec{p}_X^{\,\prime},s_X',i') | X(\vec{p}_X,s_X,i) \ra &=
    \delta_{s_X's_X^{\phantom{\prime}}}\delta_{i'i}(2\pi)^3 2p_X^0 
    \delta^{(3)}(\vec{p}_X^{\,\prime} - \vec{p}_X) \equiv \delta_X\,,
     \\
  \la \bar{X}(\vec{p}_{\bar{X}}^{\,\prime},t_{\bar{X}}',j') |
  \bar{X}(\vec{p}_{\bar{X}},t_{\bar{X}},j) \ra &= 
    \delta_{t_{\bar{X}}'t_{\bar{X}}^{\phantom{\prime}}}\delta_{j'j}(2\pi)^3 2p^0_{\bar{X}} 
    \delta^{(3)}(\vec{p}_{\bar{X}}^{\,\prime} - \vec{p}_{\bar{X}})
    \equiv \delta_{\bar{X}}\,, \quad X=q,Q\,,
  \end{aligned}
\end{equation}
$s_{q,Q}$ and $t_{\barq,\barQ}$
are spin indices and $i$ and $j$ are colour indices,
$\zeta_{1,2}\in[0,1]$ are the longitudinal momentum fractions of the
quarks, 
and we have introduced the measure
\begin{equation}
  \label{eq:measure}
  \begin{aligned}
\int d\mu_1 f(\mu_1) &\equiv  \int d^2 k_{1\perp} \int_0^1 d\zeta_1
\sum_{s_Q,t_\barq} \psi_{1\,  s_Q t_\barq}(\vec{k}_{1\perp},\zeta_1) 
f(p_{Q},s_Q;p_{\barq},t_\barq) \,,\\
\int d\mu_2 f(\mu_2) &\equiv  \int d^2 k_{2\perp} \int_0^1 d\zeta_2
\sum_{s_q,t_\barQ} \psi_{2\,  s_q t_\barQ}(\vec{k}_{2\perp},\zeta_2)  
f(p_{q},s_q;p_{\barQ},t_\barQ) \,, 
  \end{aligned}
\end{equation}
where $\psi_i$ are the mesonic wave functions. 
In order for the meson states to have relativistic normalisation,
\begin{equation}
  \la M_i(\vec{p}_i^{\,\prime}) | M_i(\vec{p}_i) \ra = (2\pi)^3 2p^0_i
    \delta^{(3)}(\vec{p}^{\,\prime}_i - \vec{p}_i)\,, \quad i=1,2\,,
\end{equation}
 we need the wave functions to be normalised as 
\begin{equation}
  \label{eq:wf_norm}
\coth\f{\chi}{2}
\int d^2 k_{\perp} \int_0^1 d\zeta \sum_{s,t} (2\pi)^3
  2\zeta(1-\zeta)  |\psi_{i\,
    s t}(\vec{k}_{\perp},
    \zeta)|^2 = 1\,,\quad i=1,2\, .
\end{equation}
For later convenience we define also
\begin{equation}
  \label{eq:wf_coord}
  \wf_{i\,st}(\vr_\perp,\zeta) = 
\sqrt{\coth\f{\chi}{2}} \sqrt{2\zeta(1-\zeta)2\pi}\int 
  d^2 k_\perp
  e^{i\vec{k}_\perp\cdot\vr_\perp}\psi_{i\,st}(\vec{k}_\perp, \zeta)\,, \quad
  i=1,2\, ,
\end{equation}
which has the simple normalisation
\begin{equation}
  \label{eq:norm_wf_coord}
   \int d^2 \vr_\perp \int_0^1 d\zeta \sum_{s,s'} |\wf_{i\,ss'}(\vr_\perp,
    \zeta)|^2 = 1\,, \quad i=1,2\, .
\end{equation}
The factor $\coth\f{\chi}{2}$ is practically $1$ at large $\chi$, and
we will often ignore it. In terms of the dipole states, the matrix
element Eq.~\eqref{eq:S_meson} is then rewritten as
\begin{equation}
  \label{eq:S_meson_2}
  \begin{aligned}
   S_{fi}=& \int d\mu_1'{}^* \int d\mu_2'{}^*
   \int d\mu_1 \int d\mu_2 \,
     \outs{d_1(\mu_1')d_2(\mu_2')} |
   \ins{d_1(\mu_1)d_2(\mu_2)} \\ 
   \equiv & \int d\mu_1'{}^* \int d\mu_2'{}^*
   \int d\mu_1 \int d\mu_2 \,
     S^{(dd)}_{fi}(\mu_1,\mu_2,\mu_1',\mu_2')\, ,
  \end{aligned}
\end{equation}
where the ``conjugate'' measure is defined to be
\begin{equation}
  \label{eq:measure*}
   \int d\mu_1'{}^* f(\mu_1') =  \int d^2 k_{1\perp}' \int_0^1 d\zeta_1'
  \sum_{s_Q', t_\barq'} \psi^*_{1\, s_Q'
      t_\barq'}(\vec{k}_{1\perp}',\zeta_1')f(p_{Q}',s_Q';p_{\barq}',t_\barq') \, ,
\end{equation}
and similarly for $d\mu_2'{}^*$. 

\subsection{LSZ reduction}

The next step is the application of the LSZ reduction formulas to
$S^{(dd)}_{fi}$. Although, as it is well known, there are no true
asymptotic quark or antiquark states, to which the LSZ reduction
scheme can be strictly applied, such an approach is reasonable in the
picture described above. Indeed, as we have already remarked, the 
size of the time-window $[-t_0,t_0]$ at interaction time is
determined by two conditions: that partons are approximately
well-defined particles inside of it (i.e., splitting and annihilation
processes can be neglected over $2t_0$); and that $t_0$ is large
enough for partons in different mesons to be distant from one another
at $\pm t_0$ (i.e., at the beginning and at the end of the
interactions), so that they can be considered non-interacting
asymptotic states at $\pm t_0$. However, this does not apply to the
$q$ and $\barq$ belonging to same dipole: as we will see, this will
require a correction ``by hand'' of the amplitudes in order to make
the result sensible. 

In what follows we will use a functional-integral
representation of the $T$-ordered vacuum expectation values of
operators appearing in the LSZ formulas, namely
\begin{equation}
  \label{eq:func_int}
  \br T\{{\cal O}_1[\boldsymbol{\psi},\mathbf{A}]\ldots
  {\cal O}_n[\boldsymbol{\psi},\mathbf{A}]\} \ke =  
\la\la {\cal O}_1[\psi, A] \ldots {\cal O}_n[\psi, A] \ra_\psi\ra_A\,,
\end{equation}
where boldface symbols denote operators, and the fermionic and gluonic
expectation values are defined as
\begin{equation}
  \label{eq:func_int_2}
  \begin{aligned}
    \la {\cal O}[\psi,A] \ra_\psi &= \f{\displaystyle\int [{\cal D}\psi {\cal
        D}\bar{\psi}]e^{iS_{\rm ferm}[\psi,A]}{\cal O}[\psi,A]}{\displaystyle\int
      [{\cal D}\psi {\cal D}\bar{\psi}]e^{iS_{\rm ferm}[\psi,A]}} \,,\\
 \la {\cal O}[A] \ra_A &= \f{\displaystyle\int [{\cal D}A] \det {\cal
     Q}[A]e^{iS_{\rm YM}[A]}{\cal O}[A]}{\displaystyle\int 
      [{\cal D}A]e^{iS_{\rm YM}[A]}\det {\cal Q}[A]}\, ,
  \end{aligned}
\end{equation}
where $S_{\rm ferm}$ and $S_{\rm YM}$ are respectively the fermionic
and pure-gauge part of the action, and $\det {\cal Q}[A]$ is the
fermion-matrix determinant, 
\begin{equation}
  \label{eq:ferm_mat}
  \det {\cal Q}[A] = \int
      [{\cal D}\psi {\cal D}\bar{\psi}]e^{iS_{\rm ferm}[\psi,A]}\, .
\end{equation}
Performing the reduction, in which we keep all the disconnected terms,
we find the following expression for the dipole-dipole $S$-matrix
element $S^{(dd)}_{fi}$ (see Eq.~\eqref{eq:S_meson_2}),
\begin{equation}
  \label{eq:S_dip_short}
    S_{fi}^{(dd)} = {\cal P}^{(dd)} + {\cal R}_1^{(dd)} + {\cal
      R}_2^{(dd)} + {\cal E}^{(dd)} \, ,
\end{equation}
where the various contributions are given by the gluonic expectation values
\begin{equation}
  \label{eq:S_mes_final2}
  \begin{aligned}
{\cal P}^{(dd)}(\mu_1,\mu_2,\mu_1',\mu_2') &=
  \la\left(\hS{Q}+\dl{Q}\right)\left(\hS{\barQ}+\dl{\barQ}\right)  
  \left(\hS{\barq}+\dl{\barq}\right)\left(\hS{q}+\dl{q}\right)\ra_A \,,\\ 
{\cal R}_1^{(dd)}(\mu_1,\mu_2,\mu_1',\mu_2')  &= 
  \la\left(\hS{Q}+\dl{Q}\right)\left(\hS{\barQ}+\dl{\barQ}\right)  
  V^+_{q\barq}V^-_{\barq q} \ra_A \,,\\ 
{\cal R}_2^{(dd)}(\mu_1,\mu_2,\mu_1',\mu_2') &=
  \la  V^+_{Q\barQ}V^-_{\barQ Q}
  \left(\hS{\barq}+\dl{\barq}\right)\left(\hS{q}+\dl{q}\right)\ra_A  \,, \\
{\cal E}^{(dd)}(\mu_1,\mu_2,\mu_1',\mu_2')  &= \la
V^+_{Q\barQ}V^-_{\barQ Q} V^+_{q\barq}V^-_{\barq q} \ra_A  \, .
\end{aligned}
\end{equation}
The symbols $\delta_X$ and $\delta_{\bar{X}}$, $X=q,Q$, have been defined in
Eq.~\eqref{eq:rel_norm}, and we have denoted with $\hS{Q}$ and
$\hS{\barQ}$ the truncated-connected propagators in momentum space for
$Q$ and $\barQ$, respectively, contracted with the appropriate Dirac
spinors, 
\begin{equation}
  \label{eq:defin}
  \begin{aligned}
(\hS{Q})_{i'i} =& \lim_{p_Q^{\prime 2}\to \m_Q^{\prime 2}} \lim_{p_Q^2\to\m_Q^2} 
\f{1}{Z_Q}\int d^4y\int d^4x\, e^{ip_Q'\cdot y -ip_Q\cdot x} \\ &
\phantom{\lim_{p_Q^{\prime 2}\to \m_Q^{\prime 2}}}
\times \bar{u}^{s_Q'}(p_Q') \, 
    \f{\slap_Q'-\m_Q'}{i}
    \la {Q}_{i'}(y){\barQ}_i(x)\ra_\psi 
    \f{\slap_Q-\m_Q}{i} \,u^{s_Q}(p_Q)  \,,    
\\
    (\hS{\barQ})_{j'j} =& \lim_{p_\barQ^{\prime 2}\to \m_\barQ^{\prime 2}} 
    \lim_{p_\barQ^2\to \m_\barQ^2}  
\f{1}{Z_Q}\int d^4y\int d^4x\, e^{ip_{\barQ}'\cdot y
  -ip_{\barQ}\cdot x}\\ & \phantom{\lim_{p_\barQ^{\prime 2}\to \m_\barQ^{\prime
      2}} } \times \bar{v}^{t_\barQ}(p_{\barQ})\,  
    \f{\slap_{\barQ}+\m_{\barQ}}{i}\la {\barQ}_{j'}(y){Q}_j(x)\ra_\psi
    \f{\slap_{\barQ}'+\m_{\barQ}'}{i}\, v^{t_\barQ'}(p_{\barQ}')\, ;
  \end{aligned}
\end{equation}
completely analogous expressions hold for the $q$
and $\barq$ propagators. 
Here $Z_Q$ is the renormalisation constant entering the
LSZ reduction formula, and we have denoted with $\m_Q$ (resp.~$\m_Q'$) the
``physical'' mass of $Q$ in the initial (resp.~final) states, which we
identify with the ``constituent masses'' in the dipole states, 
$\m_Q\equiv\zeta_Q m$, and similarly for the other terms. As usual,
$\slaA=A_\mu\gamma^\mu$, with $\gamma^\mu$ the Dirac matrices. 
The bispinors are normalised as
\begin{equation}
\label{eq:bisp_norm}
  \bar{u}^{s_X'}({p}_X)u^{s_X}({p}_X) = 2\tilde{m}_X\delta_{s_X's_X}, \quad
  \bar{v}^{t_{\bar{X}}}({p}_{\bar{X}})v^{t_{\bar{X}}'}({p}_{\bar{X}})
  = -2\tilde{m}_{\bar{X}}\delta_{t_{\bar{X}}'t_{\bar{X}}}\, . 
\end{equation}
Moreover, $V^+_{q\barq}$ and $V^-_{\barq q}$  are the terms
which describe the exchange of fermions $q$ and $\barq$ between the
two mesons, 
\begin{equation}
  \begin{aligned}
   (V^+_{q\barq})_{i'j'} =& \lim_{p_\barq^{\prime 2}\to \m_\barq^{\prime 2}}
   \lim_{p_q^{\prime 2}\to \m_q^{\prime 2}}
\f{1}{Z_q}\int d^4y\int d^4x\, e^{ip_q'\cdot
  y+ip_\barq'\cdot x} \\ & 
\phantom{\lim_{p_\barq^{\prime 2}\to \m_\barq^{\prime 2}}}
\times \bar{u}^{s_q'}(p_q') \, 
    \f{\slap_q'-\m_q'}{i}
\la q_{i'}(y)\barq_{j'}(x)\ra_\psi
     \f{\slap_\barq'+\m_\barq'}{i}\, v^{s_\barq'}(p_\barq')\,,
\\
    (V^-_{\barq q})_{ji} =&  \lim_{p_\barq^2\to \m_\barq^2}  \lim_{p_q^2\to \m_q^2}
\f{1}{Z_q}\int d^4y\int d^4x\, e^{-ip_\barq\cdot y
  -ip_q\cdot x}\\ & 
\phantom{\lim_{p_\barq^2\to \m_\barq^2} }
\times \bar{v}^{s_\barq}(p_\barq)\,  
    \f{\slap_\barq+\m_\barq}{i}
\la q_j(y) \barq_i(x) \ra_\psi
    \f{\slap_q-\m_q}{i} \,u^{s_q}(p_q)
\, ;
  \end{aligned}
\end{equation}
similar expressions hold for $V^+_{Q\barQ}$ and $V^-_{\barQ Q}$. 
In the following we will also use the notation
\begin{equation}
  \label{eq:pom_integ}
  \begin{aligned}
  {\cal P} &= \int d\mu_1\,d\mu_2\,d\mu_1'{}^*\,d\mu_2'{}^* \,{\cal
    P}^{(dd)}(\mu_1,\mu_2,\mu_1',\mu_2') \, ,  \\
 {\cal R}_{1,2} &= \int d\mu_1\,d\mu_2\,d\mu_1'{}^*\,d\mu_2'{}^* \,{\cal
    R}_{1,2}^{(dd)}(\mu_1,\mu_2,\mu_1',\mu_2') \, , 
  \end{aligned}
\end{equation}
to indicate the contributions to the scattering amplitudes obtained by
folding the dipole-dipole scattering-matrix elements with
the mesonic wave functions.\footnote{The remaining term, coming from
  the integration of ${\cal E}^{(dd)}$, and corresponding to the
  exchange of both the valence fermions between the interacting
  mesons, will not be considered in this paper.}

The two terms in Eq.~\eqref{eq:pom_integ} have a clear
interpretation. The term ${\cal P}$ describes a process in which the
interaction between the mesons is mediated by the gluon field; it
corresponds to Pomeron exchange, and it is the dominant one at high
energy (see Fig.~\ref{fig:0}). The terms ${\cal R}_i$ 
describe a process in which the mesons exchange also a $q\barq$ pair in
the $t$-channel, and they correspond to the exchange of a Reggeon with
non-vacuum quantum numbers (see Fig.~\ref{fig:1}). 
In perturbation theory, diagrams corresponding to these terms contain
fermion lines with large momentum flow, of order ${\cal
  O}(\sqrt{s})$, and are therefore suppressed with respect to
diagrams where only gluons are exchanged. Since there are at least two
such fermion lines, one expects a suppression of order  ${\cal
  O}(1/s)$ of Reggeon exchange with respect to Pomeron exchange (see
also Ref.~\cite{EN,EN2} for a similar argument in the case of $\gamma^*
p\to\gamma^* p$ scattering).

\subsection{Path-integral
representation for the fermion propagator}

It is convenient at this point to introduce the path-integral
representation for the propagators in the 
first-quantised theory~\cite{Fradkin}. Using
the proper-time representation of
propagators~\cite{Fock,Nambu,Schwinger} in the  
case of a fermion in an external non-Abelian gauge
field~\cite{Brandt,Brandt1,Polyakov,Korchemsky,Korchemsky2}, one
obtains\footnote{This 
  representation is known to be only formal, and that it requires an 
  appropriate regularisation in order to become fully
  meaningful~\cite{Brandt,Brandt1}. The regularised expression in Euclidean
  space allows for the explicit integration over
  momenta~\cite{Korchemsky,Korchemsky2}, but a similar result does not exist in
  Minkowski space. This is a very important issue, which is however
  beyond the scope of this paper. The formal manipulations of
  Minkowskian path integrals in the following Sections are therefore a
  heuristic procedure, but the resulting path integral will acquire a
  precise mathematical meaning when introducing the analytic
  continuation to Euclidean space.} 
\begin{equation}
  \label{eq:path_int_rep}
  \begin{aligned}
\la Q_{\alpha i}(y)\barQ_{\beta j}(x) \ra_\psi &=  \la y |
  \f{i}{i\sla{D}-m_Q+i\epsilon} |x\ra 
\\ &=
  \int_0^\infty d\nu
  e^{-i(m_Q-i\epsilon)\nu}\int_{X(0)=x}^{X(\nu)=y} \DX\,
  (\Sp_{0,\nu}[\dot{X}])_{\alpha\beta}\,
  ({W}_{0,\nu}[X]  )_{ij}\,,
  \end{aligned}
\end{equation}
where $D_\mu = \de_\mu + igA_\mu$ is the covariant derivative,
$\Sp_{0,\nu}$ is the ``spin factor''\footnote{The T-ordered
  exponential is defined as 
$$  \Texp\big\{\textstyle\int dt\, f(t)\big\} =
\textstyle\sum_{n=0}^{\infty} \textstyle\int 
  dt_1 \ldots \textstyle\int  dt_n \,\Theta(t_1-t_2) \ldots
  \Theta(t_{n-1}-t_n)\, 
  f(t_1)\ldots f(t_n)\,, $$
with $\Theta(x)$ the Heaviside step function, i.e., larger time
appears on the left.} 
\begin{equation}
  \label{eq:defin_2}
  \begin{aligned}
    &\Sp_{0,\nu}[\dot{X}] = \int \DP\, {\cal M}_{0,\nu}[\dot{X},\Pi]\, ,\\
    &{\cal M}_{\eta,\nu}[\dot{X},\Pi] = \Texp\left[i\int_\eta^\nu d\tau
      \left(\sla{\Pi}(\tau) - \Pi(\tau)\cdot\dot{X}(\tau)\right)\right]\, 
    ,
  \end{aligned}
\end{equation}
and ${W}_{0,\nu}$ is the Wilson line
\begin{equation}
  {W}_{\eta,\nu}[X] = \Texp\left[-ig \int_\eta^\nu d\tau
    A\big(X(\tau)\big) \cdot
    \dot{X}(\tau) \right]\, .
\end{equation}
The measure of the unconstrained path-integral over paths $X(\tau)$ in
coordinate space, $\DX$, and over paths $\Pi(\tau)$ in momentum space,
$\DP$, is defined as 
\begin{equation}
  \label{eq:PImeasdef}
  \int \DX \int \DP = \lim_{N\to\infty}\int d^4X_1\ldots \int
  d^4X_{N+1} \int \f{d^4\Pi_1}{(2\pi)^4}\ldots \int
  \f{d^4\Pi_{N-1}}{(2\pi)^4}\,; 
\end{equation}
the measure for paths satisfying $X(0)=x$, $X(\nu)=y$ is obtained by
inserting the delta functions $\dlx{4}(X(0)-x)\dlx{4}(X(\nu)-y)$ in
Eq.~\eqref{eq:PImeasdef}. 

In order to obtain the truncated propagators relevant to the
LSZ reduction formulas, we use the trick proposed in~\cite{Fabbr}
(based on a result of~\cite{Mile}) for the scalar propagator, which is
easily generalised to the case of the fermion propagator:
\begin{equation}
  \label{eq:trick}
  \begin{aligned}
Z_Q \hS{Q} =
\Lim \f{1}{\nu_f-\nu_i} \int & \DX \,
  e^{ip_Q'\cdot X(\nu_f)-ip_Q\cdot
    X(\nu_i)} \\ \times
  &\bar{u}^{s_Q'}(p') \Sp_{\nu_i,\nu_f}[\dot{X}]{W}_{\nu_i,\nu_f}[X]
  u^{s_Q}(p)    +  {\rm disc.}\,,
  \end{aligned}
\end{equation}
where $\Lim =\lim_{\nu_f\to\infty,\nu_i\to -\infty}\lim_{p_Q^2\to
  \m_Q^2,p_Q^{\prime 2}\to \m_Q^{\prime 2}}$, and where we have
omitted a disconnected term which will be 
reinserted when needed. Similar expressions hold for the other truncated
propagators and for the fermion-exchange terms. These expressions
will be given in the next Section, where we re-derive the
Pomeron-exchange amplitude in a very direct way, and in the following
Section where we derive the Reggeon-exchange amplitude. 

\begin{figure}[t]
  \centering
    \includegraphics[width=0.6\textwidth]{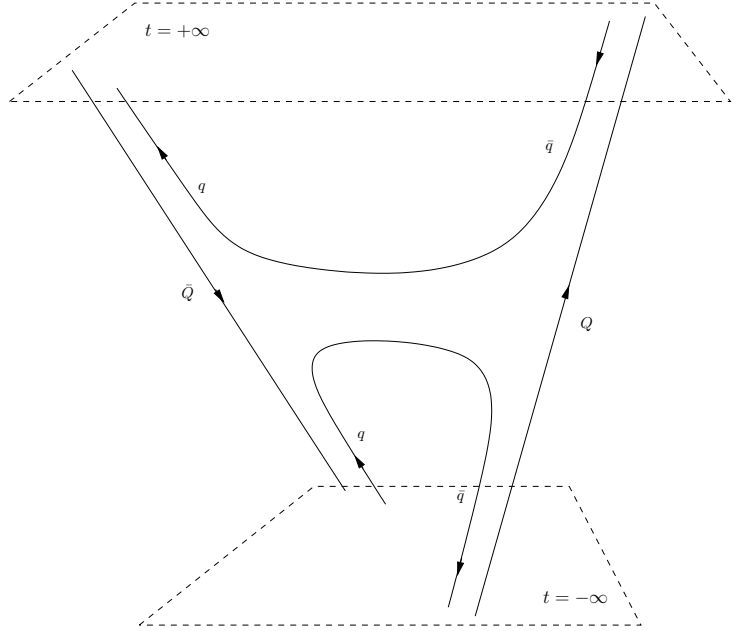}
  \caption{Space--time picture of the
    Reggeon--exchange process.} 
  \label{fig:1}
\end{figure}

\nsection{Pomeron exchange}
\label{Pomeron}

The term ${\cal P}$ of Eq.~\eqref{eq:pom_integ}, corresponding to
Pomeron exchange, has been already evaluated in the eikonal
formalism~\cite{DFK,Nachtr,BN,Dosch,LLCM1}, and it has been
investigated in many
papers~\cite{LLCM2,ILM,instantons,Jani1,adsbounds,Jani2,Janik,Kharzeev, 
lattice,lat_pomeron}. The main building block is the
truncated-connected fermion propagator in an external field, which can
be easily evaluated in an eikonal approximation using the
path-integral representation described in the previous
section. Indeed, when the initial and final momenta are almost 
lightlike, and moreover $p'\simeq p$, the classical straight-line
trajectory is expected to give the dominant contribution to the
path-integral. Consider for example $\hS{Q}$. As we show in Appendix 
\ref{app:straight}, approximating the integral with the contribution
from the classical trajectory only, 
\begin{equation}
  \label{eq:eik_traj}
  X(\tau) = b_Q + u_1\tau,\quad \Pi(\tau) = \m_Q u_1\, ,
\end{equation}
where $u_1$ has been defined in Eq.~\eqref{eq:momenta}, we obtain
\begin{equation}
  \label{eq:eikonal_3}
  \begin{aligned}
   Z_Q\hS{Q} + \delta_Q = 
 \delta_{s_Q's_Q}\, 2\sqrt{\m_Q\m_Q'}e^{i(\m_Q'+\m_Q-2m_Q)T}\int d^3b_Q\, 
  e^{iq_Q\cdot b_Q}\, {W}_{u_1}(b_Q)\, ,
\end{aligned}
\end{equation}
where $\delta_Q$ has been defined in Eq.~\eqref{eq:rel_norm}. 
The ``physical'' mass $\m_Q$ (resp.~$\m_Q'$) of quark $Q$ in the
initial (resp.~final) state is identified with the fraction of meson
mass carried by the quark, $\m_Q^{(\prime)}=\zeta^{(\prime)}m$ (see
also after Eq.~\eqref{eq:defin}). Here ${W}_{u_1}(b_Q)$ is a
straight-line Wilson line of length $2T$, parallel to $u_1$ and
centered at $b_Q$, and $q_Q=p_Q'-p_Q$ with $q_Q\cdot u_1 \simeq
0$. The length of the Wilson line is kept finite in order to
regularise IR divergencies~\cite{Verlinde}, and it has to be sent to
infinity at the end of the calculation. The integration measure
$d^3b_Q=db_Q^1d^2b_{Q\perp}$ includes only the coordinate along the
directions orthogonal to $u_1$ (in Minkowski metric); in other words,
$d^3b_Q$ are the spatial coordinates in the rest frame of the
particle. Notice that the coordinate along the direction $u_1$ of the
position of the center is irrelevant when $T$ is large. Except for the
presence of an extra phase, the difference from previous 
calculations~\cite{Nacht,Meggiolaro96,Meggiolaro00} is only apparent,  
and due to the fact that we are keeping here the trajectory of the
fermion slightly away from the light-cone. In Appendix
\ref{app:straight} we show how the two results are reconciled in the
high-energy limit.   

The appearence of the phase factor is due to the fact that we cannot
neglect completely the masses of the mesons and of the fermions:
indeed, although negligible when compared to the energy, in the phase
factor they appear multiplied by $T$, which has to be taken to
infinity at the end of the calculation. The form of the phase factor
suggests that it corresponds to the self-interaction of the
propagating fermion: when describing the scattering of mesons, this
self-interaction should play no role, since it is part of the
internal mesonic interactions, and it has therefore to be subtracted.
We will return on this point later on.

A remark is in order, concerning the identification of $\tau$ with the
proper time along the path, which is implicit in the
expression Eq.~\eqref{eq:eik_traj} for the saddle point. Although it
is not a proof, the consistency of the result Eq.~\eqref{eq:eikonal_3}
with the ones already present in the literature indicates that this
identification is actually correct for timelike paths, for which
proper time is well-defined; this is enough for our purposes, since
in this paper we deal with timelike or ``mostly timelike'' paths in
Minkowski space. In the Euclidean case, the integration
over momenta in the expression Eq.~\eqref{eq:defin_2} for the spin
factor can be explicitly performed~\cite{Korchemsky,Korchemsky2}, and the result
identifies the parameter $\tau$ as the natural parameter along the
curve, defined through $\dot{x}^2=1$ (in Euclidean metric), which is
in a sense the Euclidean analogue of proper time. Extending the
analogy to spacelike 
paths in Minkowski space, we are led to expect that $\tau$ is in that
case the ``proper-space'' defined by $\dot{x}^2=-1$; however, 
further work is needed to clarify the
meaning of the parameter $\tau$ in the general case. 

We discuss now briefly the derivation of the Pomeron-exchange
amplitude in terms of Wilson loops~\cite{DFK}, which is discussed in
detail in~\cite{Nachtr}. Since we are neglecting splitting and
annihilation processes, we have approximately $Z_Q\simeq
1$~\cite{Nachtr}. Moreover, denoting with $u_{1}^\perp$ the
longitudinal direction orthogonal to $u_1$, which in the
center-of-mass frame reads 
\begin{equation}
  u_{1}^\perp = 
\left(\sinh\f{\chi}{2},\cosh\f{\chi}{2},\vec{0}_\perp\right)\,,
\end{equation}
we have
\begin{equation}
    (q\cdot u_{1}^\perp)\, b^1_Q = \left[q\cdot\left(\coth\chi\, u_1 -
    \f{1}{\sinh\chi}\, u_2\right)\right] b^1_Q =  - (q\cdot u_2)\,  \f{
  b^1_Q}{\sinh\chi}\,, 
\end{equation}
and so, after the change of variables $z_Q=\f{b^1_Q}{\sinh\chi}$ for
the longitudinal coordinate, we can write 
\begin{equation}
  \label{eq:ferm_prop_Z=1}
  \begin{aligned}
  \hS{Q} + \delta_Q = \delta_{s_Q's_Q}\, 2\sqrt{\m_Q\m_Q'}\, & \sinh  \chi \,
  e^{i(\m_Q'+\m_Q-2m_Q)T} \\ &\times\int dz_Q \int d^2b_{Q\perp} \,
  e^{i q_Q \cdot (-z_Q u_2 + b_{Q\perp})}\,
  {W}_{u_1}(-z_Qu_2+b_{Q\perp})\, ,
  \end{aligned}
\end{equation}
where $b_{Q\perp} = (0,0,\vec{b}_{Q\perp})$.
The expressions for the other eikonal propagators are readily
obtained, and in particular we have for antiquark $\barQ$
\begin{equation}
  \label{eq:anti_ferm_prop_Z=1}
  \begin{aligned}
    \hS{\barQ} + \delta_\barQ = \delta_{t_\barQ' t_\barQ}\,
    2\sqrt{\m_\barQ\m_\barQ'}\, &\sinh \chi \,
    e^{i(\m_\barQ'+\m_\barQ-2m_Q)T} \\ &\times\int 
    dz_\barQ \int d^2b_{\barQ\perp}\,
    e^{i q_\barQ \cdot (-z_\barQ u_1 + b_{\barQ\perp})}\,
    {W}_{u_2}^*(-z_\barQ u_1+b_{\barQ\perp})\, , 
  \end{aligned}
\end{equation}
which is exactly the same expression as Eq.~\eqref{eq:ferm_prop_Z=1}
with the Wilson line changed from the fundamental to the
complex-conjugate representation, and with the roles of $u_1$ and
$u_2$ interchanged. Here $b_{\barQ\perp} = (0,0,\vec{b}_{\barQ\perp})$. 

One has now to substitute the eikonal propagators in the expression
for ${\cal P}$, and to perform the remaining integrals. The
calculation is straightforward but quite lengthy, and we skip here the
details of the derivation, which is easily adapted from~\cite{Nachtr}, 
taking into account that the Wilson lines are now timelike, rather
than lightlike. We only mention a point which will be useful in the
following discussion and in the study of the Reggeon-exchange case,
concerning the integration over the longitudinal
coordinates. Discarding the variables which are not relevant here, we
have to perform the integral
\begin{equation}
  \begin{aligned}
 I_{\cal P}=   \int dz_Q \int dz_\barq \int dz_q &\int dz_\barQ  \, 
e^{-i(q_Q \cdot  u_2 z_Q +q_\barq \cdot  u_2 z_\barq + q_q \cdot  u_1
  z_q +q_\barQ \cdot  u_1 z_\barQ)} \\ & \times \la {W}_{u_1}(-z_Qu_2)
 {W}_{u_1}^*(-z_\barq u_2) {W}_{u_2}(-z_qu_1) {W}_{u_2}^*(-z_\barQ
 u_1)\ra_A\, .
  \end{aligned}
\end{equation}
The Wilson lines are cut off at some proper-times $\pm T_i$,
$i=Q,\barq,q,\barQ$, which are {\it a priori} unrelated. Exploiting
the invariance of $W_{u_i}$ under translations along the longitudinal
coordinate parallel to $u_i$ (which, strictly speaking, holds in the
limit of infinite length), and the invariance of the expectation value
under translations, we can rewrite this integral as
\begin{equation}
  \begin{aligned}
 I_{\cal P}=    \int dz_Q \int dz_\barq & \int dz_q \int dz_\barQ \, 
e^{-i[(q_Q+ q_\barq) \cdot  u_2 z_Q +q_\barq \cdot  u_2 (z_\barq- z_Q)
  + q_q \cdot  u_1 (z_q - z_\barQ) + (q_\barQ + q_q) \cdot  u_1 z_\barQ]} \\ &
\times \la {W}_{u_1}(0 \cdot u_2) 
 {W}_{u_1}^*((z_Q-z_\barq) u_2) {W}_{u_2}((z_\barQ-z_q)u_1)
 {W}_{u_2}^*(0 \cdot u_1)\ra_A\, ,
  \end{aligned}
\end{equation}
and changing variables to $z_Q,z_\barq \to z_Q,z_1=z_\barq-z_Q$ and
$z_\barQ,z_q \to z_\barQ,z_2=z_q-z_\barQ$, we obtain
\begin{equation}
  \begin{aligned}
 I_{\cal P}= (2\pi)^2\delta\big((q_Q+ q_\barq) \cdot  u_2\big)&
 \delta\big( (q_\barQ + q_q) 
\cdot  u_1\big)\int dz_1 \int dz_2 \,e^{-i(q_\barq \cdot  u_2 z_1 
  + q_q \cdot  u_1 z_2)} \\ & \times \la {W}_{u_1}(0 \cdot u_2) 
 {W}_{u_1}^*(z_1 u_2) {W}_{u_2}(z_2 u_1)
 {W}_{u_2}^*(0 \cdot u_1)\ra_A\, .
  \end{aligned}
\end{equation}
Taking into account that in our approximation $q_\barQ \cdot  u_2 =
q_q \cdot  u_2 = q_Q \cdot  u_1 = q_\barq \cdot  u_1 = 0$, and that
$u_1\cdot u_2 = \cosh\chi$, we have 
\begin{equation}
  \begin{aligned}
 I_{\cal P}=
 \f{1}{\sinh\chi}(2\pi)^2
 \delta^{(2)}(\vec p_{f\parallel}- \vec p_{i\parallel}) 
&\int dz_1
 \int dz_2 \,
e^{-im\cosh\chi[(\zeta_1-\zeta_1') z_1  + (\zeta_2'-\zeta_2)z_2]} \\
&
\times \la {W}_{u_1}(0 \cdot u_2)  
 {W}_{u_1}^*(z_1 u_2) {W}_{u_2}(z_2 u_1)
 {W}_{u_2}^*(0 \cdot u_1)\ra_A\,,
  \end{aligned}
\end{equation}
where $\vec p_\parallel=(p^0,p^1)$ are the longitudinal components of the
four-vector $p$. Finally, rescaling $\cosh\chi z_i = \tilde{z}_i$, we
obtain in the limit $\chi\to\infty$
\begin{equation}
  \begin{aligned}
 I_{\cal P}=
 \f{1}{m^2\sinh\chi\cosh^2\chi}(2\pi)^2&
 \delta^{(2)}(\vec p_{f\parallel}- \vec p_{i\parallel})   
 \delta(\zeta_1'-\zeta_1)\delta(\zeta_2'-\zeta_2) \\ &
 \times \la {W}_{u_1}(0 \cdot u_2)  
 {W}_{u_1}^*(0 \cdot u_2) {W}_{u_2}(0 \cdot u_1)
 {W}_{u_2}^*(0 \cdot u_1)\ra_A\,.
  \end{aligned}
\end{equation}
The resulting configuration of Wilson lines is such that the center of
each line is fixed, and lies at the origin of the longitudinal
plane. At this point, 
we have to recall that we do not want to describe the propagation of four
independent fermions, but rather that of two mesons represented in
terms of colourless $q\barq$ dipoles. In the high-energy limit, the
mesons, and therefore the dipoles, extend in the transverse plane
only, due to Lorentz contraction. If we want to recover this physical
picture in the amplitude, we need that the length of the Wilson lines
corresponding to fermions in the same dipole be the same, i.e.,
$T_Q=T_\barq$ and $T_q=T_\barQ$, and moreover we need to connect them
with straight-line ``links'' in the transverse plane, in order to
obtain a gauge-invariant object. The quantities relevant to the
description of the interacting dipoles are therefore two rectangular
Wilson loops, whose precise definition is given below.

\begin{figure}[t]
  \centering
  \includegraphics{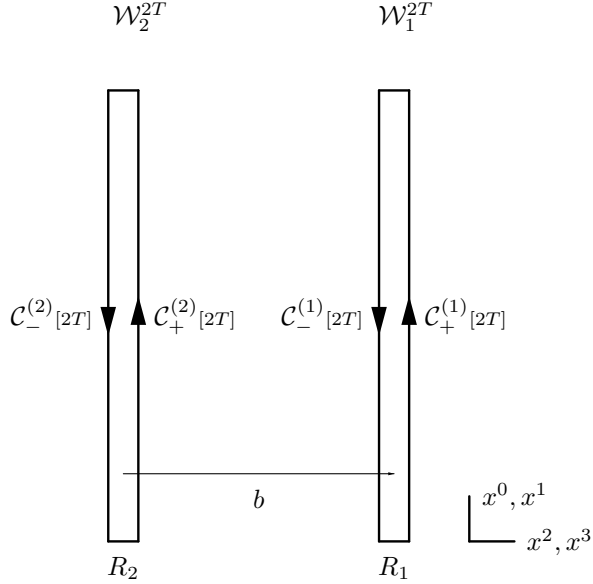}
  \caption{Schematic representation of the Wilson loops 
    $\W_{1,2}^{2T}$, relevant to Pomeron exchange, defined by the
    paths ${\cal C}^{(1,2)}_{\pm}$ of Eq.~\eqref{eq:rect_WL}. The
    length of each component of the path is indicated inside square
    brackets.} 
  \label{fig:wloop_pom}
\end{figure}
Performing the remaining integrals, one obtains for ${\cal P}$ the
following expression:
\begin{equation}
  \label{eq:pomeron}
  \begin{aligned}
  {\cal P} =&~ 2 s (2\pi)^4
  \dlx{4}(p_f-p_i) e^{i2(m-m_Q-m_q)2T} \int_0^1 d\zeta_1 \int_0^1
  d\zeta_2 \int d^2R_{1\perp} \int  d^2R_{2\perp} \\ &\times 
  \rho_1(\vec{R}_{1\perp},\zeta_1) \rho_2(\vec{R}_{2\perp},\zeta_2)
  \int d^2b_\perp    e^{i\vq_\perp\cdot 
    \vec{b}_\perp} \la \W_1^{2T}(\vec{b}_\perp,\vec{R}_{1\perp})
  \W_2^{2T}(\vec{0}_\perp,\vec{R}_{2\perp}) 
  \ra_A \, , 
  \end{aligned}
\end{equation}
where we have introduced the notation
\begin{equation}
   \rho_1(\vec{R}_{1\perp},\zeta_1) = \sum_{s_Q t_\barq} |\wf_{1\,s_Q
    t_\barq}(\vec{R}_{1\perp},\zeta_1)|^2 \,, \qquad
  \rho_2(\vec{R}_{2\perp},\zeta_2) = \sum_{s_q t_\barQ} |\wf_{2\,s_q
    t_\barQ}(\vec{R}_{2\perp},\zeta_2)|^2 \, .
\end{equation}
Note that $\int_0^1 d\zeta \int
d^2R_{\perp}\,\rho_{1,2}(\vec{R}_{\perp},\zeta)=1$ due to the
normalisation of the wave functions. 
Here  $\W_{1,2}^{2T}$ are the rectangular Wilson loops mentioned
above,
\begin{equation}
  \label{eq:rect_WL0}
\W_{i}^{2T} = \f{1}{N_c} \,\tr\, \Texp\left\{-ig\oint_{{\cal C}^{(i)}}
  A(x)\cdot dx\right\}\,,  
\end{equation}
which run along the paths ${\cal C}^{(1,2)}= {\cal C}^{(1,2)}_+ \circ
{\cal C}^{(1,2)}_- $ (see Fig.~\ref{fig:wloop_pom}),  
\begin{equation}
  \label{eq:rect_WL}
  \begin{aligned}
    {\cal C}^{(1)}_\pm :&\quad X_{1\pm}(\tau) = \pm u_1\tau + b \pm \f{R_1}{2}\,
    ,&& &  \tau\in [-T,T]\, ,\\
    {\cal C}^{(2)}_\pm :&\quad X_{2\pm}(\tau) = \pm u_2\tau \pm
    \f{R_2}{2}\, ,  && & \tau\in [-T,T]\, ,
  \end{aligned}
\end{equation}
where
\begin{equation}
  \label{eq:trans_vect}
  R_i = (0,0,\vec{R}_{i\perp})\, , \qquad b= (0,0,\vec{b}_{\perp})\, ,
\end{equation}
with $X_{i+}$ (resp.~$X_{i-}$) travelled forward (resp.~backward)
along the direction $u_i$ (hence the $\pm$ sign in front of $u_i$), 
and closed at $\tau=\pm T$ by straight-line paths in the transverse
plane in order to ensure gauge invariance. Since the two loops are
independent objects, and their lengths have to be sent to infinity at
the end of the calculation, there is no obstacle to choose the same
length $2T$. Notice that all values of $\zeta_1$, $\zeta_2$ are
involved in Eq.~\eqref{eq:pomeron}, and that the fraction of
longitudinal momentum carried by a parton is the same in the initial
and final state.   

The result above coincides with the one given in~\cite{Nachtr},
differing only by a phase.\footnote{In \eqref{eq:pomeron} we have
  neglected multiplicative factors which tend to 1 in the high-energy
  limit.} 
Nevertheless, this expression cannot be the complete  
answer. One reason is that an $S$-matrix element has to be
renormalisation-group invariant, and this is not the case for this
expression, since the rectangular Wilson loops get multiplicatively
renormalised due to the presence of cusps~\cite{Brandt2}. Another,
more physical reason, is that for large distances 
one expects the impact-parameter amplitude to vanish, since it
corresponds to a process where the mesons undergoing the scattering
process are very far away: since at large distances the
Wilson loop correlator is expected to factorise, the impact-parameter
amplitude could vanish only if the Wilson loop expectation value were 1
independently of its size.\footnote{We note in passing that although
  this is generally not true, this is
  actually the case for lightlike loops in the Stochastic Vacuum
  Model~\cite{BN,Nachtr}.} 

To understand the origin of the problem, one can consider the
amplitude for an isolated stable 
meson to remain unchanged. According to the LSZ approach, $in$ and
$out$ state should coincide in this case, namely, considering for
definiteness $M_1$, 
\begin{equation}
  \label{eq:mes_inout}
 \outs{M_1(p_1')} | \ins{M_1(p_1)} = (2\pi)^3 2p_1^0
    \delta^{(3)}(\vec{p}_1{}' - \vec{p}_1) \, .
\end{equation}
However, using the same approximation for the fermion propagator in
order to compute this quantity, one finds instead
\begin{equation}
  \label{eq:mes_inout_2}
  \begin{aligned}
 &\outs{M_1(p_1')} | \ins{M_1(p_1)} =
 \\ & \phantom{carlo} (2\pi)^3 2p_1^0
    \delta^{(3)}(\vec{p}_1{}' - \vec{p}_1) e^{i(m-m_Q-m_q)2T}  \int_0^1 d\zeta \int 
    d^2R_{\perp}\,    \rho_1(\vec{R}_{\perp},\zeta)
    \la \W_1^{2T}(\vec{b}_\perp,\vec{R}_\perp) \ra_A \, .   
  \end{aligned}
\end{equation}
This result is the same obtained in~\cite{Nachtr}, again up to a phase
factor. 
Notice that the expectation value is actually independent of the
position and orientation of the Wilson loop due to translation and
Lorentz invariance, and depends only on the longitudinal and trasverse
sizes, i.e., 
\begin{equation}
\label{eq:recloop}
\la \W_1^{2T}(\vec{b}_\perp,\vec{R}_\perp) \ra_A= \la
\W_2^{2T}(0,\vec{R}_\perp) \ra_A \equiv 
\W(2T,|\vec{R}_\perp|) \,.
\end{equation}
The reason for this discrepancy is probably that our description of
the meson in terms of $q\barq$ dipoles is too na\"ive. In particular,
we have completely neglected the fact that the fermions in each dipole
are both self-interacting and interacting with each other: these
interactions should actually be part of the description of the meson, 
and should play no role in scattering processes. In the approach
described above the fermions are effectively independent: as a
consequence, the internal interactions of the mesons appear as part of
the scattering process. As we have already pointed out, the phase
factor is due to the self-interactions of quarks and antiquarks. 
On the other hand, it is reasonable to identify the Wilson-loop
expectation value in Eq.~\eqref{eq:mes_inout_2} as the consequence of 
the interaction between the quark and the antiquark forming the dipole. We
see therefore that over a propagation proper-time $T_p$, the
contribution of the internal interactions for a freely-propagating
dipole of transverse size $R_t$ amounts to a factor 
\begin{equation}
 {\cal B}_{\rm internal}(T_p,R_t) =  e^{i(m-m_Q-m_q)T_p}\,
\W(T_p,R_t)\,. 
\end{equation}
In order to restore the
correct description of mesons, we have to divide out the contributions
from the internal interactions, and so we adopt the following
prescription: for a dipole of size $R_t$ propagating over a
proper-time $T_p$, we multiply by a factor $[{\cal B}_{\rm
  internal}(T_p,R_t)]^{-1}$. Using it in Eq.~\eqref{eq:mes_inout_2}, we
obviously recover the desired result, thanks to the normalisation of
$\rho_1$. However, this prescription has 
now to be used also in the case of 
interacting dipoles. 
This is done straightforwardly for the Pomeron-exchange amplitude,
where the size of the dipoles is the same in the initial and final
state, and we obtain
\begin{equation}
  \label{eq:pomeron_final}
\begin{aligned}
  \A_{\cal P}(s,t) = -i 2 s &
   \int_0^1 d\zeta_1 \int_0^1 d\zeta_2 \int d^2R_{1\perp}\int
  d^2R_{2\perp}\,
  \rho_1(\vec{R}_{1\perp},\zeta_1)
  \rho_2(\vec{R}_{2\perp},\zeta_2)
  \\ & \times  \int d^2b_\perp e^{i\vq_\perp\cdot
    \vec{b}_\perp} \left[\f{\la
      \W_1^{2T}(\vec{b}_\perp,\vec{R}_{1\perp}) 
      \W_2^{2T}(\vec{0}_\perp,\vec{R}_{2\perp}) \ra_A }
    {\la \W_1^{2T}(\vec{b}_\perp,\vec{R}_{1\perp})\ra_A 
      \la\W_2^{2T}(\vec{0}_\perp,\vec{R}_{2\perp}) 
    \ra_A }-1\right]\, ,
  \end{aligned}
\end{equation}
where $t=-\vq^{\,2}$, and we have used the notation 
\begin{equation}
  \label{eq:ampl_not}
  \begin{aligned}
    {\cal P} = \delta_{fi} + i(2\pi)^4\dlx{4}(p_f-p_i)\A_{\cal P}\, .
  \end{aligned}
 \end{equation}
The limit $T\to\infty$ is understood to be taken at the end of the
calculation. 
This is the expression usually found in the recent literature (see,
e.g., Refs.~\cite{LLCM1,Meggiolaro05}), which possesses the properties
discussed above.

\nsection{Reggeon exchange}
\label{amplitude}

In this Section we want to derive a nonperturbative expression for the 
Reggeon-exchange amplitude, using the space-time picture of the
process as a guideline. According to Feynman's picture of high-energy
scattering~\cite{FeynH}, the interaction between two colliding hadrons
is mediated by those partons which carry a small
fraction of longitudinal momentum, and which can therefore be
considered as belonging to the wave function of both hadrons. In the
case of Reggeon exchange in meson-meson scattering, the
coordinate-space picture of the process in the longitudinal plane is
then the following (see Fig. \ref{fig:1}). From the 
point of view of an observer in the center-of-mass frame, in the
initial stage of the process a ``wee'' (but fast) valence parton of
meson 1, say, the quark, and a ``wee'' valence parton of meson 2,
say, the antiquark, enter the interaction region along the classical
straight-line trajectories of the mesons, then ``bend'' their
trajectory, and annihilate producing gluons; in the final stage of the
process, these gluons produce a ``wee'' $q\barq$ pair, whose
components rejoin the ``spectator'' partons to form the mesons in the
final state. 
Things can go also in the reverse order, with the
production of a fermion-antifermion pair preceeding the
annihilation.\footnote{Notice that the interaction region does not
  allow for a classically allowed description, since there is either
  faster-than-light propagation, or violation of energy
  conservation.}  
As for the ``spectator'' partons, which carry a
relevant fraction of longitudinal momentum, they travel almost
undisturbed along their eikonal trajectories.

The physical picture given above will be a useful guideline in the
derivation of the Reggeon-exchange amplitude. 
 Consider for definiteness the term ${\cal R}_1$. 
The ``spectator'' partons $Q$ and $\barQ$ are treated as in the
Pomeron-exchange case, and so the corresponding truncated-connected
propagators in the external gluon field are evaluated in the eikonal
approximation, thus giving Eqs.~\eqref{eq:ferm_prop_Z=1} and
\eqref{eq:anti_ferm_prop_Z=1}. As regards the 
exchanged partons $q$ and $\barq$, the path-integral representation of
the quantities describing their propagation is
\begin{equation}
  \label{eq:V-_PI}
  \begin{aligned}
    V^-_{\barq q}   =& -\Lim \f{1}{\Delta\nu} \int d^4 x_i \int
    d^4x_f \,e^{-ip_\barq\cdot x_f -ip_q\cdot  x_i}\, {\cal V}^-(x_i,x_f)
    \,, \\
  {\cal V}^-(x_i,x_f)  =& \int_{x_i}^{x_f} \DX
  e^{-i\Delta\nu (m_q-i\epsilon)}   \bar{v}(p_\barq,t_\barq)
    \Sp_{\nu_i,\nu_f}[\dot{X}]u(p_q,s_q) 
    W_{\nu_i,\nu_f}[X]\,, 
  \end{aligned}  
\end{equation}
for the annihilation part, and
\begin{equation}
  \label{eq:V+_PI}
  \begin{aligned}
    V^+_{q \barq}    =& -\Lim' \f{1}{\Delta\nu'} \int d^4 x_i'\int
    d^4x_f'\, e^{ip_\barq'\cdot x_i' + ip_q'\cdot  x_f'}\, {\cal V}^+(x_i',x_f')
    \,,\\
  {\cal V}^+(x_i',x_f')   =& \int_{x_i'}^{x_f'} \DXp  
    e^{-i\Delta\nu' (m_q-i\epsilon)}   \bar{u}(p_q',s_q')
    \Sp_{\nu_i',\nu_f'}[\dot{X}']v(p_\barq',t_\barq')   
    W_{\nu_i',\nu_f'}[X']\,, 
  \end{aligned}
\end{equation}
for the creation part, where $\Delta\nu=\nu_f-\nu_i$ and
$\Delta\nu'=\nu_f'-\nu_i'$, and 
\begin{equation}
  \label{eq:Lim}
  \begin{aligned}
     &\Lim=\lim_{\nu_f\to\infty,\nu_i\to -\infty}\lim_{p_q^2\to \m_q^2,p_\barq^2\to
       \m_\barq^2}\, ,\quad
     \Lim'=\lim_{\nu_f'\to\infty,\nu_i'\to
       -\infty}\lim_{p_q^{\prime 2}\to \m_q^{\prime 2},p_\barq^{\prime 2}\to 
       \m_\barq^{\prime 2}} \, . 
  \end{aligned}
\end{equation}
Here and in the following we use the notation $\int_{x_i}^{x_f} \DX$ for 
integrals over paths with fixed endpoints, $X(\nu_i)=x_i$, $X(\nu_f)=x_f$.   
We have neglected the disconnected terms since
$p_q\neq p_\barq$, $p_q'\neq p_\barq'$. 
It is now convenient to change variables as follows, 
\begin{equation}
  \label{eq:ch_var_reg}
  \begin{aligned}
    x_i &= x_0 - T_i u_2 + x_{i\perp}\, ,  &  x_f &= x_0 - T_f u_1
    + x_{f\perp}\, , \\
    x_i' &= x_0' + T_i' u_1 + x_{i\perp}'\, ,   & x_f' &= x_0' + T_f' u_2
    + x_{f\perp}'\, ,
  \end{aligned}
\end{equation}
where
\begin{equation}
  \label{eq:ch_var_reg_2}
  \begin{aligned}
    x_0 &= (x_0^0,x_0^1,\vec{0}_\perp)\, , &&&
    x_0' &= 
    (x_0^{\prime 0},x_0^{\prime 1},\vec{0}_\perp) \, ,\\
    x_{i\perp} &= (0,0,\vec{x}_{i\perp})\, , &&& x_{i\perp}' &=
    (0,0,\vec{x}_{i\perp}^{\,\prime})\, , \\
    x_{f\perp} &= (0,0,\vec{x}_{f\perp})\, , &&& x_{f\perp}' &=
    (0,0,\vec{x}_{f\perp}^{\,\prime}) \, ,
  \end{aligned}
\end{equation}
so that the integration measure becomes
\begin{equation}
  \label{eq:ch_var_reg_3}
  \begin{aligned}
  d^4x_i d^4x_f &= \sinh\chi \, d^2x_0\, dT_i\, dT_f\, d^2x_{i\perp}\,
  d^2x_{f\perp}\, ,\\
  d^4x_i' d^4x_f' &= \sinh\chi\, d^2x_0'\, dT_i'\, dT_f'\, d^2x_{i\perp}'\,
  d^2x_{f\perp}'   \, .
  \end{aligned}
\end{equation}
Plugging everything in the expression for ${\cal R}_1$, we obtain
\begin{equation}
  \label{eq:R_0}
  \begin{aligned}
    {\cal R}^{(dd)}_1 &=  \Lim \f{1}{\Delta\nu}\, \Lim' \f{1}{\Delta\nu'}
    \f{1}{N_c^2}4(\sinh\chi)^4  
    \sqrt{\m_Q\m_Q'\m_\barQ\m_\barQ'}   \delta_{s_Q' s_Q} 
    \delta_{t_\barQ' t_\barQ}  
    \\ &\times \int dz_Q \int dz_\barQ \int d^2x_0 \int  dT_i \int dT_f
    \int d^2x_0' \int dT_i' \int dT_f'  
    \int d^{12}x_\perp
    \\ &\times  e^{i\psi(T)} e^{i\phi(x_\perp)}  e^{-i(p_Q'-p_Q)\cdot
      u_2 z_Q} e^{i(p_\barQ'-p_\barQ)\cdot u_1 
      z_\barQ}   e^{i(p_\barq'+ p_q')\cdot x_0'} 
    e^{-i(p_\barq+ p_q)\cdot x_0} 
    \\ &\times \la \tr\big\{W_{u_1}(-z_Qu_2+ b_{Q\perp}){\cal
      V}^-(x_i,x_f)  W_{u_2}(z_\barQ u_1+ b_{\barQ\perp}){\cal
      V}^+(x_i',x_f')    \ra_A \, ,
  \end{aligned}
\end{equation}
where the trace is over colour
indices, and where we have introduced the phases
\begin{equation}
\label{eq:phases}
  \begin{aligned}
    \psi(T) &=  T_Q (\m_Q + \m_Q' - 2m_Q)  + T_\barQ(\m_\barQ
    + \m_\barQ'-2m_Q) \\
    & \phantom{auiuiuiuiuiuiuiuiuiuiuiuiuiui}+ T_i\m_q + T_f\m_\barq
    + T_i'\m_\barq' 
    + T_f'\m_q'\,,\\ 
    \phi(x_\perp) &= (p_Q'-p_Q)\cdot b_{Q\perp}
    + (p_\barQ'-p_\barQ)\cdot b_{\barQ\perp} \\
    &\phantom{auiuiuiuiuiuiuiuiuiuiuiuiuiui} 
    +p_\barq'\cdot x_{i\perp}' + p_q'\cdot x_{f\perp}'  -p_\barq\cdot
    x_{f\perp} - p_q\cdot x_{i\perp}\,,
  \end{aligned}
\end{equation}
and the compact notation
$d^{12}x_\perp\!=\!d^2b_{Q\perp}
 d^2b_{\barQ\perp} d^2x_{i\perp}d^2x_{f\perp} d^2x_{i\perp}'  d^2x_{f\perp}'$
for the integration over the transverse variables. 
We recall that 
\begin{equation}
  \m_Q = \zeta_1 m\,, \qquad
  \m_\barq = (1-\zeta_1) m\,, \qquad
  \m_q = \zeta_2 m\,, \qquad \m_\barQ = (1-\zeta_2) m\,,
\end{equation}
and similarly for primed quantities. Notice that we have kept 
different the lengths $2T_Q$ and $2T_\barQ$ of the two eikonal Wilson
lines $W_{u_1}$ and $W_{u_2}$.

\subsection{Integration over longitudinal variables: $z_Q,z_\barQ,x_0,x_0'$}
\label{subsec:x0}

The next step is
to take care of the integration over the longitudinal coordinates. In
order to do so, we have to take into account that the expectation
value is again invariant under translations, as can be directly
checked:\footnote{Note that the ``spin factor'' terms inside ${\cal
    V}^\pm$ are unaffected by a translation, since they depend only
  on the paths' tangent vectors, see Eq.~\eqref{eq:defin_2}.} 
\begin{equation}
  \begin{aligned}
    &  \la \tr\big\{W_{u_1}(b_Q){\cal
      V}^-(x_i,x_f) W_{u_2}(b_\barQ){\cal
      V}^+(x_i',x_f')
    \ra_A = \\  &\la \tr\big\{W_{u_1}(b_Q+a){\cal
      V}^-(x_i+a,x_f+a) W_{u_2}(b_\barQ+a){\cal
      V}^+(x_i'+a,x_f'+a)
    \ra_A    \,.
  \end{aligned}
\end{equation}
In particular, a translation in the longitudinal plane affects only
$x_0$ and $x_0'$, and not $T_{i,f},T_{i,f}'$. Exploiting this fact and
the invariance of the (very long) eikonal Wilson lines under
translations along their directions, we can write (with a small abuse
of notation, and discarding
variables which are not relevant here)
\begin{equation}
  \begin{aligned}
    &  \la \tr\big\{W_{u_1}(-z_Q u_2){\cal
      V}^-(x_0) W_{u_2}(-z_\barQ u_1){\cal
      V}^+(x_0')
    \ra_A = \\  &\la \tr\big\{W_{u_1}(0\cdot u_2 ){\cal
      V}^-(x_0+z_Q u_2+z_\barQ u_1)
    W_{u_2}(0 \cdot u_1){\cal
      V}^+(x_0'+z_Q u_2+z_\barQ u_1)
    \ra_A    \,,
  \end{aligned}
\end{equation}
and changing variables from $z_Q,z_\barQ,x_0,x_0'$  to
$z_Q,z_\barQ,y_0=x_0+z_Q u_2+z_\barQ u_1,y_0'=x_0'+z_Q u_2+z_\barQ
u_1$ we can explicitly integrate over $z_Q,z_\barQ$, obtaining the
factor ($q_X\equiv p_X'-p_X$)
\begin{figure}[t]
  \centering
  \includegraphics[width=0.35\textwidth]{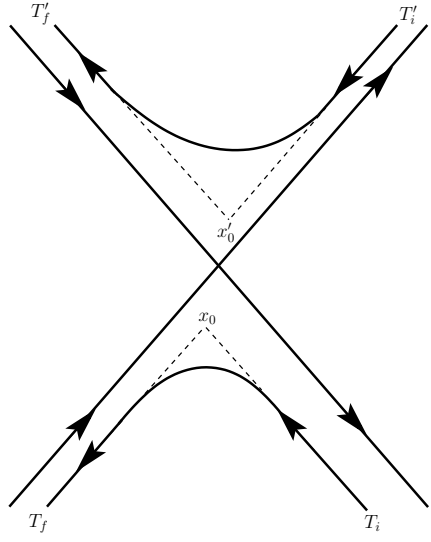}
  \caption{Longitudinal projection of the partons' trajectories (solid
    lines): the straight lines correspond to the ``spectator'' fermions,
    while the curved lines (of ``proper-time'' length
    $\Delta\nu,\Delta\nu'$) correspond to the exchanged fermions. The 
    dashed lines (of ``proper-time'' length $T_i+T_f,T_i'+T_f'$)
    depict the eikonal trajectories for the exchanged fermions.}
  \label{fig:long_proj}
\end{figure}
\begin{equation}
  \label{eq:delta_long}
  \begin{aligned}
 \int dz_Q & \int dz_\barQ \,  e^{-i(q_Q+q_\barq + q_q)\cdot u_2 z_Q}
  e^{i(q_\barQ+q_\barq + q_q)\cdot u_1 z_\barQ} = \\ &
(2\pi)^2
  \delta\big((q_Q+q_\barq + q_q)\cdot u_2\big) \delta\big((q_\barQ+q_\barq +
  q_q)\cdot u_1\big) \simeq \\ & \phantom{ioboja} (2\pi)^2
  \delta\big((q_Q+q_\barq + q_q + 
  q_\barQ)\cdot u_2\big) \delta\big((q_\barQ+q_\barq +  q_q + q_Q)\cdot u_1\big) =
  \\ & \phantom{iobojacciacciacciaiobojacciacciacciacciaccia}
   \f{1}{\sinh\chi} (2\pi)^2\dlx{2}(\vec p_{f\para}- \vec p_{i\para})\,.
 \end{aligned}
\end{equation}
We consider next the integration over $d^2y_0$ and $d^2y_0'$.
Since $p_{1,2}'\simeq p_{1,2}$, we have approximately for the relevant part of
the phase 
\begin{equation}
  \label{eq:zeta_0}
  \begin{aligned}
    (p_\barq'+ p_q')\cdot y_0' &\simeq
    m\left[y_0^{\prime 0}\cosh\f{\chi}{2}(1-\zeta_1' + \zeta_2')
    + y_0^{\prime 1}\sinh\f{\chi}{2}(1-\zeta_1' - \zeta_2')\right]\, ,\\
    (p_\barq+ p_q)\cdot y_0 &\simeq
    m\left[y_0^0\cosh\f{\chi}{2}(1-\zeta_1 + \zeta_2)
    + y_0^1\sinh\f{\chi}{2}(1-\zeta_1 - \zeta_2)\right]\,,
  \end{aligned}  
\end{equation}
and changing variables to $z^0=\cosh\f{\chi}{2} y_0^0$, $z^1=
\sinh\f{\chi}{2} y_0^1$, $z'{}^0=\cosh\f{\chi}{2} y_0'{}^0$, $z'{}^1=
\sinh\f{\chi}{2} y_0'{}^1$, we have 
\begin{equation}
  \label{eq:deltazeta_2}
  \begin{aligned}
I_{\cal R} =    &  \int d^2y_0\int d^2y_0'
e^{i(p_\barq'+ p_q')\cdot y_0'} e^{-i(p_\barq+ p_q)\cdot
  y_0}f(y_0^0,y_0^1,y_0^{\prime 0},y_0^{\prime 1}) =\\ &
 \left(\f{2}{\sinh\chi}\right)^2 \int d^2z \int d^2z'
 e^{im[z_0^{\prime 0}(1-\zeta_1' + \zeta_2')
    + z_0^{\prime 1}(1-\zeta_1' - \zeta_2')]}\\ &
\times e^{im[z_0^0(1-\zeta_1 + \zeta_2)
    + z_0^1(1-\zeta_1 - \zeta_2)]}
    f\left(\f{z^0}{\cosh\f{\chi}{2}},\f{z^1}{\sinh\f{\chi}{2}},
      \f{z'{}^0}{\cosh\f{\chi}{2}},\f{z'{}^1}{\sinh\f{\chi}{2}}\right)\, ,
  \end{aligned}
\end{equation}
where we have denoted 
\begin{equation}
\label{eq:short_f}
\begin{aligned}
  &f(y_0^0,y_0^1,y_0'{}^0,y_0'{}^1) 
  = \\ &\phantom{aaaaa}\la \tr\big\{W_{u_1}(0\cdot u_2 ){\cal
      V}^-(x_0+z_Q u_2+z_\barQ u_1)
    W_{u_2}(0 \cdot u_1){\cal
      V}^+(x_0'+z_Q u_2+z_\barQ u_1)
    \ra_A \,.
\end{aligned}
\end{equation}
If we now take naively the infinite-energy limit, $y_0,y_0'$ are
fixed to zero, and moreover we obtain delta-functions which fix to
zero the longitudinal-momentum fractions of the exchanged partons,
namely 
\begin{equation}
\label{eq:deltazeta_3}
I_{\cal R} \mathop\to_{\chi\to\infty}   \left(\f{(2\pi)^2}{m^2\sinh\chi}\right)^2
\delta(1-\zeta_1) 
    \delta(1-\zeta_1')\delta(\zeta_2)\delta(\zeta_2') f(0,0,0,0)\, .
\end{equation}
The delta functions in Eq.~\eqref{eq:deltazeta_3} make us run into
problems: if we take for the wave functions the usual form proportional to
$\zeta^{\beta}(1-\zeta)^{\gamma}$, unless
${\beta}={\gamma}=0$ we obtain either exactly zero or a
divergence when setting $\zeta=0$ or $\zeta=1$. For example, in the
phenomenological Wirbel-Stech-Bauer ansatz~\cite{WSB} one has
${\beta}={\gamma}=1/2$, and so the 
meson-meson Reggeon-exchange amplitude would be zero. However, the
delta-functions are obtained only in the strict $\chi\to\infty$
limit, and while this implies of 
course that in the high-energy limit $\zeta\to 0$, it says nothing
about $\zeta$ when the energy is large but finite. Moreover, the
consideration above shows that the way in which $\zeta$ approaches
zero as the energy increases is relevant in the determination of the
energy dependence of the Reggeon-exchange amplitude: this requires a
careful analysis of the integral above. Before doing that, we complete
the derivation of the expression for the scattering amplitude,
understanding that the limit $\zeta\to 0$ has to be taken in order to
obtain the high-energy expression for the amplitude, but delaying the
discussion on how this limit has to be taken.

\subsection{Integration over longitudinal variables: $T_{i,f}$, $T_{i,f}'$}

Up to here, we have not exploited yet the physical picture of the
process, described at the beginning of this Section. 
According to
this picture, we expect that the relevant contributions to the path
integrals come from those paths which at early and late proper-times
coincide with the straight (timelike) lines which describe the
propagation of the fast partons before and after the interaction. This
suggests that the integration range for $T_i$, $T_f$, $T_i'$ and
$T_f'$ can be limited to positive values only, so that $x_i^0, x_f^0 <
x_0^0$, $x_i^{\prime 0}, x_f^{\prime 0} > x_0^{\prime 0}$. 
Moreover, we expect the main contribution to come from
those paths which depart from the eikonal trajectories only in the 
time window corresponding to the duration of the interaction, 
which is much smaller than the total time of
the process: for these ``mostly timelike'' paths one has approximately
$T_i+T_f \sim \Delta\nu-L_0$, $T_i'+T_f' \sim \Delta\nu'-L_0$, with
$|L_0|\ll \Delta\nu,\Delta\nu'$. Here $L_0$ is the difference between
the characteristic ``proper-time'' duration of the fermion-exchange
process, and the proper-time corresponding to the free eikonal
propagation of fermions, as depicted in Fig.~\ref{fig:long_proj}. 
These paths are therefore expected to contain two long
straight-line timelike segments at early and late proper-times,
corresponding to the propagation of the partons $q$ and $\barq$ before
and after the interaction between the two colliding mesons. At this
stage of the process the mesons are not yet or no more interacting
with each other, so that the contribution of these straight-line
segments to the scattering amplitude should depend only weakly on the
actual position of the endpoints, i.e., on the values of $T_{i,f}$,
$T_{i,f}'$, after the subtraction of internal interactions (see the 
discussion at the end of Section \ref{Pomeron}).

\begin{figure}[t]
  \centering
  \includegraphics[width=1\textwidth]{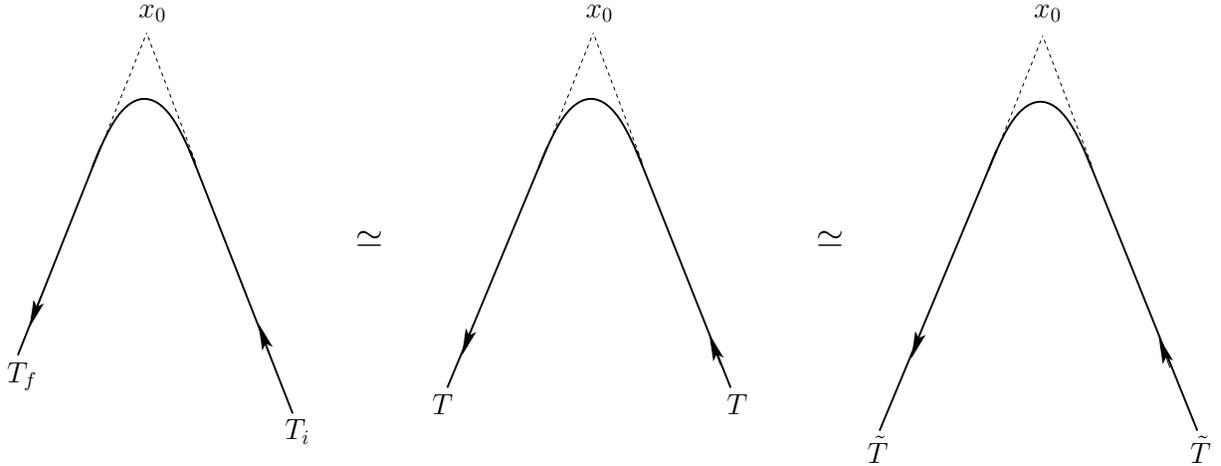}
  \caption{Paths giving approximately the same contribution to the
    Reggeon-exchange scattering amplitude, expressed as an integral
    over the trajectories of the exchanged fermions. The length of the
    first two paths is the same and equal to $\Delta\nu$, with
    $T=(T_i+T_f)/2$, while the third path is of length $L=\Delta\nu +
    2(\tilde T -T)$, with $|\tilde T -T| \ll \Delta\nu$.} 
  \label{fig:equiv_path}
\end{figure}
Consider for definiteness a typical path $X(\tau;T_i,T_f)$, where we
have explicitated the dependence on the initial and final points, which
contributes to ${\cal V}^-$; the same argument works also for the
paths $X'(\tau';T_i',T_f')$ contributing to ${\cal V}^+$. The
approximate independence of $T_i,T_f$ of the contribution 
of $X(\tau;T_i,T_f)$ means that it is
approximately the same as that of $X(\tau;T,T)$ with $2T =
T_f+T_i$, where the non-straight-line part of the paths
$X(\tau;T_i,T_f)$ and $X(\tau;T,T)$ coincide (see
Fig.~\ref{fig:equiv_path}). For the paths which we 
expect to be relevant $2T\sim \Delta\nu- L_0$, and the multiplicity of
each of these contributions is therefore of the order of $2\Delta\nu$: 
since we have to divide by $\Delta\nu$ when taking the limit
$\Delta\nu\to\infty$, these are the contributions which are expected
to survive. For definiteness, we take $T$ in the interval 
$\Delta\nu - L_0 \le 2T \le \Delta\nu + L_0$, for some fixed (but for
the moment unspecified) value of $L_0>0$. This parameter sets the
``tolerance'' for the deviation of the relevant paths of the exchanged
fermions from their eikonal trajectories: we will return on this point
at the end of Section \ref{sub_ampli}. 
Changing variables in the integral to $T_i,T$ we obtain
\begin{equation}
\label{eq:path_appr_1}
  \begin{aligned}
&  \int dT_i \int  dT_f \, {\cal I}(T_i,T_f) \simeq 
 2 \int_{0}^{\Delta\nu+L_0} dT_i \int_{\f{1}{2}(\Delta\nu-L_0)}^{\f{1}{2}(\Delta\nu+L_0)} dT 
   \, {\cal I}(T,T)\\ &
  \simeq 2 \Delta\nu \int_{\f{1}{2}(\Delta\nu-L_0)}^{\f{1}{2}(\Delta\nu+L_0)} dT\, 
  {\cal I}(T,T) \,, 
  \end{aligned}
\end{equation}
where ${\cal I}$ is a shortcut notation for the integrand, 
and similarly, setting $2T'=T_f'+T_i'$, 
\begin{equation}
\label{eq:path_appr_2}
  \begin{aligned}
&  \int dT_i' \int dT_f' \, {\cal I}(T_i',T_f') 
  \simeq 2 \Delta\nu' \int_{\f{1}{2}(\Delta\nu'-L_0)}^{\f{1}{2}(\Delta\nu'+L_0)} dT'\, 
  {\cal I}(T',T') \,.  
  \end{aligned}
\end{equation}
The factors $\Delta\nu$, $\Delta\nu'$ are cancelled by corresponding
factors in Eq.~\eqref{eq:R_0}, and the only dependence left on these
quantities is in the integration range for $T$ and $T'$, and in the
phase factors contained in ${\cal V}^\pm$, see Eqs.~\eqref{eq:V-_PI}
and \eqref{eq:V+_PI}.

At this point we can discuss how the picture of transverse dipoles
can be implemented in the Reggeon-exchange case, paralleling the
discussion of the Pomeron-exchange case, although some subtleties
have to be taken into account. Apart from setting the
longitudinal-momentum fractions of the exchanged fermions to
vanishing values, the result of the integration over $x_0,x_0'$ displayed
in Eq.~\eqref{eq:deltazeta_3} constrains the tips of the ``wedges''
formed by the longitudinal projection of the eikonal trajectories of
$q$ and $\barq$ to coincide, in particular putting them at the origin
of coordinates. Since now $T$ and $T'$ are large in the limit of large
$\Delta\nu$, $\Delta\nu'$, we can ``tune'' the lengths of the eikonal
trajectories of the ``spectator'' partons, i.e., we can choose them to
be not fixed but equal to $T+T'$, without changing much the result due
to the weak dependence on the position of the endpoints. Notice that
we are also moving the position of the center of the eikonal lines,
which should not affect much the result for the same reason. Since at
this point the trajectories of the partons are properly paired, the
picture of transverse dipoles emerges, and we can identify and divide
out the contribution of the internal interactions. Defining the
transverse sizes 
\begin{equation}
  \begin{aligned}
        \vec{R}_{1\perp} &= \vec{b}_{Q\perp} - \vec{x}_{f\perp} &
    \vec{R}_{2\perp} &= -\vec{b}_{\barQ\perp} + \vec{x}_{i\perp} \,,\\
    \vec{R}_{1\perp}' &= \vec{b}_{Q\perp} - \vec{x}_{i\perp}' &
    \vec{R}_{2\perp}' &= -\vec{b}_{\barQ\perp} + \vec{x}_{f\perp}'\,,
  \end{aligned}
\end{equation}
the process which we are describing is that of two incoming dipoles of
sizes $|\vec{R}_{1,2\perp}|$ propagating over a proper-time $T$ until
the nominal interaction point, which is chosen to lie at the origin of
coordinates in the longitudinal plane, and two outgoing dipoles of
sizes $|\vec{R}_{1,2\perp}^{\,\prime}|$ propagating over a
proper-time $T'$ after the interaction. Applying the prescription
discussed in the previous Section, we have to divide the integrand by
the factor (see Eq.~\eqref{eq:recloop} for the notation)
\begin{equation}
  \label{eq:reg_norm_fac}
  \begin{aligned}
  {\cal B}_{\rm internal}^{\cal R} =&~ e^{i2T(m - m_Q - m_q)}e^{i2T'(m -
    m_Q - m_q)} \,\W(T,|\vec{R}_{1\perp}|)\,\W(T,|\vec{R}_{2\perp}|)
  \\ & \times 
\W(T',|\vec{R}_{1\perp}^{\,\prime}|)\,
  \W(T',|\vec{R}_{2\perp}^{\,\prime}|) \,,
  \end{aligned}
\end{equation}
involving the expectation value of four rectangular Wilson loops. 
On the other hand, the combination of the phase factors coming from
the ``spectator'' and from the exchanged partons is
\begin{equation}
  \begin{aligned}
    &e^{i[T(\m_Q  + \m_\barQ-2m_Q) + T'(\m_Q'
      +\m_\barQ'-2m_Q)]}e^{i[T(\m_q+ \m_\barq) +
      T'(\m_\barq' + \m_q')]}  e^{-i\Delta\nu m_q}
    e^{-i\Delta\nu' m_q} = \\
    &     e^{i2(T+T')(m-m_Q-m_q)}    e^{-i(\Delta\nu-2T)m_q}
    e^{-i(\Delta\nu'-2T')m_q}\,,
  \end{aligned}
\end{equation}
which reduces to 
\begin{equation}
  e^{-i(\Delta\nu-2T)m_q} e^{-i(\Delta\nu'-2T')m_q}\,,
\end{equation}
after the subtraction of self-interactions.

It is useful to remark that if we add (or subtract) a segment of straight
line to a given path $X$, $\Delta\nu$ and $T$ are increased (or
decreased) by the same amount, so that the difference $\Delta\nu-2T$
remains unchanged. Such a segment does not change the contribution of
the path if it is not too large, and so this allows us to add to each
path a straight-line segment of variable (possibly negative) length,
in order the set the initial and final points to fixed values $x_i\to 
-\tilde{T}u_2$, $x_f\to -\tilde{T}u_1$, with $\tilde{T}$ large,
without changing the integrand appreciably (see
Fig.~\ref{fig:equiv_path}). The integration over $T$ becomes then an
integration over the total length $L=\Delta\nu + 2(\tilde{T}-T)$ of
the new path,   
\begin{equation}
 2  \int_{\f{1}{2}(\Delta\nu-L_0)}^{\f{1}{2}(\Delta\nu+L_0)}  dT \to
 \int_{2\tilde{T}-L_0}^{2\tilde{T}+ L_0} dL\,,
\end{equation}
with $\tilde{T}\to\infty$ at the end of the calculation. 
The dependence on $\Delta\nu$ has therefore been replaced by that on
$\tilde{T}$; more precisely, we have replaced the
integration over the endpoints at fixed total length $\Delta\nu$,
which is sent to infinity at the end of the calculation when
$\nu_f\to\infty,\nu_i\to-\infty$, with
an integration over the length of the path while keeping the endpoints
fixed at $-u_{1,2}\tilde{T}$, and sending them 
to infinity, at the end of the calculation, along the directions
$-u_{1,2}$ corresponding to the eikonal trajectories of the incoming
partons. The argument can be repeated for the path $X'$, fixing its
endpoints at $u_{1,2}\tilde{T}'$ and replacing
\begin{equation}
 2  \int_{\f{1}{2}(\Delta\nu'-L_0)}^{\f{1}{2}(\Delta\nu'+L_0)} dT' \to
 \int_{2\tilde{T}'-L_0}^{2\tilde{T}'+ L_0} dL'\,,
\end{equation} 
with $L'$ the total length of the new path, and $\tilde{T}'\to\infty$
at the end of the calculation. 
There is no problem at this point in taking $\tilde{T}$ and
$\tilde{T}'$ to be equal, and we therefore set
$\tilde{T}=\tilde{T}'=T$.

\subsection{Integration over transverse variables}

We are left now with the integration over the transverse positions of
the partons, which can be partially 
performed exploiting again the invariance of the expectation value
under translations. The integral is of the form 
\begin{equation}
  \label{eq:inte_transvo}
 I_\perp =  \int d^{12}x_\perp e^{i\phi(x_\perp)}
  F(\vec{b}_{Q\perp},\vec{b}_{\barQ\perp},\vec{x}_{i\perp}',\vec{x}_{f\perp}',
  \vec{x}_{i\perp},\vec{x}_{f\perp})\,,
\end{equation}
where $d^{12}x_\perp$ is a compact notation for the integration over
the transverse variables, and the phase $\phi(x_\perp)$ has been given
in Eq.~\eqref{eq:phases}. Changing variables to
\begin{equation}
  \label{eq:transv_vect}
  \begin{aligned}
    \vec{C}_\perp &= \f{\vec{b}_{Q\perp}+\vec{b}_{\barQ\perp}}{2}\,,
    &\vec{b}_\perp &= \f{\vec{b}_{Q\perp} + \vec{x}_{f\perp}
      -\vec{b}_{\barQ\perp} - \vec{x}_{i\perp}  }{2}\,,\\ 
    \vec{R}_{1\perp} &= \vec{b}_{Q\perp} - \vec{x}_{f\perp}\,, &
    \vec{R}_{2\perp} &= -\vec{b}_{\barQ\perp} + \vec{x}_{i\perp}\,, \\
    \vec{R}_{1\perp}' &= \vec{b}_{Q\perp} - \vec{x}_{i\perp}'\,, &
    \vec{R}_{2\perp}' &= -\vec{b}_{\barQ\perp} + \vec{x}_{f\perp}'\, ,
  \end{aligned}
\end{equation}
and exploiting translation invariance of the expectation value to
eliminate the variable $\vec{C}_\perp$ from the integrand, we obtain
\begin{equation}
  \label{eq:transv_int_2}
  \begin{aligned}
I_\perp =   & (2\pi)^2 \dlx{2}(\vec p_{f\perp}-\vec p_{i\perp})
  \int d^2b_\perp \int    d^2R_{1\perp}\int d^2R_{2\perp} \int
  d^2R_{1\perp}'\int d^2R_{2\perp}' \\ & \times 
    e^{i(\vec{q}_\perp    \cdot\vec{b}_\perp -  \vec{k}_{1\perp}'\cdot
      \vec{R}_{1\perp}' 
    +\vec{k}_{1\perp}\cdot\vec{R}_{1\perp} -\vec{k}_{2\perp}'\cdot
      \vec{R}_{2\perp}' + \vec{k}_{2\perp}\cdot \vec{R}_{2\perp}
    )}\\ & \textstyle \times F\left(\vec{b}_\perp +\f{\vec{R}_{1\perp}}{2}, -
    \f{\vec{R}_{2\perp}}{2}, \vec{b}_\perp + \f{\vec{R}_{1\perp}}{2}-
    \vec{R}_{1\perp}',  \vec{R}_{2\perp}'- \f{\vec{R}_{2\perp}}{2},
    \f{\vec{R}_{2\perp}}{2}, \vec{b}_\perp -
    \f{\vec{R}_{1\perp}}{2}\right)\, ,
  \end{aligned}
\end{equation}
where  $\vec{q}_\perp =\vec{p}_{2\perp}^{\,\prime}-\vec{p}_{2\perp}$
(see Eq.~\eqref{eq:momenta}). For future utility, we introduce the
notation 
\begin{equation}
  \label{eq:path_norm3}
  \Delta \vec R_{i\perp} = \f{1}{2}(\vec R_{i\perp}^{\,\prime} -\vec R_{i\perp})\,,
\end{equation}
for the variation of the dipole sizes between initial and final state.

\subsection{Reggeon-exchange amplitude}
\label{sub_ampli}

\begin{figure}[t]
  \centering
  \includegraphics{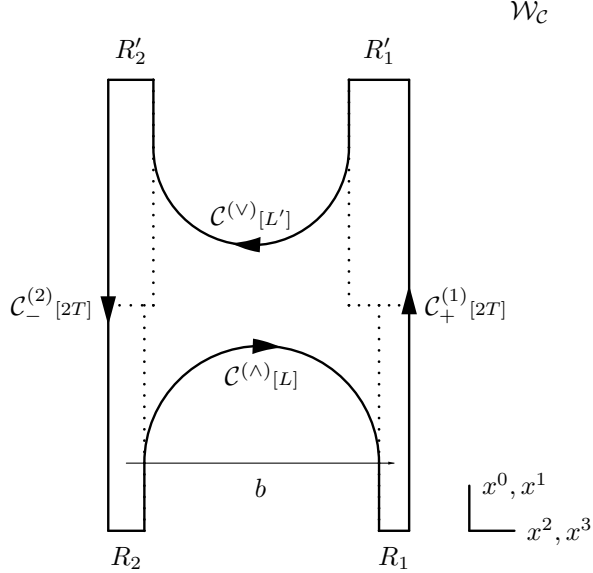}
  \caption{Schematic representation of the Wilson loop 
     $\W_{\cal C}$, relevant to Reggeon exchange, defined by the
     path ${\cal C} = {\cal
  C}^{(1)}_+\circ\, {\cal C}^{(\wedge)}\circ\, {\cal  C}^{(2)}_-\circ\, {\cal
  C}^{(\vee)}$, see Eqs.~\eqref{eq:curved} and \eqref{eq:straight}. The
    length of each component of the path is indicated inside square
    brackets. The contours of the Wilson loops contributing to the
    normalisation factor, running along the paths $\bar{\cal
      C}^{(i)}_\pm$ of
    Eq.~\eqref{eq:path_norm}, are also drawn with dotted lines (note
    that they have been displaced in the longitudinal plane, without
    changing their expectation value, in order
    to fit into the path ${\cal C}$).}  
  \label{fig:wloop_reg}
\end{figure}

The final result for the meson-meson Reggeon-exchange amplitude is
obtained after folding the corresponding dipole-dipole amplitude with
the mesonic wave functions. This step is straightforward, and we thus quote 
only the final result. To this extent, we introduce the notation 
\begin{equation}
  \label{eq:wl}
  \begin{aligned}
  W_\wedge^{j_1 i_2}[X,L;T,x_{i\perp},x_{f\perp}] &= 
    \left(\Texp\left\{-ig\int_{{\cal C}^{(\wedge)}} 
    A(X)\cdot dX \right\}\right)_{j_1 i_2}\, ,\\ 
    W_\vee^{i_2' j_1'}[X',L';T,x_{i\perp}',x_{f\perp}'] &=
    \left(\Texp\left\{-ig\int_{{\cal C}^{(\vee)}} 
    A(X')\cdot d{X}'\right\}\right)_{i_2' j_1'}\, ,
  \end{aligned}
\end{equation}
for the Wilson line running along the paths $X$ and $X'$, of
``proper-time'' length $L$ and $L'$, corresponding to the
trajectories of the exchanged partons, which are integrated over in
the path integral,
\begin{equation}
  \label{eq:curved}
  \begin{aligned}
    {\cal C}^{(\wedge)} :~&X(\tau), \qquad
    \tau\in[-T,-T+L]\,, \\ 
    ~& X(-T) = -u_2T + x_{i\perp} = -u_2T + \f{R_2}{2} \equiv
    x^{(\wedge)}_i \,, \\ ~& 
    X(-T+L) = -u_1T + x_{f\perp} = -u_1T + b 
    - \f{R_1}{2} \equiv
    x^{(\wedge)}_f\,, \\ 
    {\cal C}^{(\vee)} :~&X'(\tau), \qquad
    \tau\in[-T,-T+L']\,, \\ 
    ~&X'(-T) =  u_1T +  x_{i\perp}' = u_1T + b -R_1' + \f{R_1}{2}  \equiv
    x^{(\vee)}_i\,, \\
    ~&X'(-T+L') = u_2T + x_{f\perp}' = u_2T + R_2'
    -\f{R_2}{2} \equiv
    x^{(\vee)}_f\, .
  \end{aligned}
\end{equation}
where we have set
\begin{equation}
    b= (0,0,\vec{b}_\perp) \,, \qquad
    R_i=(0,0,\vec{R}_{i\perp})\,, \qquad
    R_i'=(0,0,\vec{R}_{i\perp}^{\,\prime})\,, \quad i=1,2\,.
\end{equation}
We also write
\begin{equation}
  \label{eq:sf}
  \begin{aligned}
\Sps_\wedge^{t_\barq s_q}[\dot{X},L;p_\barq,p_q] &=
\f{1}{2\sqrt{\m_q\m_\barq}}
\bar{v}(p_\barq,t_\barq)\Sp_{-T,-T+L}[\dot{X}]u(p_q,s_q)\, ,\\  
\Sps_\vee^{s_q' t_\barq'}[\dot{X}',L';p_q',p_\barq'] &=
\f{1}{2\sqrt{\m_q'\m_\barq'}}
\bar{u}(p_q',s_q')\Sp_{-T,-T+L'}[\dot{X}']v(p_\barq',t_\barq')
\, ,
  \end{aligned}
\end{equation}
where $\Sps_{\wedge,\vee}$ are the spin factors corresponding to the
paths $X$ and $X'$, contracted with the appropriate bispinors
and normalised in order to be dimensionless. 
We then define the Wilson loop
\begin{equation}
  \label{eq:loop_def}
  \begin{aligned}
  \W_{\cal C}[X,L,X',&L'] =
 \f{1}{N_c} \,\tr\, \Texp\left\{-ig\oint_{\cal C} A(x)\cdot dx\right\} 
 \equiv  \\ &
\f{1}{N_c}\, \tr\, \big\{ W_{u_1}\left(b+ \textstyle\f{R_1}{2}\right)     
    W_\wedge\left[X,L;T,\textstyle\f{R_{2}}{2},
      b-\textstyle\f{R_1}{2}\right]   \\ & 
    \phantom{dioboia}\times  
    W_{u_2}^\dag\left(-\textstyle\f{R_2}{2}\right)
    W_\vee\left[X',L';T,b+
      \textstyle\f{R_1}{2}-R_{1}',R_{2}'-\textstyle\f{R_2}{2}\right]\big\}\,
    ,  
  \end{aligned}
\end{equation}
running along the path ${\cal C}$ defined as ${\cal C} = {\cal
  C}^{(1)}_+\circ\, {\cal C}^{(\wedge)}\circ\, {\cal  C}^{(2)}_-\circ\, {\cal
  C}^{(\vee)}$ (with the parameter along the path increasing from right
to left), with
\begin{equation}
\label{eq:straight}
  \begin{aligned}
    {\cal C}^{(1)}_+ :&~X_+^{(1)}(\tau) = u_1\tau + b + \f{R_1}{2}, &&&
    \tau&\in[-T,T] \,, \\ 
    {\cal C}^{(2)}_- :&~ X_-^{(2)}(\tau) = -u_2\tau - \f{R_2}{2} , &&&
    \tau&\in[-T,T]\, , \\ 
  \end{aligned}
\end{equation}
corresponding to the trajectories of the ``spectator''
partons (which are the same as in the Pomeron-exchange case,
Eq.~\eqref{eq:rect_WL}), and ${\cal C}^{(\wedge)}$ 
and ${\cal C}^{(\vee)}$ defined in 
Eq.~\eqref{eq:curved}. The minus sign in front of $u_2$ in ${\cal
  C}^{(2)}_-$ reflects the fact that it is travelled backward along
the direction $u_2$.
The four pieces above are connected by straight-line 
paths in the transverse plane (not explicitly written in
Eq.~\eqref{eq:loop_def}), in order to make the expression
gauge-invariant (see Fig. \ref{fig:wloop_reg}). Introducing the
normalisation factor 
Eq.~\eqref{eq:reg_norm_fac}, we define also the normalised Wilson-loop
expectation value 
\begin{multline}
  \label{eq:loop_norm}
 {\cal U}_{\cal C}[X,L,X',L'] \equiv 
{\la \W_{\cal
     C}[X,L,X',L'] 
     \ra_A} 
   \Big[\la \W_1^T(\vec{b}_\perp,\vec{R}_{1\perp})\ra_A
       \la \W_2^T(\vec{0}_\perp,\vec{R}_{2\perp})\ra_A
      \\ \times \la
      \W_1^T(\vec{b}_\perp-\Delta\vec{R}_{1\perp},\vec{R}_{1\perp}')\ra_A 
       \la \W_2^T(\Delta\vec{R}_{2\perp},\vec{R}_{2\perp}')\ra_A \Big]^{-1}
\,.
\end{multline}
For definiteness, we have expressed the normalisation factor in terms
of the Wilson loops $\W_{i}^T(\vec{d}_\perp,\vec{D}_\perp)$, 
\begin{equation}
  \label{eq:rect_WL_norm}
\W_{i}^{T}(\vec{d}_\perp,\vec{D}_\perp) = \f{1}{N_c} \,\tr\,
\Texp\left\{-ig\oint_{\bar{{\cal C}}^{(i)}(\vec{d}_\perp,\vec{D}_\perp)} 
  A(x)\cdot dx\right\}\,,  
\end{equation}
running along the paths $\bar{\cal
  C}^{(i)}(\vec{d}_\perp,\vec{D}_\perp)=\bar{\cal 
  C}^{(i)}_+(\vec{d}_\perp,\vec{D}_\perp)\circ \bar{\cal
  C}^{(i)}_-(\vec{d}_\perp,\vec{D}_\perp)$, 
\begin{equation}
  \label{eq:path_norm}
  \begin{aligned}
  \bar{\cal C}^{(i)}_\pm(\vec{d}_\perp,\vec{D}_\perp):&~
\bar{X}^{(i)}_\pm(\tau) = \pm u_i\tau +
  d \pm \f{D}{2}\,, \qquad \tau\in[-\textstyle\f{T}{2},\textstyle\f{T}{2}]\, ,
\\ &~   d = (0,0,\vec{d}_\perp)\,, \qquad D=(0,0,\vec{D}_\perp)\,,
\end{aligned}
\end{equation}
properly closed by straight-line paths in the transverse plane. 
Finally, folding with the wave functions and extracting the scattering
amplitude from ${\cal R}_1$, 
\begin{equation}
  \label{eq:reg_ampl}
  {\cal R}_1 = i(2\pi)^4\dlx{4}(p_f-p_i)\A_{{\cal R}_1} \, ,
\end{equation}
we obtain (up to multiplicative factors that tend to 1 in the
high-energy limit)
\begin{equation}
  \label{eq:regge_ampl}
  \begin{aligned}
    \A_{{\cal R}_1}(s,t) = &
    \lim_{\zeta_1\to 1,\zeta_2\to 0}  
    \int  d^2R_{1\perp}\int  d^2R_{2\perp} \int d^2R_{1\perp}' \int
    d^2R_{2\perp}' \,
      \rho^{(\barq)}_{1\,t_\barq't_\barq}(\vr_{1\perp},\vr_{1\perp}',\zeta_1)\\
      & \phantom{carlotto} \times  
      \rho^{(q)}_{2\,s_q's_q} (\vr_{2\perp},\vr_{2\perp}',\zeta_2)\,
    \A_{{\cal R}_1}^{(dd)\, s_q' t_\barq';\,t_\barq
      s_q}(s,t;\vr_{1\perp},\vr_{1\perp}',\vr_{2\perp},\vr_{2\perp}')\,, 
  \end{aligned}
\end{equation}
where we have denoted
\begin{equation}
  \label{eq:rhos}
  \begin{aligned}
\rho^{(\barq)}_{1\,t_\barq't_\barq}(\vr_{1\perp},\vr_{1\perp}',\zeta_1)
&= \sum_{s_Q}  \wf_{1\,s_Q t_\barq'}^*(\vr_{1\perp}',\zeta_1) \wf_{1\,s_Q
  t_\barq}(\vr_{1\perp},\zeta_1)   \,,\\ 
\rho^{(q)}_{2\,s_q's_q}(\vr_{2\perp},\vr_{2\perp}',\zeta_2) &=
\sum_{t_\barQ}\wf_{2\,s_q' t_\barQ}^*(\vr_{2\perp}',\zeta_2) 
    \wf_{2\,s_q t_\barQ}(\vr_{2\perp},\zeta_2)\,, 
\end{aligned}
\end{equation}
and we have introduced the dipole-dipole Reggeon-exchange amplitude 
\begin{equation}
  \label{eq:regge_ampl_dip}
  \begin{aligned}
&    \A_{{\cal R}_1}^{(dd)\, s_q' t_\barq';\,t_\barq
  s_q}(s,t;\vr_{1\perp},\vr_{1\perp}',\vr_{2\perp},\vr_{2\perp}') =
-i2s  \left(\f{2\pi}{m}\right)^2  \f{1}{N_c}
\int d^2b_\perp e^{i\vec{q}_\perp \cdot\vec{b}_\perp} 
\\ & \phantom{carlacaralcaacaa} \times
    \int dL \int_{x_i^{(\wedge)}}^{x_f^{(\wedge)}} \DX \int dL'
    \int_{x_i^{(\vee)}}^{x_f^{(\vee)}} \DXp 
    e^{-i(m_q-i\epsilon)(L + L')} e^{i4m_qT}
    \\ & \phantom{carlacaralcaacaa}\times 
    \Sps_\wedge^{t_\barq s_q}[\dot{X},L;p_\barq,p_q] \,
    \Sps_\vee^{s_q' t_\barq'}[\dot{X}',L';p_q',p_\barq']  \,
    {\cal U}_{\cal C}[X,L,X',L']\,,
  \end{aligned}
\end{equation}
where $L$ and $L'$ lie in the range $[2T-L_0,2T+L_0]$; the limit
$T\to\infty$ has to be taken at the end of the calculation. An
expression analogous to Eq.~\eqref{eq:regge_ampl} is obtained 
for $\A_{{\cal R}_2}$, substituting $\A_{{\cal R}_1}^{(dd)}$ with
$\A_{{\cal R}_2}^{(dd)}$, which in turn is obtained by replacing
$R_i\to -R_i$, $R_i'\to -R_i'$ and $m_q\to m_Q$ in the
right-hand side of Eq.~\eqref{eq:regge_ampl_dip},\footnote{In
  principle one should also replace $\zeta_1\to 
  1-\zeta_1$, $\zeta_2\to 1-\zeta_2$ in Eq.~\eqref{eq:regge_ampl_dip},
  but as we discuss in subsection \ref{subsec:van_mom} its dependence 
  on $\zeta_{1,2}$ can be neglected in the {soft} high-energy limit.}  
and changing the limit to $\lim_{\zeta_1\to 0,\zeta_2\to 1}$ in
Eq.~\eqref{eq:regge_ampl}. Notice that at large $N_c$ the
Reggeon-exchange amplitude is of order ${\cal O}(1/N_c)$, as expected.

The dipole-dipole Reggeon-exchange amplitude
Eq.~\eqref{eq:regge_ampl_dip} calls for a few important remarks.
\begin{itemize}
\item The dipole-dipole Reggeon-exchange amplitude is independent of
  the longitudinal-mo\-men\-tum fractions, and thus is not affected by the 
  problem of taking the limit $\zeta_1\to 1,\zeta_2 \to 0$, 
  mentioned above in subsection \ref{subsec:x0}. Indeed,
  as we show in subsection \ref{subsec:van_mom}, the dependence of
  $\Sps_\wedge$ and $\Sps_\vee$ on $\zeta_{1,2}$ can be neglected in
  the {soft} high-energy limit. Therefore, the basic contribution
  to the Reggeon-exchange meson-meson scattering amplitude is of
  universal nature, i.e., independent of the kind of mesons involved
  in the scattering process.  
\item Since we are dealing here with a physical scattering amplitude,
  involving colour-neutral states, IR-divergencies are expected to
  be absent, and so the limit $T\to\infty$ of Eq.~\eqref{eq:regge_ampl_dip} 
  should be finite. Indeed, the normalised
  expectation value Eq.~\eqref{eq:loop_norm} is expected to be
  independent of $T$ at large $T$, since the contributions to the
  integral over gauge fields of regions
  far away from the interaction region, where the relevant paths
  coincide with the eikonal trajectories,  should cancel between
  numerator and denominator. Moreover, in these regions the contracted
  spin factor is expected to be dominated by the classical trajectory,
  thus reducing to unity. As for the phase factors $e^{im_q(L-2T)}$ and
  $e^{im_q(L'-2T)}$, a simple change of variables shows that they are
  actually independent of $T$.
\item Notice that the final expression, even in
  the $T\to\infty$ limit, still depends on $\LL$, whose value has not
  been specified yet. In order to see how it can be fixed, recall that
  $\pm\LL$ are the endpoints of the integration range in the variable
  $L-2T$, which provides a measure of the deviation of the paths of
  the exchanged fermions from their eikonal trajectories. If our
  picture is correct and only paths which do not deviate too much from
  the eikonal trajectories give relevant contributions, the
  integration over $L-2T$ should not be very sensitive to the
  integration range as soon as all the relevant paths have been
  included. Stated differently, as a function of $\LL$ the path
  integral is expected to approach a constant for $\LL \gtrsim L_{\rm
    char}$, for some characteristic $L_{\rm char}$. In this case, it
  would not matter too much if we set $\LL = L_{\rm char}$ at the end
  of the calculation, or if we take $\LL\to\infty$, thus removing the
  problem of determining the correct value of $L_{\rm char}$. However,
  to keep the discussion more general, especially as regards the
  derivation of the analytic continuation relations in Section
  \ref{an_cont}, we keep $\LL$ as an adjustable parameter.   
\item Finally, we stress the
  fact that this expression gives the contribution leading in energy in
  the given, parton-inelastic channel: other contributions of the same
  order in $s$ can come from  subleading contributions to the
  Pomeron-exchange amplitude, i.e., to the parton-elastic process, but
  they are clearly not entering here. 
\end{itemize}
As we have pointed out above, the dipole-dipole Reggeon-exchange
amplitude gives the basic, universal contribution to the mesonic
amplitude. Nevertheless, as we have already mentioned, the fact that
we have to take the limit of vanishing longitudinal-momentum fraction
of the exchanged partons affects the energy dependence of the mesonic
scattering amplitude. The next subsection is devoted to this issue.

\subsection{Limit of vanishing longitudinal-momentum fraction}
\label{subsec:van_mom}

We have now to discuss the important issue of how the limit $\zeta_1\to
1$, $\zeta_2\to 0$ has to be implemented. Recall that we have to deal
with the integral 
\begin{equation}
  \label{eq:deltazeta_2_bis}
  \begin{aligned}
I_{\cal R}    &=  \int d^2y_0 d^2y_0'
e^{i(p_\barq'+ p_q')\cdot y_0'} e^{-i(p_\barq+ p_q)\cdot
  y_0}f(y_0^0,y_0^1,y_0'{}^0,y_0'{}^1) \,,
  \end{aligned}
\end{equation}
where $f$ has been defined in Eq.~\eqref{eq:short_f}. 
We have already shown that taking directly the
high-energy limit, this integral gives the product of
delta-functions Eq.~\eqref{eq:deltazeta_3}, 
which sets $\zeta_2=1-\zeta_1=0$. This suggests that the
typical values of the longitudinal-momentum fractions of the
exchanged fermions, i.e., $\zeta_2$ and 
$1-\zeta_1$, decrease with energy until they reach zero in
the strict infinite-energy limit. 

In order to understand how such typical values depend on energy, or
equivalently how the delta-functions in Eq.~\eqref{eq:deltazeta_3} 
are approached in the high-energy limit, it is useful to notice 
that due to the short-range nature of strong interactions, one
expects only a finite region of space-time to be important in the
integral Eq.~\eqref{eq:deltazeta_2_bis}. More precisely, it is known
that the QCD vacuum is characterised by a ``vacuum correlation
length''
$a$~\cite{corr_length,corr_length2,corr_length3,corr_length4,corr_length5, 
corr_length6,corr_length7,corr_length8,corr_length9},
which sets the scale for the 
gauge-invariant two-point field-strength correlator in Euclidean
space. In a broader sense, we expect also that the ``vacuum correlation
length'' determines the distance beyond which parts of a nonlocal
operator, such as a Wilson loop, do not ``feel'' each other. In the
case at hand, it should then lead to an estimate of the relevant
region in $(y_0,y_0')$-space, in which the Wilson lines corresponding
to the exchanged fermions interact non-negligibly with both the
Wilson lines corresponding to the ``spectator'' fermions.

The first step is to translate the meaning of the ``vacuum correlation
length'' from Euclidean to Minkowski space. In Euclidean space, $a$
sets the scale for the exponential damping of correlators $\sim
e^{-\sqrt{\Delta X_E^2}/a}$, where $\Delta X_E$ is the separation between two
points and the metric is Euclidean. Performing
the inverse Wick-rotation, the relevant sphere $|\Delta X_E|\lesssim a$
is transformed into the region $|(\Delta X_M^0)^2 -
  \Delta\vec{X}_M^{\,2}|^{\f{1}{2}} \lesssim a$.  
For spacelike separation $\Delta X_M$ one
has still exponential damping, while for timelike separation the
exponential becomes a phase. Nevertheless, when the timelike
separation is larger than $a$, this phase varies rapidly, so that one
expects destructive interference between the contributions. The
relevant region is therefore expected to be given by $|(\Delta
  X_M^0)^2 -  \Delta\vec{X}_M^{\,2}|^{\f{1}{2}} \lesssim a$.  
Moreover, the lightlike ``branches'' of this region become smaller and
smaller as the separation along the lightcone increases, thus making
the damping or the phase variation more rapid. In a first
approximation, one can therefore focus on the ``box'' $|\Delta
X_M^0|\lesssim a$, $|\Delta\vec{X}_M|\lesssim a$ as the relevant
region. 

\begin{figure}[t]
  \centering
  \includegraphics[width=0.55\textwidth]{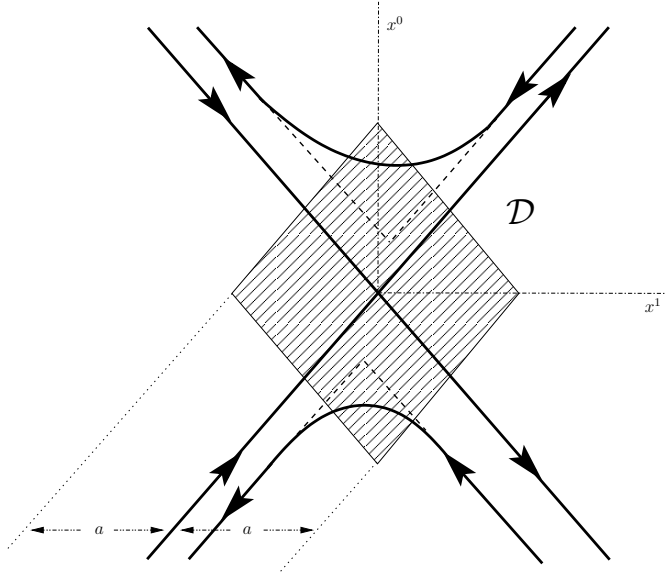}
  \caption{Spacetime region ${\cal D}$ relevant for the integration over
     the positions $y_0$ and $y_0'$ of the tips of the lower and upper
     ``wedge'' in Eq.~\eqref{eq:deltazeta_2_bis}.}  
  \label{fig:domain}
\end{figure}

The second step is to determine the region of spacetime which affects 
the Wilson lines corresponding to the ``spectator'' fermions. It is
easy to see that each line ``feels'' a strip of spatial width $\sim a$
(up to numerical factors), so that the region which affects both is
approximately given by the domain ${\cal D}$ depicted\footnote{To be
  precise, the domain ${\cal D}$ is defined as the intersection of two
  strips of width $2a$ in the spatial $x^1$-direction, centered along
  the straight-line Wilson lines.} in Fig.~\ref{fig:domain}. Here we
are considering the most favourable case, in which the spatial transverse  
separation is small compared to the longitudinal one.

Taking into account that the typical paths of the exchanged fermions
contributing to the path integral are expected not to depart too much
from the ``wedges'' depicted in Fig.~\ref{fig:long_proj}, we can thus
estimate the relevant region of integration for $y_0$ and $y_0'$ as
the domain ${\cal D}$ in Fig.~\ref{fig:domain}. Indeed, for
$y_0,y_0'\in {\cal D}$ each of the curved Wilson lines interacts with
both the straight-line ones: more precisely, in this case there is a
non-negligible contribution 
from the interaction region near the origin of coordinates, beside the
contributions at early and late times, that are cancelled by the
normalisation factor and do not contribute to the scattering amplitude.
This is necessary if we want that the
exchanged partons be constituents of one meson, ``before'' the
exchange, and of the other meson ``after'' the exchange.\footnote{Here
``before'' and ``after'' do not refer to the temporal evolution, but
to the evolution of the process as seen from the exchanged partons'
point of view.} In the representation of the process in terms of
Wilson lines, this is the counterpart of Feynman's picture of the
exchanged partons as being part of the wave functions of both the
interacting hadrons.  

The final step is to estimate the integral
Eq.~\eqref{eq:deltazeta_2_bis}. The simplest approximation is to take
$f(y_0^0,y_0^1,y_0'{}^0,y_0'{}^1)$ as a constant inside of ${\cal D}$
and zero outside, i.e.,
\begin{equation}
  \label{eq:deltazeta_2_bis_2}
  \begin{aligned}
   I_{\cal R} &\simeq  \int d^2y_0 d^2y_0'
e^{i(p_\barq'+ p_q')\cdot y_0'} e^{-i(p_\barq+ p_q)\cdot
  y_0}f(0,0,0,0)\chi_{\cal D}(y)\chi_{\cal D}(y')\,,
  \end{aligned}
\end{equation}
where $\chi_{\cal D}$ is the  characteristic function of ${\cal D}$,
which can be conveniently expressed as
\begin{equation}
\label{eq:characteristic_f}
  \chi_{\cal D}(y)= \Theta\left(|y\cdot u_1^\perp| - a\cosh\f{\chi}{2}\right)
  \Theta\left(|y\cdot u_2^\perp| - a\cosh\f{\chi}{2}\right) \,,
\end{equation}
with $\Theta(x)$ the Heaviside step function, and with $u_i^\perp$ the
vector orthogonal to $u_i$ in Minkowski metric,
\begin{equation}
  u_1^\perp = \left(\sinh\f{\chi}{2},\cosh\f{\chi}{2},\vec{0}_\perp\right)\,,
  \qquad u_2^\perp =
  \left(\sinh\f{\chi}{2},-\cosh\f{\chi}{2},\vec{0}_\perp\right)\,. 
\end{equation}
The integral can be easily evaluated, and for large $\chi$ it gives
\begin{equation}
  \label{eq:IR_estimate}
  \begin{aligned}
  I_{\cal R} \simeq & \, f(0,0,0,0)\f{1}{(\sinh\chi)^2}\left(\f{2\pi}{m}\right)^4 
  \delta_{aE}(1-\zeta_1) \delta_{aE}(\zeta_2)
  \delta_{aE}(1-\zeta_1') \delta_{aE}(\zeta_2')\,, \\
\delta_\Lambda(x) &\equiv \f{\sin \Lambda x}{\pi x}\,.
  \end{aligned}
\end{equation}
For large energy we recover the delta functions of
Eq.~\eqref{eq:deltazeta_3}, since, as it is well 
known, $\delta_\Lambda(x) \to \delta(x)$ for $\Lambda\to\infty$. 
However, the expression above gives us the possibility to determine how the
limits $\zeta_1\to 1$, $\zeta_2\to 0$ have to be taken: since
$\sin\Lambda x / x \simeq \Lambda$ when $\Lambda x \ll 1$, we have to
set $\zeta_2\simeq 1/aE$ and $1-\zeta_1\simeq 1/aE$ when the energy is
large. Indeed, as we will see in a moment, the integrals in the
$\zeta$-variables are essentially of the form 
\begin{equation}
  \label{eq:zeta_dep_E}
  \begin{aligned}
  I_\alpha &= \int_0^1 d\zeta \zeta^\alpha \f{\sin aE\zeta}{\pi\zeta}
  \simeq \f{aE}{\pi}\int_0^{\f{1}{aE}} d\zeta \zeta^\alpha  =
  \f{(aE)^{-\alpha}}{\pi} \int_0^1 dx x^\alpha \\ &=
  \f{(aE)^{-\alpha}}{\pi(\alpha+1)} =
  \f{1}{\pi(\alpha+1)}\int_0^1d\zeta \zeta^\alpha
  \delta\left(\zeta-(aE)^{-1}\right) \, ,
\end{aligned}
\end{equation}
so that actually their evaluation is equivalent to setting
$\zeta=1/aE$, up to numerical factors. The case of $\zeta\to 1$ is
completely analogous, and the result is shown to be equivalent to setting
$1-\zeta=1/aE$. In order to see how this kind
of integrals comes about, we have to discuss the dependence on the
$\zeta$-variables of the various quantities.\footnote{ It is worth
  mentioning that the same conclusions are obtained exploiting the
  interpretation of $a$ as the typical linear size of the domains
  where colour fields are highly correlated in the QCD
  vacuum~\cite{NR,BHN}. Adopting this point of view, the relevant region
  of integration for $y_0,y_0'$ is determined by requiring that the
  incoming and outgoing partons spend some time in the same colour
  domain $|y_0^2|,|y_0^{\prime 2}|\lesssim a^2$. This excludes the
  case where $y_0,y_0'$ are in the ``tails'', since one of the sides
  of the ``wedges'' would lie almost entirely outside of the domain.
  The relevant region reduces therefore to $|y_0^{0,1}|,|y_0^{\prime
    0,1}|\lesssim a$, which yields the same estimate
  Eq.~\eqref{eq:IR_estimate} for
  $I_{\cal R}$, up to the replacement
  $\delta_{aE}(1\!-\zeta_1^{(\prime)}) \delta_{aE}(\zeta_2^{(\prime)})\!\to\!
  2\delta_{aE}(1\!-\zeta_1^{(\prime)}\!+\zeta_2^{(\prime)})
  \delta_{aE}(1\!-\zeta_1^{(\prime)}\!-\zeta_2^{(\prime)})$.}

A first possible source of $\zeta$ factors are the ``contracted spin
factors'' $\Sps_\wedge^{t_\barq s_q}$ and $\Sps_\vee^{s_q' t_\barq'}$,
defined in Eq.~\eqref{eq:sf}. In order to see that they are
independent of the $\zeta$-variables in a first approximation, let us
write down explicitly the bispinors corresponding to $q$ and $\barq$
(in the Dirac basis), 
\begin{equation}
  \label{eq:Dspinors}
  \begin{aligned}
    u^{(s_q)}(p_q) &= \sqrt{\zeta_2}\sqrt{E+m}
    \left(
      \begin{array}{c}
        \phi^{(s_q)}\\
        \f{-\zeta_2 p \sigma_1 +
          \vec{p}_{q\perp}\cdot\vec{\sigma}_\perp}{\zeta_2(E+m)}
        \phi^{(s_q)}
      \end{array}\right)     \,,\\
v^{(t_\barq)}(p_\barq) &= \sqrt{1-\zeta_1}\sqrt{E+m}
\left(
      \begin{array}{c}
        \f{(1-\zeta_1) p \sigma_1 +
          \vec{p}_{\barq\perp}\cdot\vec{\sigma}_\perp}{(1-\zeta_1)(E+m)}       
        \tilde{\phi}^{(t_\barq)}\\  
        \tilde{\phi}^{(t_\barq)}        
      \end{array}\right)  \,,
  \end{aligned}
\end{equation}
where $\phi^{(s_q)}$ and $\tilde{\phi}^{(t_\barq)}$ are two-component
spinors. The dependence on $\zeta$ is indeed of the form considered in
Eq.~\eqref{eq:zeta_dep_E}. Moreover, since the square-root factors
are canceled by the denominators of $\Sps_\wedge^{t_\barq s_q}$ and
$\Sps_\vee^{s_q' t_\barq'}$, the only dependence comes from the
terms involving the transverse momentum of the partons
$\vec{p}_{q,\barq\perp} = \vec{p}_{1,2\perp}/2 \pm \vec{k}_\perp$. However,
the transverse momentum $\vec{k}_\perp$ of a parton inside the meson
is typically distributed around zero with a width approximately equal
to the mass of the meson, and moreover the transverse momentum of the
scattered mesons is small, due to the softness of the
process. Recalling from Eq.~\eqref{eq:zeta_dep_E} that $\zeta\sim
1/aE$, and that the vacuum correlation length is of the order of
$a\sim 0.2\div 0.3\,{\rm fm} \sim 1\div 1.5\,{\rm
  GeV}^{-1}$~\cite{corr_length4,corr_length5,corr_length6,corr_length7,
corr_length8,corr_length9}, we can estimate   
\begin{equation}
  \f{|\vec{p}_{q\perp}|}{\zeta_2(E+m)} 
\lesssim   \f{|\f{\vec{p}_{\perp}}{2}| + |\vec{k}_\perp|}{\f{1}{aE}(E+m)} 
\lesssim \f{a(|\vec{q}_\perp| + m)}{2}\,, 
\end{equation}
which is less than one in the considered range of $t$ and for not too
heavy mesons. This estimate is quite conservative, and actually we
expect that the typical transverse momentum of a parton involved in
the process is of the same order of $\sqrt{-t}$: for very small
transferred momentum $\sqrt{-t}\ll 1\, {\rm GeV}$, we expect therefore
that we can neglect the terms involving the transverse momentum of the
partons, at least in a first approximation, so that the ``contracted spin
factors'' are independent of $\zeta$.

The second source of $\zeta$-factors are the coordinate-space wave
functions $\wf_{i}(\zeta)$, where we have dropped the dependence on
irrelevant variables: indeed, they contain a ``kinematical'' factor 
$\sqrt{\zeta(1-\zeta)}$, as well as any possible dependence on $\zeta$
coming from the momentum-space wave functions $\psi_{i}(\zeta)$. What
matters here is the dependence on $\zeta$ near the endpoints, which we
take of the usual form $\wf_{i}(\zeta) =
\zeta^{\f{1}{2}+\beta_i}(1-\zeta)^{\f{1}{2}+\gamma_i} g(\zeta)$ with
$g(0),g(1)\ne 0$.\footnote{Note that with respect to the discussion in 
  subsection \ref{subsec:x0} we have redefined $\beta\to
  1/2+\beta$ and $\gamma\to 1/2+\gamma$. } 
We assume for simplicity that $\beta_i$, $\gamma_i$ are independent of
the dipole size. For $\zeta\lesssim 1/(aE)\to 0$ one has that
$\wf_{i}(\zeta) \mathop\simeq_{\zeta\to 0} \zeta^{\f{1}{2}+\beta_i}
g(0)$, and in turn $\rho_i\mathop\simeq_{\zeta\to 0}
\zeta^{1+2\beta_i} [g(0)]^2$ (see Eq.~\eqref{eq:rhos}), and we see 
therefore that the form considered in Eq.~\eqref{eq:zeta_dep_E} is
actually correct, with $\alpha=1+2\beta_i$. Similarly, for
$1-\zeta\lesssim 1/(aE)\to 0$ one has that $\wf_{i}(\zeta)
\mathop\simeq_{\zeta\to 1} (1-\zeta)^{\f{1}{2}+\gamma_i} g(1)$, and
$\rho_i \mathop\simeq_{\zeta\to 1} (1-\zeta)^{1+2\gamma_i} [g(1)]^2$, 
and Eq.~\eqref{eq:zeta_dep_E} is obtained by changing variables to
$1-\zeta \to \zeta$, and setting $\alpha=1+2\gamma_i$. 

A few comments are now in order. 
\begin{itemize}
\item The restriction $\zeta\lesssim (aE)^{-1}$
is in accordance with Feynman's picture of high-energy
scattering~\cite{FeynH}, 
implying that only ``wee'' partons participate to the interaction. 
In our setting, this can be understood qualitatively in terms of
uncertainty relations in the following way. Assuming that the
interaction takes place in a spatial region of extension $\sim a$ in
the direction of flight, one has $\Delta x\sim a$; since an exchanged
parton, say, the quark $q$, belongs to the wave functions of both the
interacting hadrons, its momentum can be both $+p_q$ and $-p_q$, so
that $\Delta p_x \sim 2 p_q \sim 2 \zeta_2 E$; finally, from $\Delta
x\Delta p_x \sim 1$ one gets $\zeta_2 \sim 1/2aE$.
\item The approximation considered here is
rather crude, and a more detailed study is needed to check if the
estimate of the relevant region of integration is correct. Indeed, a
different dependence of the domain ${\cal D}$ on the angle $\chi$
could change the way in which $\zeta\to 0$ as a function of energy. On
the other hand, only the way in which the relevant region depends on
$\chi$ is relevant to this extent, and not the detailed
functional form of $f$: for example, modifying the characteristic functions
$\chi_{\cal D}$ in Eq.~\eqref{eq:deltazeta_2_bis_2}, by substituting
the Heaviside functions in Eq.~\eqref{eq:characteristic_f} with
damping exponentials, would yield again $\zeta\lesssim (aE)^{-1}$,
while changing of course the numerical prefactors. 
\item Finally, notice that the
behaviour of the wave functions near the endpoints affects the
dependence on energy of the Reggeon-exchange amplitude, but only
through an overall power-law factor which does not depend on $t$. In
the language of Regge theory, this corresponds to a constant
shift of the Regge trajectory. 
\end{itemize}
This last point requires to be developed in details. We
have that the Reggeon-exchange amplitude is proportional to 
\begin{equation}
\label{eq:sdep_1}
  \A_{{\cal R}_1} \propto s I_{1+2\gamma_1}I_{1+2\beta_2} \propto s
  (aE)^{-2(1+\gamma_1+\beta_2)} = s\left( 
\f{4}{a^2 s}\right)^{1+\gamma_1+\beta_2} \sim s^{-\gamma_1-\beta_2}\,,
\end{equation}
where $I_\alpha$ is defined in Eq.~\eqref{eq:zeta_dep_E}, and 
where in principle $\beta_i,\gamma_i$, which appear in the meson wave
function, depend on the type of meson, but not on the transferred
momentum $t$. 
By the same token, we have for the other Reggeon-exchange amplitude
\begin{equation}
\label{eq:sdep_2}
  \A_{{\cal R}_2} \propto s I_{1+2\gamma_2}I_{1+2\beta_1}  \propto s
  (aE)^{-2(1+\gamma_2+\beta_1)} = s\left( 
\f{4}{a^2 s}\right)^{1+\gamma_2+\beta_1} \sim s^{-\gamma_2-\beta_1}\,,
\end{equation}
and the only thing that changes is the flavour of the exchanged
fermion-antifermion pair. 
Notice that the right-hand side of the equations above does not
contain the whole dependence on energy of the amplitude, but that
nevertheless the remaining dependence is a universal function of
$E/m$. Therefore, universality and degeneracy of the subleading Regge
trajectories, as observed experimentally, hint to a universal
behaviour of the wave functions near the endpoints. Indeed, 
one can immediately see that $\beta_1=\gamma_2$ and
$\beta_2=\gamma_1$, by using the behaviour of the wave functions under
charge conjugation, $\wf(\vr,\zeta)\to \wf^{(C)}(\vr,\zeta)= \eta_c
\wf(-\vr,1-\zeta)$, with $|\eta_c|=1$. Therefore, universality would
give $\beta_1=\beta_2\equiv \beta$, and so $\wf_{i}(\zeta) = 
[\zeta(1-\zeta)]^{\f{1}{2}+\beta} g_i(\zeta)$, with the same $\beta$
independently of the flavours $q$ and $Q$ of the valence partons,
i.e., independently of the meson. 
On the other hand, it would be interesting to investigate to what extent
the universality of the contribution of the subleading Regge
trajectory, i.e., its independence of the specific scattering process,
is confirmed by experiments. We note in passing that the issue of
universality of the leading contribution in the Wilson-loop formalism
has been recently discussed in~\cite{lat_pomeron}, where strong
indications are found from the lattice results
of~\cite{lattice,instantons} for a universal behaviour of the relevant
Wilson-loop correlation function, and therefore for the hadron-hadron
total cross section.  

Beside universality, another important issue is the understanding of
the relation between our results and the usual picture of Regge
poles in the crossed channel, which is not explicit in our
formalism. A first hint is obtained through the use of gauge/gravity
duality~\cite{Jani,reggeon_duality}, as we will discuss below in Section
\ref{saddle}.

\nsection{Analytic continuation into Euclidean space}
\label{an_cont}

As it is well known, path integrals are difficult to treat 
in Minkowski space-time outside of perturbation theory, due to the
wild fluctuations of the phase factor. 
A more precise definition of path integrals is given by formulating
them in Euclidean space, and by subsequently performing the inverse Wick
rotation $x_{E4}\to ix^0$ to obtain a Minkowskian quantity. Moreover,
a variety of techniques is available to evaluate them
nonperturbatively in Euclidean space, most notably through the lattice
regularisation, and, in recent times, by means of the gauge/gravity
correspondence. However, physical processes happen in Minkowski
space-time, and thus a Euclidean formulation can be provided only
when one has established what is the Minkowskian quantity of
interest. The aim of the derivation of the path-integral
representation for the Reggeon-exchange amplitude, given in the
previous Section, was therefore to identify such a quantity in the
case of interest, and although we have not been completely rigorous
from a mathematical point of view, the resulting expression should
reflect the main properties of the desired amplitude.

The next step is to perform the Wick rotation of this amplitude into
Euclidean space, or, more precisely, to find the Euclidean
quantity whose inverse Wick rotation coincides with the given
amplitude. Moreover, from the practical point of view, it is better to
have a formulation of the Wick rotation of the amplitude in terms of
analytic continuation relations for the ``external parameters'' (e.g.,
in the case at hand, the hyperbolic angle $\chi$ or the length
parameter $T$). For this purpose, we write the dipole-dipole
Reggeon-exchange amplitude in the following form:  
\begin{equation}
  \label{eq:ampl_F}
    \begin{aligned}
    \A_{{\cal R}_1}^{(dd)} =  
-i2s  \left(\f{2\pi}{m}\right)^2  \f{1}{N_c} 
\int d^2b_\perp e^{i\vec{q}_\perp \cdot\vec{b}_\perp}  &
(\bar{v}^{t_\barq}(p_\barq))_{\alpha} (u^{s_q}(p_q))_{\beta} 
(\bar{u}^{s_q'}(p_q'))_{\alpha'} (v^{t_\barq'}(p_\barq'))_{\beta'}
\\ &
\times 
 {\cal F}_{\alpha\beta;\alpha'\beta'}
    (\chi,T;\vec{b}_{\perp},\vec{R}_{1\perp},\vec{R}_{2\perp},
    \vec{R}_{1\perp}',\vec{R}_{2\perp}')  
    \, , 
  \end{aligned}
\end{equation}
where we have introduced the quantity
\begin{equation}
  \label{eq:piece}
  \begin{aligned}
&    {\cal F}_{\alpha\beta;\alpha'\beta'}
    (\chi,T ;\vec{b}_{\perp},\vec{R}_{1\perp},\vec{R}_{2\perp},
    \vec{R}_{1\perp}',\vec{R}_{2\perp}')= 
\\ & \phantom{l}
    \int_{2T-\LL}^{2T+\LL} dL \int_{x_i^{(\wedge)}}^{x_f^{(\wedge)}}
    \DX \int \DP \int_{2T-\LL}^{2T+\LL} dL' 
    \int_{x_i^{(\vee)}}^{x_f^{(\vee)}}  \DXp \int \DPp
    e^{-i(m_q-i\epsilon)(L + L'-4T)} \\ &  \phantom{dioladronacciacci} \times
    (\M_{-T,-T+L}[\dot{X},\Pi])_{\alpha\beta}
    (\M_{-T,-T+L'}[\dot{X}',\Pi'])_{\alpha'\beta'} \,
{\cal U}_{\cal C}[X,L,X',L']
 \,.
  \end{aligned}
\end{equation}
As we will show below, this is the quantity which admits a convenient
Euclidean representation, from which it could be obtained by means of
rather simple analytic continuation relations. Since spinor indices
play no role in the discussion, we will drop them in the following. 
Moreover, Lorentz invariance allows to restrict to the case of
positive hyperbolic angle, without any loss of
information~\cite{crossing}: in the 
following we will therefore take $\chi>0$.

What we expect is that the analytic continuation into Euclidean space
can be achieved by performing the appropriate analytic continuation in
the variables $\chi$ and $T$, as it happens in the case of the
Pomeron-exchange amplitude~\cite{Meggiolaro97,Meggiolaro98,Meggiolaro02,Meggiolaro05,crossing,crossing2,EMduality}.
To show that this is essentially the case, we follow the approach of
\cite{EMduality}, which we briefly recall. The main idea is to perform
an appropriate rescaling of fields and coordinates, in order to show
explicitly the dependence on these variables in the action, while
removing it from the other terms. In the case of the Pomeron-exchange
amplitude there is no integration over trajectories, but only the
functional integration over the gluonic and fermionic fields. In that
case the procedure succeeds completely, and it is possible to
give a nonperturbative justification to the analytic continuation
relations by inspecting the domain of convergence of the functional
integral. In the case at hand, we have to deal also with the
integration over the trajectories and over the momenta of the exchanged
particles: as we will show below, the dependence on the relevant
variables cannot be completely removed from the spin factor. As a
consequence, the derivation of the analytic continuation relation that
we give is formal, and its validity relies on the assumption of an
appropriate analyticity domain.

We proceed now with the derivation. We define $T=\xi\tilde{T}$,
and rescale the gluon fields as follows:
\begin{equation}
  \label{eq:rescale}
  \begin{aligned}
    A_\mu(x) &= \phi_\nu(z) M^\nu_{\phantom{\nu}\mu}(\chi,\xi)\, , \qquad
    z^\mu = M^\mu_{\phantom{\mu}\nu}(\chi,\xi) x^\nu\\
    M^\mu_{\phantom{\mu}\nu}(\chi,\xi) &=
    \diag\left(\f{1}{\sqrt{2}\xi\cosh\f{\chi}{2}},
    \f{1}{\sqrt{2}\xi\sinh\f{\chi}{2}},1,1\right) \, .
 \end{aligned}
\end{equation}
After the rescaling we obtain $\chi,\xi$-dependent expressions for
the Yang-Mills action and for the fermion-matrix determinant,
expressed as functionals of the new gauge field $\phi(z)$, i.e.,
$S_{\rm YM}[A(x)] = S_{M\chi,\xi}[\phi(z)]$, 
${\cal Q}[A(x)] = {\cal Q}_{M\chi,\xi}[\phi(z)]$, whose
explicit forms are not relevant 
here, and which can be found in~\cite{EMduality}. We make this
explicit by rewriting the expectation value with respect to the
rescaled fields as 
$\la\ldots\ra_A = \la\ldots\ra_{{M\chi,\xi}}$.
Moreover, we define $\LL=\xi\tLL$, and we rescale the integration
variables in the path integrals as follows, 
\begin{equation}
  \label{eq:rescale_2}
  \begin{aligned}
    Z^\mu = M^\mu_{\phantom{\mu}\nu}(\chi,\xi) X^\nu\, , \qquad
    \Pi_\mu = \rho_\nu M^\nu_{\phantom{\nu}\mu}(\chi,\xi)\, , \qquad L
    = \xi\tilde{L}\, , 
  \end{aligned}
\end{equation}
and similarly for primed quantities. 
After these transformations, we obtain for ${\cal F}$ the
expression\footnote{\label{foot:ieps} We drop the $-i\epsilon$ term,
  which as we will see is correctly recovered when going back from
  Euclidean to Minkowski space (see footnote \ref{foot:ieps2}).}
\begin{equation}
  \label{eq:piece_2}
  \begin{aligned}
{\cal F}&(\chi,T;\vec{b}_{\perp},\vec{R}_{1\perp},\vec{R}_{2\perp},
\vec{R}_{1\perp}',\vec{R}_{2\perp}' 
) = \tilde{\cal
  F}(\chi,\xi;\tilde{T};\vec{b}_{\perp},\vec{R}_{1\perp},\vec{R}_{2\perp}
,\vec{R}_{1\perp}',\vec{R}_{2\perp}')   
   \equiv 
   \\ &\left(\f{1}{\xi\sinh\chi}\right)^2
   \int_{2\tilde{T}-\tLL}^{2\tilde{T}+\tLL}
    d\tilde{L} \int_{z_i^{(\wedge)}}^{z_f^{(\wedge)}} \DZ \int \DR 
      \int_{2\tilde{T}-\tLL}^{2\tilde{T}+\tLL} d\tilde{L}' 
    \int_{z_i^{(\vee)}}^{z_f^{(\vee)}} \DZp \int  \DRp
     \\ &    \phantom{diosp}
    \times e^{-im_q\xi(\tilde{L}+\tilde{L}'-4\tilde{T})}
    \tilde{\M}_{-\tilde{T},-\tilde{T}+\tilde{L}}[\dot{Z},\rho]
    \tilde{\M}_{-\tilde{T},-\tilde{T}+\tilde{L}'}[\dot{Z}',\rho']
     \tilde{\cal U}_{\tilde{\cal
         C}}[\chi,\xi;Z,\tilde{L},Z',\tilde{L}'] \,,
  \end{aligned}
\end{equation}
where we have denoted the spin factors expressed in terms of the new
variables as
\begin{equation}
  \label{eq:red_M}
  \begin{aligned}
    \tilde{\M}_{-\tilde{T},-\tilde{T}+\tilde{L}}[\dot{Z}, \rho] =
    \Texp &\left[i\int_{-\tilde{T}}^{-\tilde{T}+\tilde{L}} d\tau
      \left(\xi \rho_\mu (\tau) M^\mu_{\phantom{\mu}\nu}(\chi,\xi)\gamma^\nu  -
        \rho_\mu(\tau)\dot{Z}^\mu(\tau)\right)\right]\, ,
 \end{aligned}
\end{equation}
and similarly for
$\tilde{\M}_{-\tilde{T},-\tilde{T}+\tilde{L}'}[\dot{Z}', \rho']$, 
and where the normalised expectation value $\tilde{\cal U}_{\tilde{\cal C}}$,
\begin{multline}
  \label{eq:tildeW}
\tilde{\cal U}_{\tilde{\cal C}}[\chi,\xi;Z,\tilde{L},Z',\tilde{L}']
= {\la \tilde{\W}_{\tilde{\cal
     C}}[Z,\tilde{L},Z',\tilde{L}'] 
     \ra_{{M\chi,\xi}}} \Big[
       \la\tilde{\W}_1^{\tilde T}(\vec{b}_\perp,\vec{R}_{1\perp})\ra_{{M\chi,\xi}}
       \la\tilde{\W}_2^{\tilde
         T}(\vec{0}_\perp,\vec{R}_{2\perp})\ra_{{M\chi,\xi}} \\ \times
       \la\tilde{\W}_1^{\tilde
         T}(\vec{b}_\perp-\Delta\vec{R}_{1\perp},\vec{R}_{1\perp}')
       \ra_{{M\chi,\xi}}   
       \la\tilde{\W}_2^{\tilde
         T}(\Delta\vec{R}_{2\perp},\vec{R}_{2\perp}')\ra_{{M\chi,\xi}} 
\Big]^{-1}\, ,
\end{multline}
is defined in terms of the Wilson loops
\begin{equation}
  \label{eq:tilde_W}
  \begin{aligned}
    \tilde{\W}_{\tilde{\cal
     C}}[Z,\tilde{L},Z',\tilde{L}'] &= 
 \f{1}{N_c} \,\tr\,\Texp\left\{-ig\oint_{\tilde{\cal C}} \phi_\mu(z) 
  dz^\mu\right\}\,,\\
\tilde{\W}_i^{\tilde T}(\vec{d}_{i\perp},\vec{D}_{i\perp}) &=
\f{1}{N_c} \,\tr\,\Texp\left\{-ig\oint_{\tilde{\bar{{\cal
          C}}}^{(i)}(\vec{d}_{\perp},\vec{D}_{\perp})}  
  \phi_\mu(z)    dz^\mu\right\}\,.
  \end{aligned}
\end{equation}
The paths entering Eq.~\eqref{eq:tilde_W} are defined as follows,
\begin{equation}
  \label{eq:red_W}
\tilde{\cal C} = \tilde{\cal C}^{(1)}_+\circ \tilde{\cal
        C}^{(\vee)}\circ \tilde{\cal C}^{(2)}_-\circ \tilde{\cal
        C}^{(\wedge)}\, , \quad
\tilde{\bar{{\cal C}}}^{(i)}(\vec{d}_{\perp},\vec{D}_{\perp}) = \tilde{\bar{{\cal
  C}}}^{(i)}_+(\vec{d}_{\perp},\vec{D}_{\perp})\circ \tilde{\bar{{\cal 
  C}}}^{(i)}_-(\vec{d}_{\perp},\vec{D}_{\perp})\, ,
\quad i=1,2\,, 
\end{equation}
and the various pieces are given by the following expressions: for the
straight-line parts, 
\begin{equation}
  \label{eq:red_W_2}
\tilde{\cal C}^{(1)}_+ :\,\, Z^{(1)}_+(\tau) = n_1\tau + b +
\f{R_1}{2}\,, 
 \qquad \tilde{\cal C}^{(2)}_- :\,\, Z^{(2)}_-(\tau) = -n_2\tau - \f{R_2}{2}\, ,
\end{equation}
with $\tau \in [-\tilde{T},\tilde{T}]$, and
\begin{equation}
  \label{eq:red_W_2bis}
\tilde{\bar{{\cal C}}}^{(i)}_\pm(\vec{d}_{\perp},\vec{D}_{\perp})
:\,\, \tilde{\bar{Z}}^{(i)}_\pm(\tau) = \pm n_i\tau + d \pm
\f{D}{2}\,, 
\end{equation}
with $\tau \in [-\f{\tilde{T}}{2},\f{\tilde{T}}{2}]$, where
\begin{equation}
  \label{eq:red_W_2_ext}
n_1 =
 \f{1}{\sqrt{2}}(1,1,\vec{0}_\perp)\, , \quad 
n_2 = \f{1}{\sqrt{2}}(1,-1,\vec{0}_\perp)\, , \quad
d = (0,0,\vec{d}_\perp)\,, \quad
D=(0,0,\vec{D}_\perp)\,,
\end{equation}
while for the curved parts 
\begin{equation}
  \label{eq:red_W_3}
  \begin{aligned}
\tilde{\cal C}^{(\wedge)} :&~Z(\tau),
\qquad \phantom{'}\tau \in[-\tilde{T},-\tilde{T}+\tilde{L}]\,, \\ 
&~Z(-\tilde{T}) = -n_2\tilde{T}  + \f{R_2}{2} = z_i^{(\wedge)}\,, &&&
&Z(-\tilde{T}+\tilde{L}) = -n_1\tilde{T} + b - \f{R_1}{2} = z_f^{(\wedge)}\, ,\\
\tilde{\cal C}^{(\vee)} :&~Z'(\tau), 
\qquad \tau \in[-\tilde{T},-\tilde{T}+\tilde{L}']\,, \\
 &~Z'(-\tilde{T}) =
n_1\tilde{T} + b  -R_1' + \f{R_1}{2} = z_i^{(\vee)}\,, &&& 
&Z'(-\tilde{T}+\tilde{L}') = n_2\tilde{T} + R_2' -
\f{R_2}{2}=z_f^{(\vee)}\,.
  \end{aligned}
\end{equation}
The various pieces are connected by the appropriate straight-line
paths in the transverse plane at $\tau=\pm \tilde T$ or $\tau=\pm
\tilde T/2$, which we are not writing down explicitly. As we have already
mentioned, the only 
dependence left on $\chi$ and $\xi$ is in the rescaled action, and in
the $\gamma$-matrix term in the spin factor. Also, additional
dependence could appear when introducing the appropriate
regularisation, which is required to make the spin factor a
mathematically meaningful quantity~\cite{Korchemsky,Korchemsky2}. For
the time being we are therefore unable to give a complete proof for
the analytic continuation, including the determination of a
sufficiently wide analyticity domain. Here we limit ourselves to the
determination of the appropriate relation which allows to go from
Minkowski to Euclidean space, and viceversa, assuming the existence of
such a domain. 

The stage is now set to determine the form of the analytic continuation
into Euclidean space. We know from \cite{EMduality} that performing 
the analytic continuation $\xi\to -i\eta, \chi\to i\theta$, the
Yang-Mills action and the fermion-matrix determinant go over into 
rescaled versions of the Euclidean Yang-Mills action and of the
Euclidean fermion-matrix determinant, respectively.  
More precisely, the rescaled Euclidean action $S_{E{\rm
    YM}}[A_E(x_E)]=S_{E\theta,\eta}[\phi(z)]$, and the rescaled
Euclidean fermion-matrix determinant ${\cal Q}_E[A_E(x_E)] = {\cal
  Q}_{E\theta,\eta}[\phi(z)]$, are obtained by performing
the following transformation of fields and coordinates,
\begin{equation}
  \label{eq:rescale_E}
  \begin{aligned}
   z^\mu &= M_{E\mu\nu} P_{\nu\rho} x_{E\rho}\, , \qquad
   A_{E\mu}(x_E) = \phi_\rho(z) P_{\rho\nu} M_{E\nu\mu}\\
   M_{E\mu\nu} &=
    \diag\left(\f{1}{\sqrt{2}\eta\cos\f{\theta}{2}},
    \f{1}{\sqrt{2}\eta\sin\f{\theta}{2}},1,1\right)\, ,
 \end{aligned}
\end{equation}
where the matrix $P$ permutes the components of the Euclidean
coordinates in order to put them in the order $4123$, and the values
0 and 4 of the spacetime index are identified. The use of a 
contravariant index for the Euclidean coordinate causes no ambiguity.
The relation between the Minkowskian and Euclidean action and
fermion-matrix determinant is expressed as
\begin{equation}
  \label{eq:actions}
  S_{E\theta,\eta}[\phi(z)] = S_{M i\theta,-i\eta}[\phi(z)]\,, \qquad 
  {\cal Q}_{E\theta,\eta}[\phi(z)] = {\cal Q}_{M i\theta,-i\eta}[\phi(z)]\,.
\end{equation}
The analytic continuation Eq.~\eqref{eq:actions} is valid for
$\theta\in(0,\pi)$, and starting from the analytic expression 
at $\chi>0$ in Minkowski space. The restriction on $\theta$ does not
cause any loss of information, due to the $O(4)$ invariance of the
Euclidean theory~\cite{crossing}. Performing now the analytic
continuation $\xi\to -i\eta, \chi\to i\theta$ in \eqref{eq:piece_2},
we obtain  
\begin{equation}
  \label{eq:piece_3}
  \begin{aligned}
    \tilde{\cal F}&
    (i\theta,-i\eta;\tilde{T};\vec{b}_{\perp},\vec{R}_{1\perp},\vec{R}_{2\perp}
    ,\vec{R}_{1\perp}',\vec{R}_{2\perp}')   
    = \tilde{\cal F}_E
    (\theta,\eta;\tilde{T};\vec{b}_{\perp},\vec{R}_{1\perp},\vec{R}_{2\perp}
    ,\vec{R}_{1\perp}',\vec{R}_{2\perp}')
    \equiv\\ &     \left(\f{1}{\eta\sin\theta}\right)^2 
\int_{2\tilde{T}-\tLL}^{2\tilde{T}+\tLL}
    d\tilde{L} \int_{z_i^{(\wedge)}}^{z_f^{(\wedge)}} \DZ \int \DR 
      \int_{2\tilde{T}-\tLL}^{2\tilde{T}+\tLL} d\tilde{L}' 
    \int_{z_i^{(\vee)}}^{z_f^{(\vee)}} \DZp \int \DRp
     \\ & \phantom{trem}
\times e^{-m_q\eta(\tilde{L}+\tilde{L}'-4\tilde{T})}
    \tilde{\M}^{(E)}_{-\tilde{T},-\tilde{T}+\tilde{L}}[\dot{Z},\rho]
    \tilde{\M}^{(E)}_{-\tilde{T},-\tilde{T}+\tilde{L}'}[\dot{Z}',\rho']
    \tilde{\cal U}^{(E)}_{\tilde{\cal
        C}}[\theta,\eta;Z,\tilde{L},Z',\tilde{L}'] \,,
  \end{aligned}
\end{equation}
where we have used the notation 
\begin{equation}
  \label{eq:red_M_E}
  \begin{aligned}
    \tilde{\M}^{(E)}_{-\tilde{T},-\tilde{T}+\tilde{L}}[\dot{Z},\rho]
    &=  \Texp  \left[i\int_{-\tilde{T}}^{-\tilde{T}+\tilde{L}} d\tau
      \left(\eta \rho_\mu (\tau) M_{E\mu\nu}(\theta,\eta) \gamma_{E\nu} -
        \rho_\mu(\tau)\dot{Z}^\mu(\tau)\right)\right]\, ,
 \end{aligned}
\end{equation}
and similarly for $
\tilde{\M}^{(E)}_{-\tilde{T},-\tilde{T}+\tilde{L}'}[\dot{Z}',\rho']$,  
for the analytically-continued spin factor, with $\gamma_{E\mu}$ the
Euclidean gamma-matrices $\gamma_{E0}=\gamma^0$,
$\gamma_{Ej}=-i\gamma^j$, and  
\begin{multline}
  \label{eq:tildeW_E}
\tilde{\cal U}^{(E)}_{\tilde{\cal
    C}}[\theta,\eta;Z,\tilde{L},Z',\tilde{L}'] 
= \la \tilde{\W}_{\tilde{\cal
     C}}[Z,\tilde{L},Z',\tilde{L}'] 
     \ra_{{E \theta,\eta}} \Big[\la
       \tilde{\W}_1^{\tilde T}(\vec{b}_\perp,\vec{R}_{1\perp})\ra_{{E \theta,\eta}}\la
       \tilde{\W}_2^{\tilde T}(\vec{0}_\perp,\vec{R}_{2\perp})\ra_{{E
           \theta,\eta}}
       \\ \times \la 
       \tilde{\W}_1^{\tilde
         T}(\vec{b}_\perp-\Delta\vec{R}_{1\perp},\vec{R}_{1\perp}')\ra_{{E
           \theta,\eta}} 
       \la \tilde{\W}_2^{\tilde
         T}(\Delta\vec{R}_{2\perp},\vec{R}_{2\perp}')\ra_{{E
           \theta,\eta}}\Big]^{-1}\, , 
\end{multline}
for the analytically-continued normalised expectation value. 
Notice that the Wilson loops in Eq.~\eqref{eq:tildeW_E} are exactly the
same defined above in Eq.~\eqref{eq:tilde_W}. The expectation 
value obtained using the rescaled Euclidean Yang-Mills action and
rescaled Euclidean fermion-matrix determinant has been denoted with
$\la\ldots\ra_{E \theta,\eta}$. 

Now, we already know from~\cite{EMduality} that
the expectation values $\la \tilde{\W}_i^{\tilde T}\ra_{{E \theta,\eta}}$
in Eq.~\eqref{eq:tildeW_E} are simply the expression
in terms of the rescaled Euclidean action of the expectation values
$\la\W^{(E)\,T_E}_{i}\ra_E$, where $\la\ldots\ra_E$ is the expectation
value in the sense of the usual Euclidean functional integral, and the
Wilson loops $\W^{(E)\,T_E}_{i}$ are defined as follows, 
\begin{equation}
  \label{eq:loop_def_E0}
 \W^{(E)\,T_E}_{i}(\vec{d}_{\perp},\vec{D}_{\perp})= \f{1}{N_c} \,\tr\,
 \Texp\left\{-ig\oint_{\bar{{\cal
       C}}_{E}^{(i)}(\vec{d}_{\perp},\vec{D}_{\perp})} A_{E}(x_E) \cdot 
   dx_{E}\right\}\,,
\end{equation}
where the dot stands for the Euclidean scalar product, and the paths
are defined as 
\begin{equation}
  \label{eq:def_W_E0}
\bar{{\cal C}}_{E}^{(i)}(\vec{d}_{\perp},\vec{D}_{\perp}) = \bar{{\cal
  C}}^{(i)}_{E+}(\vec{d}_{\perp},\vec{D}_{\perp})\circ \bar{{\cal 
  C}}^{(i)}_{E-}(\vec{d}_{\perp},\vec{D}_{\perp})\, , \quad i=1,2\,,
\end{equation}
with the various pieces being given by the straight lines
\begin{equation}
  \label{eq:def_W_E_2}
  \begin{aligned}
\bar{{\cal C}}^{(i)}_{E\pm}(\vec{d}_{\perp},\vec{D}_{\perp}) &:\,
\bar{X}^{(i)}_{E\pm}(\tau) = \pm u_{Ei}\tau + d_E \pm \f{D_{E}}{2}\,, 
  \end{aligned}
\end{equation}
with $\tau\in[-\textstyle\f{{T}_E}{2},\textstyle\f{{T}_E}{2}]$, having set
$T_E=\eta\tilde{T}$, and with
\begin{equation}
  \label{eq:def_W_E_2_ext}
  \begin{aligned}
 u_{E1} &= (\sin\theta,\vec{0}_\perp,\cos\theta)\, , &&&
u_{E2} &= (-\sin\theta,\vec{0}_\perp,\cos\theta)\, ,
\\  
d_E &= (0,\vec{d}_\perp,0)\, , &&& D_{E} &=
(0,\vec{D}_{\perp},0)\, ,
  \end{aligned}
\end{equation}
and moreover closed by appropriate straight-line paths in the
transverse plane at $\pm T_E/2$.
In particular, $d_E$ and $D_E$ take the following values,
$d_E=0$, $b_E$, $b_E-\Delta R_{E1}$, $\Delta R_{E2}$ and
$D_E=R_{E1,2}$, $R_{E1,2}'$, with 
\begin{equation}
  \label{eq:eucl_trans_fourvect}
    b_E = (0,\vec{b}_\perp,0)\, , \quad R_{Ei} =
    (0,\vec{R}_{i\perp},0)\, , \quad R_{Ei}' =
    (0,\vec{R}_{i\perp}',0)\, ,   
    \quad \Delta R_{Ei} =
    (0,\Delta\vec{R}_{i\perp},0)\, .
\end{equation}
To see what the other terms correspond to in the Euclidean theory
expressed through the usual variables, we have to rescale back
coordinates and momenta in the path integral according to the
following transformations, 
\begin{equation}
  \label{eq:rescale_E_2}
  \begin{aligned}
    Z^\mu = M_{E\mu\nu} P_{\nu\rho} X_{E\nu}\, , \qquad
\Pi_{E\mu} = \rho_\rho P_{\rho\nu} M_{E\nu\mu}\,,
  \end{aligned}
\end{equation}
and moreover to set ${L}_{E0} =\eta\tilde{L}_0$, ${L}_E
=\eta\tilde{L}$, and similarly for primed quantities. It is then
immediate to see that the analytically-continued spin factor is simply
the rescaled version of the usual Euclidean spin factor, 
\begin{equation}
  \label{eq:red_M_E_2}
  \begin{aligned}
    {\M}^{(E)}_{-T_E,T_E+L_E}[\dot{X}_E,\Pi_E] &= 
    \Texp \left[i\int_{-T_E}^{T_E+{L}_E} d\tau
      \left(\sla{\Pi}_E (\tau) -
        \Pi_{E}(\tau)\cdot \dot{X}_{E}(\tau)\right)\right]\, , 
 \end{aligned}
\end{equation}
with $\sla{\Pi}_E=\Pi_{E\mu}\gamma_{E\mu}$, and similarly for
${\M}^{(E)}_{-T_E,T_E+L'_E}[\dot{X}'_E,\Pi_E']$; also, the expectation
value $\la \tilde{\W}_{\tilde{\cal
    C}}[Z,\tilde{L},Z',\tilde{L}']\ra_{{E \theta,\eta}}$ is equal to  
the usual expectation value $\la\W^{(E)}_{{\cal
    C}_E}[X_E,{L}_E,X_E',{L}_E']\ra_E$ of the following Euclidean
Wilson loop, 
\begin{equation}
  \label{eq:loop_def_E}
  \begin{aligned}
  \W^{(E)}_{{\cal C}_E}[X_E,{L}_E,X_E',{L}_E'] =&
 \f{1}{N_c} \,\tr\, \Texp\left\{-ig\oint_{{\cal C}_E} A_{E}(x_E)\cdot
   dx_{E}\right\}\,,
\end{aligned}
\end{equation}
where the path ${\cal C}_E$ is defined as
\begin{equation}
  \label{eq:def_W_E}
  \begin{aligned}
{\cal C}_E &= {\cal C}^{(1)}_{E+}\circ {\cal
        C}_E^{(\vee)}\circ {\cal C}^{(2)}_{E-}\circ {\cal
        C}_E^{(\wedge)}\, ,
  \end{aligned}
\end{equation}
with ${\cal C}^{(1)}_{E+}$, ${\cal C}^{(2)}_{E-}$ given by 
\begin{equation}
  \label{eq:def_W_E_2_rep}
  \begin{aligned}
{\cal C}^{(1)}_{E+} &:\, X^{(1)}_{E+}(\tau) = u_{E1}\tau + b_E + \f{R_{E1}}{2}\,, 
&&&  
{\cal C}^{(2)}_{E-} &:\, X^{(2)}_{E-}(\tau) = -u_{E2}\tau -
\f{R_{E2}}{2}\, ,
  \end{aligned}
\end{equation}
with $\tau\in[-{T}_E,{T}_E]$, and moreover
\begin{equation}
  \label{eq:def_W_E_3}
  \begin{aligned}
{\cal C}_E^{(\wedge)} :&~X_E(\tau)\,, 
\qquad\phantom{'}\tau\in[-{T}_E,-{T}_E+{L}_E] \\ 
&~X_E(-{T}_E) = -u_{E2}{T}_E + \f{R_{E2}}{2}  &&&
&X_E(-{T}_E+{L}_E) = -u_{E1}{T}_E + b_E - \f{R_{E1}}{2} \\
&~\phantom{X_E(-{T}_E)}=  x_{Ei}^{(\wedge)}\,, &&& &\phantom{X_E(-{T}_E+{L}_E)}=
x_{Ef}^{(\wedge)}\, , \\
{\cal C}_E^{(\vee)} :&~X'_E(\tau)\,,
\qquad\tau\in[-{T}_E,-{T}_E+ {L}_E']
\\
&~X'_E(-{T}_E) = u_{E1}{T}_E + b_E - R_{E1}'+ \f{R_{E1}}{2} 
&&& &X'_E(-{T}_E+{L}_E') = u_{E2}{T}_E + R_{E2}' - 
\f{R_{E2}}{2}  \,,\\ 
&~\phantom{X'_E(-{T}_E)}= x_{Ei}^{(\vee)}\,, &&&
&\phantom{X'_E(-{T}_E+{L}_E')}= x_{Ef}^{(\vee)} \,. 
  \end{aligned}
\end{equation}
It turns out therefore that
$\tilde{\cal F}_E$ is simply the rescaled version of ${\cal F}_E$, 
\begin{equation}
  \label{eq:piece_3.5}
      \tilde{\cal
      F}_E(\theta,\eta;\tilde{T};\vec{b}_{\perp},\vec{R}_{1\perp},\vec{R}_{2\perp}
    ,\vec{R}_{1\perp}',\vec{R}_{2\perp}')
    = {\cal
      F}_E(\theta,T_E;\vec{b}_{\perp},\vec{R}_{1\perp},\vec{R}_{2\perp}
    ,\vec{R}_{1\perp}',\vec{R}_{2\perp}')\,,
\end{equation}
where
\begin{equation}
  \label{eq:piece_4}
  \begin{aligned}
   &{\cal
      F}_E(\theta,T_E;\vec{b}_{\perp},\vec{R}_{1\perp},\vec{R}_{2\perp}
    ,\vec{R}_{1\perp}',\vec{R}_{2\perp}')
 = \\ & 
    \int_{2T_E-\LLe}^{2T_E+\LLe}
    d{L}_E \int_{x_{Ei}^{(\wedge)}}^{x_{Ef}^{(\wedge)}} \DXE \int \DPE 
      \int_{2T_E-\LLe}^{2T_E+\LLe} d{L}_E' 
    \int_{x_{Ei}^{(\vee)}}^{x_{Ef}^{(\vee)}} \DXpE  \int \DPpE 
     \\ & \phantom{madiaaa}
\times e^{-m_q({L}_E+{L}_E'-4T_E)}
    {\M}^{(E)}_{-T_E,-T_E+L_E}[\dot{X}_E,\Pi_E]
    {\M}^{(E)}_{-T_E,T_E+L_E'}[\dot{X}_E',\Pi_E']\,  
\\ & \phantom{madiaaa}
\times
    {\cal U}^{(E)}_{{\cal
        C}_E}[X_E,L_E,X_E',L_E']\, ,
  \end{aligned}
\end{equation}
where the Euclidean normalised expectation value is given by
\begin{multline}
  \label{eq:loop_norm_E}
 {\cal U}^{(E)}_{{\cal
     C}_E}[X_E,L_E,X_E',L_E'] \equiv
 \la \W^{(E)}_{\cal 
     C}[X_E,L_E,X_E',L_E'] 
     \ra_E \Big[\la
     \W^{(E)\,T_E}_{1}(\vec{b}_\perp,\vec{R}_{1\perp})\ra_E
      \\ 
       \times\la
       \W^{(E)\,T_E}_{2}(\vec{0}_\perp,\vec{R}_{2\perp})\ra_E \la
       \W^{(E)\,T_E}_{1}(\vec{b}_\perp-\Delta\vec{R}_{1\perp},\vec{R}_{1\perp}')\ra_E
       \la   
       \W^{(E)\,T_E}_{2}(\Delta\vec{R}_{2\perp},\vec{R}_{2\perp}')\ra_E\Big]^{-1}\, .
\end{multline}
In conclusion, comparing Eqs.~\eqref{eq:piece_2}, \eqref{eq:piece_3}
and \eqref{eq:piece_3.5}, we obtain the desired analytic continuation 
relations connecting the Minkowskian quantity ${\cal F}$, entering the 
expression for the Reggeon-exchange amplitude, and its Euclidean
counterpart ${\cal F}_E$,
\begin{equation}
  \label{eq:an_cont}
  \begin{aligned}
  {\cal F}_E(\theta,T_E,\LLe;
  \vec{b}_{\perp},\vec{R}_{1\perp},\vec{R}_{2\perp}
  ,\vec{R}_{1\perp}',\vec{R}_{2\perp}' ) &= {\cal
    F}(i\theta,-iT_E,-i\LLe;\vec{b}_{\perp},\vec{R}_{1\perp},\vec{R}_{2\perp}
  ,\vec{R}_{1\perp}',\vec{R}_{2\perp}')\,
  ,\\
  {\cal F}(\chi,T,\LL;\vec{b}_{\perp},\vec{R}_{1\perp},\vec{R}_{2\perp} ,\vec{R}_{1\perp}',\vec{R}_{2\perp}') &=
  {\cal
    F}_E(-i\chi,iT,i\LL;\vec{b}_{\perp},\vec{R}_{1\perp},\vec{R}_{2\perp}
  ,\vec{R}_{1\perp}',\vec{R}_{2\perp}' )
  \, .
  \end{aligned}
\end{equation}
The relations Eq.~\eqref{eq:an_cont}, which allow to reconstruct
the physical Reggeon-exchange amplitude from a calculation in
Euclidean space, call for a few remarks.\footnote{\label{foot:ieps2}
  As anticipated in footnote \ref{foot:ieps}, one sees by direct
  inspection that when performing the analytic continuation from
  Euclidean to Minkowski spacetime, the exponential term in
  Eq.~\eqref{eq:piece_4} contains a negative real part in the 
  exponent, so reproducing the $-i\epsilon$ term of the original
  Minkowskian expression.}  
\begin{itemize}
\item The analytic continuation relations Eq.~\eqref{eq:an_cont} are
  similar to those obtained in the case of the Pomeron-exchange amplitude,
  the only modification being that we also have to perform $\LLe\to
  i\LL$. This can be a non trivial task, since it requires the
  determination of the precise analytic dependence on $\LLe$.
  However, if the Euclidean path integral saturates at a certain
  characteristic value of $L_E$, at which all the relevant
  contributions have been included, it would be possible to take
  $\LLe\to\infty$ without changing appreciably the result.
\item The dependence on $T_E$ is expected to become trivial in the
  limit $T_E\to\infty$, the argument being the same given in the
  Minkowskian case. If ${\cal F}$, ${\cal F}_E$ have finite limits as 
  $T,T_E\to\infty$, as we expect, and moreover if they satisfy
  appropriate analyticity assumptions, it is possible to prove that in
  the limit $T,T_E\to\infty$ the analytic continuation relations
  simplify, reducing to the analytic continuation $\theta\to -i\chi$,
  $\LLe\to i\LL$ only. The argument is based on the
  Phragm\'en-Lindel\"of theorem, and can be easily adapted from
  Ref.~\cite{Meggiolaro05}. The same argument can be applied to the
  dependence on $\LLe$: if we can take the limits $\LLe\to\infty$ and
  $\LL\to\infty$ without changing appreciably the results, with the
  appropriate analyticity assumptions it is possible to prove in the
  same way that in this limit the analytic continuation reduces simply
  to $\theta\to -i\chi$. 
\item As we have already said, at the present stage we are not in the
  position to determine the domain of analyticity of ${\cal F}$ and
  ${\cal F}_E$, and thus the domain of validity of
  Eq.~\eqref{eq:an_cont}. Consider ${\cal F}_E$ for definiteness. In
  order to make these relations meaningful, it is necessary that the
  analyticity domain ${\cal D}_E$ of ${\cal F}_E$ contains the real segment
  $(0,\pi)$ at $\Re{\eta}>0,\Im{\eta}=0$, and the negative imaginary half-axis
  $i\theta\in(0,\infty)$ at $\Re{\eta}=0,\Im{\eta}>0$.~\footnote{We
    understand that they have to lie in the same connected 
    component of ${\cal D}_E$.} This has been tacitly assumed when
  deriving the analytic continuation relations
  Eq.~\eqref{eq:an_cont}. Under this hypothesis, the analytic
  continuation relations would therefore allow to reconstruct the
  physical Minkowskian amplitude at $\chi>0$ starting from the
  Euclidean quantity ${\cal F}_E$ in the interval $\theta\in(0,\pi)$. 
\item The main obstacle in carrying over here the
  discussion of Ref.~\cite{EMduality} is the dependence on
  $\theta,\eta$ (resp.~$\chi,\xi$) of the rescaled Euclidean
  (resp.~Minkowskian) spin factor. Since the integrand is an analytic
  function of the relevant variables, the criterion for analyticity is
  the convergence\footnote{Strictly speaking, uniform convergence
    would be a sufficient condition.} of the functional integral over
  gauge fields, and of the path integrals over the trajectories $X,X'$
  and the momenta $\Pi,\Pi'$. It is
  known~\cite{Brandt,Brandt1,Korchemsky,Korchemsky2} 
  that, as it stands, the integral over momenta is not converging even
  in Euclidean space, and that an appropriate regularisation is
  needed. Multiplying the integrand by a factor
  $e^{-\int_{\nu_i}^{\nu_f} d\nu\,\epsilon(\nu)
    \sqrt{\Pi_E(\nu)^2}}$~\cite{Strominger}, it can be shown 
  that~\cite{Korchemsky,Korchemsky2}   
  \begin{equation}
    \label{eq:eucl_spin_expl}
    \Sp_{\nu_i,\nu_f}^{(E)}[\dot{X}_E] \equiv
    \int \DPE \, {\M}^{(E)}_{\nu_i,\nu_f}[\dot{X}_E,\Pi_E] =
    \prod_{\nu=\nu_i}^{\nu_f}\delta\left(1-\big(\dot{X}_E(\nu)\big)^2\right) 
    \f{1+\sladx(\nu)}{2}\,,
  \end{equation}
  when the regularisation is removed, i.e., $\epsilon(\nu)\to 0$.  
  A detailed study of the analyticity domain of ${\cal F}$
  should therefore start from the study of convergence of the
  integration over momenta of the regularised spin factor, for complex
  values of the variables $\theta,\eta$. This requires further work,
  which is outside the scope of the present paper.
\end{itemize}
The expression Eq.~\eqref{eq:piece_4}, together with the analytic
continuation relations Eq.~\eqref{eq:an_cont}, provide the basis for the
Euclidean approach to the Reggeon-exchange amplitude suggested in
Ref.~\cite{Jani}. In the next Section we make contact with the
proposal of~\cite{Jani}, and with the more recent investigations on
the same line discussed in~\cite{reggeon_duality}.

\nsection{Reggeon exchange and gauge/gravity duality}
\label{saddle}

In this Section we want to make contact with the previous analysis of
the Reggeon-exchange amplitude in a Euclidean
setting~\cite{Jani,reggeon_duality}. In particular, we want to discuss
how the basic expression proposed there for the Reggeon-exchange
amplitude is related to the one obtained in this paper. We will refer
in particular to the more recent and more detailed analysis contained
in Ref.~\cite{reggeon_duality}. 

The process considered in Ref.~\cite{reggeon_duality} is the
elastic scattering of two heavy-light mesons $M_{1,2}$ of large mass 
$m_{1,2}$, i.e., $M_1 = Q\barq$ and $M_2=\barQ' q$, where $Q$ and
$\barQ'$ are heavy and of different flavours, while $q$ and $\barq$
are light and of the same flavour. This choice was made in order to
have a single type of Reggeon exchange, namely the one in which the
interacting mesons exchange the valence $q$ and $\barq$
partons. Moreover, the choice of heavy mesons is made so that the
typical sizes of the constituent dipoles are small, $|\vr_{i\perp}| \sim
m_i^{-1} \ll \Lambda_{\rm QCD}^{-1}$. In this way, in a first
approximation one can focus directly on the dipole-dipole
Reggeon-exchange amplitude, ignoring the integration over the dipole
size and orientation.

It is straightforward to adapt the calculations of the previous
Sections to this case. First of all, the relation between $\chi$ and
$s$ at high energy is modified to $\chi \simeq \log(s/m_1 m_2)$;
moreover, Eq.~\eqref{eq:S_dip_short} simplifies to 
\begin{equation}  
  \label{eq:S_dip_short_2}
    S_{fi}^{(dd)} = {\cal P}^{(dd)} + {\cal R}_1^{(dd)} \, .
\end{equation}
The rest of the derivation is not modified, in particular the
expressions Eqs.~\eqref{eq:regge_ampl} and \eqref{eq:regge_ampl_dip}
for the Reggeon-exchange amplitude and the analytic continuation
relations Eq.~\eqref{eq:an_cont} remain unchanged. 

Introducing now the following shorthand notation for the normalised
Wilson-loop expectation value and for the spin factor, 
\begin{equation}
  \label{eq:short_path_int}
  \begin{aligned}
{\cal U}^{(E)}[{\cal C}_E] &= {\cal U}^{(E)}_{{\cal C}_E}[X_E,L_E,X_E',X_E']\,,\\
{\cal I}[{\cal C}_E] &=  {\Sp}^{(E)}_{-T_E,-T_E+L_E}[\dot{X}_E]
    {\Sp}^{(E)}_{-T_E,T_E+L_E'}[\dot{X}_E']\,,
  \end{aligned}
\end{equation}
where ${\Sp}^{(E)}_{\nu_i,\nu_f}$ has been defined in
Eq.~\eqref{eq:eucl_spin_expl}, and moreover denoting the path
integrals as follows, 
\begin{equation}
  \label{eq:short_path_int2}
  \begin{aligned}
\int {\cal DC}_{E}^{(\wedge)} &=  \int_{2T_E-\LLe}^{2T_E+\LLe}
    d{L}_E \int \DXE 
\,, \\
   \int {\cal DC}_{E}^{(\vee)} &=  \int_{2T_E-\LLe}^{2T_E+\LLe} d{L}_E' 
    \int \DXpE 
\,,
  \end{aligned}
\end{equation}
we can write
\begin{multline}
  \label{eq:notation_contact_0}
   {\cal F}_E(\theta,T_E;\vec{b}_{\perp},\vec{R}_{1\perp},\vec{R}_{2\perp}
    ,\vec{R}_{1\perp}',\vec{R}_{2\perp}') = \\  \int {\cal DC}_{E}^{(\vee)} \int
    {\cal DC}_{E}^{(\wedge)} \, e^{-m_q({L}_E+{L}_E'-4T_E)} {\cal U}^{(E)}[{\cal
        C}_E] {\cal I}[{\cal C}_E]\,.
\end{multline}
One can now easily be convinced by a simple comparison that ${\cal
  F}_E$ is the same quantity\footnote{The only difference is the
  inclusion of the spin factor of the ``spectator'' partons in
  $\tilde{a}$, which is however equivalent to the identity when
  contracted with the corresponding bispinors, and can thus be
  discarded.}  as  the ``Euclidean amplitude'' $\tilde{a}$ of
Eq.~(3.3) in Ref.~\cite{reggeon_duality}. 

It is then possible, at this point, to carry out the analysis
performed in that paper. Here we will not repeat the
analysis in details, but simply summarise the main points.
The basic idea is to employ the gauge/gravity duality in a confining
background in order to evaluate the Wilson loop expectation values
entering Eq.~\eqref{eq:notation_contact_0}. 
The first precise formulation of this duality, the well known AdS/CFT
correspondence~\cite{adscft1,adscft2,adscft3}, relates the weak-coupling,
supergravity limit of type IIB string theory in $AdS_5\times S^5$, to
four-dimensional ${\cal N}=4$ SYM theory, which is a conformal (and
thus non confining) field theory, in the limit of large number of
colours $N_c$ and strong 't Hooft coupling $\lambda=g_{\rm
  YM}^2N_c$. In particular, the AdS/CFT correspondence gives the
following area-law prescription for the expectation value of   
a Wilson loop running along the path ${\cal C}_E$ in Euclidean
space~\cite{Wilson,Wilson2,GO,DGO}, 
\begin{equation}
  \label{eq:area_law}
  \la \W[{\cal C}_E] \ra = {\cal F}l[{\cal C}_E]
  e^{-\f{1}{2\pi\alpha'}A_{\rm min}[{\cal C}_E]}\, .
\end{equation}
Here $A_{\rm min}$ is the area of the minimal surface in the Euclidean 
version of the $AdS_5$ metric (i.e., in hyperbolic space),
$1/(2\pi\alpha')=\sqrt{\lambda}/(2\pi)$ 
is the string tension, and ${\cal F}l[{\cal C}_E]$ stands for the
contribution of quantum fluctuations around the minimal surface.

Although various attempts have been made, a precise
formulation of the duality for QCD is not known yet (assuming it
exists). Nevertheless, a few general properties of the gravity dual of
a confining theory have been established: in particular, the presence
of a confinement scale in the gauge theory translates into a
characteristic scale $R_0$ in the metric of the gravity theory,
associated for example to the horizon of a black hole~\cite{AdSBH}, or
to the position of a hard wall~\cite{Polchinski}, or to the scale
associated to a soft wall~\cite{Karch}. Such a scale essentially
separates the regions of small and large $z$, where $z$ is the fifth
coordinate of AdS space: while for small $z$ the metric diverges as  
some inverse power of $z$, for $z$ of the order of $R_0$ the metric
turns out to be effectively flat. Moreover, the area-law prescription
Eq.~\eqref{eq:area_law} carries over to the confining
case, substituting the AdS metric with an appropriate confining
background~\cite{Sonn0,Sonn1,Sonn2} and replacing $1/(2\pi\alpha')$ with
an effective string tension $1/(2\pi\alpha'_{\rm eff})$. 

At this point, one has to substitute the area-law expression
Eq.~\eqref{eq:area_law}, with the minimal surface determined in the
appropriate confining metric, into the path integral
Eq.~\eqref{eq:notation_contact_0}. The resulting expression is still
too difficult to deal with, not to mention the fact that the exact
metric to be used is not yet known. In order to obtain an estimate, a
few approximations are therefore needed. 

The general features of the metric described above suggest a
convenient approximation scheme to evaluate the Wilson-loop
expectation value in a generic confining background~\cite{Jani2}. The
small-$z$ behaviour suggests that, in order to minimise the area, it
is convenient for the surface to rise almost vertically from the
boundary, without appreciable motion in the other directions, at least
when the typical size of the Wilson loop is not too small. The
presence of a horizon puts an upper bound on this vertical rise;
moreover, when $z\sim R_0$, the surface lives effectively in flat
space. As a result, the minimal surface is expected to be constituted
by two parts: an almost vertical wall rising from the boundary up to
the horizon, and transporting there the boundary conditions, and a
solution of the Plateau problem in flat space.  

The solution of the Plateau problem in the general case is not
known even in flat space. Nevertheless, the particular configuration
in the case at hand suggests that the relevant contributions to the
path integral come from those trajectories of the exchanged fermions
which lie on the helicoid determined by the eikonal trajectories of
the ``spectator'' fermions~\cite{Jani,reggeon_duality}. In particular,
the small dipole size makes the dependence on the dipole orientation
trivial in a first approximation. This leads to the following
approximation for the path integral Eq.~\eqref{eq:notation_contact_0}, 
\begin{equation}
  \label{eq:notation_contact_1}
   {\cal F}_E(\theta,T_E;\vec{b}_{\perp},\vec{R}_{1\perp},\vec{R}_{2\perp}
    ,\vec{R}_{1\perp}',\vec{R}_{2\perp}') \approx  \int {\cal DC}_{E}^{(\vee)} \int
    {\cal DC}_{E}^{(\wedge)} \, e^{-S_{\rm eff}[{\cal
        C}_{E}^{(\vee)},\,{\cal C}_{E}^{(\wedge)}]} 
    {\cal I}[{\cal C}_E]\,,
\end{equation}
where the ``effective action'' $S_{\rm eff}$ is given by
\begin{equation}
  \label{eq:eff_area}
  S_{\rm eff} = \f{1}{2\pi\alpha'_{\rm 
      eff}}A_{\rm min}[{\cal C}_E^{(\vee)},\,{\cal C}_E^{(\wedge)}] +
  \hat{m}_q\big(L_E[{\cal C}_E^{(\vee)}]+L_E[{\cal
    C}_E^{(\wedge)}]-4T_E\big) \, , 
\end{equation}
and the Euclidean paths of the exchanged fermions ${\cal C}_E^{(\vee)}$,
${\cal C}_E^{(\wedge)}$ are constrained to lie on the helicoid determined
by the paths ${\cal C}_{E+}^{(1)}$ and ${\cal
  C}_{E-}^{(2)}$. Here $L[{\cal C}_E^{(\vee),(\wedge)}]$ denote the length
of the Euclidean paths of the exchanged fermions. Ultraviolet
divergencies require a renormalisation of the quark mass to
$\hat{m}_q$~\cite{reggeon_duality}. 

The final step is a saddle-point approximation of
Eq.~\eqref{eq:notation_contact_1}. The saddle point is determined by
minimising the functional Eq.~\eqref{eq:eff_area}. The detailed
calculations are reported in~\cite{Jani}, for the massless quark case
$\hat{m}_q=0$, and in~\cite{reggeon_duality} for the more general case
of a massive quark, and will not be discussed here. We simply mention
that an exact solution to the saddle-point equations can be found in
implicit form for $b\le b_c = 4\pi \alpha'_{\rm eff}\hat{m}_q$, with
$b=|\vec{b}_\perp|$ the impact-parameter distance. An explicit
solution can be obtained in the case of small angles $\theta$, and we
report here the 
corresponding result for the ``effective action''
Eq.~\eqref{eq:eff_area}: 
\begin{equation}
  \label{eq:effact_smallth}
  S_{\rm eff} =
  \f{b^2}{2\pi\alpha_{\rm eff}'\theta}\acosh\f{b_c}{b} + 
  2\pi^2\alpha_{\rm eff}'\hat{m}_q^2 -
  \f{2b\hat{m}_q}{\theta}\sqrt{\left(\f{b_c}{b}\right)^2-1}\,.
\end{equation}
At this point one has to perform the analytic continuation to 
Minkowski space. Although the expression Eq.~\eqref{eq:effact_smallth}
is valid only at small $\theta$, it is nevertheless worth to
investigate what it leads to. Notice that there is no dependence on
$T_E$, as expected.\footnote{Since we are considering only the
saddle-point, we have no control on the dependence on $\LLe$. This
is not a problem, however, if we are allowed to take
$\LLe\to\infty$ without changing the result, see discussion at the end
of Section \ref{an_cont}.}  After analytic continuation, the resulting
expression can be extended to $b>b_c$, as it is explained
in~\cite{reggeon_duality}, so that it is possible to take the limit of
small quark mass. Up to order ${\cal O}(\alpha_{\rm eff}'\hat{m}_q^2)$,   
\begin{equation}
  \label{eq:small_m}
  S_{{\rm eff},\,M} \simeq \f{b^2}{4\alpha_{\rm eff}'\chi} - \f{4b\hat{m}_q}{\chi} 
     + 2\pi^2\alpha_{\rm eff}' \hat{m}_q^2\, ,
\end{equation}
where we have denoted with $S_{{\rm eff},\,M}$ the analytic
continuation of $S_{{\rm eff}}$ to Minkowski spacetime. 
Rewriting now the dipole-dipole Reggeon-exchange scattering
amplitude Eq.~\eqref{eq:regge_ampl_dip} in the impact-parameter
representation,
\begin{equation}
  \label{eq:regge_ampl_dip_imp_par}
  \A_{{\cal R}_1}^{(dd)}(s,t) = -i2s \int d^2b_\perp e^{i\vec{q}_\perp
    \cdot\vec{b}_\perp} a_{{\cal R}_1}^{(dd)}(\chi,\vec{b}_\perp)
\end{equation}
where we have dropped the dependence on the dipole sizes and the
spin indices, and substituting the result
Eq.~\eqref{eq:small_m} in it, one finds (to first order in
$\sqrt{\alpha_{\rm eff}'}\hat{m}_q$)
\begin{equation}
  \label{eq:regge_ampl_dip_imp_par_2}
  a_{{\cal R}_1}^{(dd)}(\chi,\vec{b}) \approx 
  e^{-\f{b^2}{4\alpha_{\rm eff}'\chi}}\left(1+
    \f{4b\hat{m}_q}{\chi}\right) \times {\cal K}\, . 
\end{equation}
Here we have put in ${\cal K}$ all the remaining factors,
including the contribution of the spin factor, those of the 
quantum fluctuations ${\cal F}l[{\cal C}_E]$ (see
Eq.~\eqref{eq:area_law}) of the string around the minimal surface, 
and the determinant coming from the integration of quadratic
fluctuations of the boundary around the saddle-point. 
As discussed in~\cite{reggeon_duality}, this expression leads to a
linear Reggeon trajectory $\alpha(t)=\alpha_0+ \alpha_{\rm eff}' t$,
although the Regge singularity is not simply a pole when $\hat m_q\neq 0$,
but contains also a logarithmic branch point. This result is
independent of possible prefactors $s^{\delta\alpha}\chi^{n_\chi}
b^{n_b}$ (with $n_\chi,n_b\in\mathbb{N}$), which could be present in
${\cal K}$, but which are not under control at the present stage. In
particular, a factor $s^{\delta\alpha}$ simply shifts the trajectory
by a constant amount, while $\chi^{n_\chi} b^{n_b}$ can change the
order of the pole but neither the Regge trajectory nor the presence of
a logarithmic branch point~\cite{reggeon_duality}.

Some important remarks are in order.
\begin{itemize}
\item The extra factor of $s$ in  Eqs.~\eqref{eq:regge_ampl_dip} and 
\eqref{eq:regge_ampl_dip_imp_par} is removed  by the integration over
the longitudinal momentum fraction, as can be seen from
Eqs.~\eqref{eq:sdep_1} and \eqref{eq:sdep_2}. More precisely, the
overall power of $s$ at $t=0$, i.e., the ``Reggeon intercept'' $\alpha_0$,
depends on the end-point behaviour of the mesonic wave functions, as
discussed at the end of Section \ref{amplitude}. The simplest choice,
corresponding to the phenomenological Wirbel-Stech-Bauer
ansatz~\cite{WSB} where the dependence on $\zeta_1$ and $\zeta_2$
is purely ``kinematical'' (i.e., $\alpha_{1,2}=\beta_{1,2}=0$ in 
Eqs.~\eqref{eq:sdep_1} and \eqref{eq:sdep_2}; cfr. also
Eq.~\eqref{eq:wf_coord}), gives an intercept $\alpha_0=0$.

\item The spin factor ${\cal I}[{\cal C}_E]$ (see
  Eq.~\eqref{eq:short_path_int}) has been evaluated in exact but
  implicit form in~\cite{reggeon_duality}, but the corresponding
  small-$\theta$ approximation has been shown not to lead to a fully
  reliable analytic continuation into Minkowski space-time. A more
  detailed study is needed in order to clarify its possible
  effects. We mention however that, independently of the
  small-$\theta$ approximation, it contains a factor which behaves
  as $s^{-1}$, after analytic continuation to Minkowski space-time:
  this factor cancels the (four) factors $\sqrt{E+m}$ appearing in
  Eq.~\eqref{eq:Dspinors}, when contracting with the Dirac bispinors.

\item The quantity ${\cal F}l[{\cal C}_E]$ has been evaluated
  in~\cite{Jani} in the massless-quark case, where it leads to a
  factor $s^{\f{n_\perp}{24}}$, with $n_\perp$ the number of
  transverse directions in which the string can fluctuate, which
  increases the Reggeon intercept. The
  corresponding calculation in the case of massive quarks is more
  difficult, due to the nontrivial form of the resulting minimal
  surface, and it has not been
  performed yet. 

\item The evaluation of the effect of fluctuations of the boundary
  around the saddle point solution requires first of all a precise
  formulation of the saddle-point approximation for the
  path-integral Eq.~\eqref{eq:notation_contact_0}, which is not
  available at the moment.

\item The fact that the
  Regge slope is equal to the inverse of the string tension, which
  appears in the confining potential, is a first indication in order to
  understand the relation between our formalism based on Wilson loops
  and the usual picture of Regge poles. Indeed, in this picture the
  Regge trajectory $\alpha(t)$ at $t>0$ provides the relation between
  the mass $M$ and the spin $J$ of the ``families'' of particles
  exchanged in the scattering process, i.e.,
  $J=\alpha(M^2)=\alpha_0+\alpha_1 M^2$; in turn, in the 
  QCD-string picture the slope $\alpha_1$ is exactly the inverse
  of the string tension. Combining these results, one is led to expect
  that the same $\alpha_1$ appears in the Regge trajectory and
  in the static potential, an expectation that is met by the above
  result, with $\alpha_1=\alpha_{\rm eff}'$. 

\end{itemize}
The investigation of these issues is beyond the scope of this paper,
and more work is needed in order to complete the dual gravity picture
of {soft} high-energy scattering.

\nsection{Conclusions and Outlook}
\label{concl}

In this paper we have proposed a derivation of a nonperturbative
expression for the scattering amplitude of the Reggeon-exchange
process in high-energy elastic meson-meson scattering. Using a
partonic description of hadrons, along the lines of~\cite{Nacht}, such
a process is identified with the exchange between the mesons of a 
(Reggeised) pair of valence fermions, as in
Refs.~\cite{Jani,reggeon_duality}. Exploiting a path-integral
representation of the various fermionic propagators, and retaining
only the paths which are expected to give relevant contributions at
high energy, we have been able to express the Reggeon-exchange
amplitude in terms of a path-integral of the (properly normalised)
expectation value of a certain Wilson loop, over the trajectories of
the exchanged fermions. The relevant trajectories are determined by
the constraint that they coincide with the eikonal trajectories far
away from the interaction region. Moreover, under certain analyticity 
assumptions, we have shown how the Reggeon-exchange amplitude can be
reconstructed from a Euclidean quantity by means of an appropriate
analytic continuation, which is very similar to the
one~\cite{Meggiolaro05,crossing,crossing2,EMduality} employed in the case of
the leading, Pomeron-exchange amplitude. We have also shown that the 
expression derived in this paper is essentially the same one proposed 
in~\cite{Jani} and recently reconsidered in~\cite{reggeon_duality}, and
we have briefly discussed how a saddle-point 
approximation can be qualitatively performed in Euclidean space,
making use of gauge/gravity duality for a confining background and
restricting the trajectories of the exchanged fermions to a special
class, namely trajectories lying on the helicoid determined by the
``spectator'' partons' trajectories. The results obtained in this
approximation are in qualitative agreement with the phenomenology.

Let us now briefly summarise the main open issues of the approach
discussed in this paper.
\begin{itemize}
\item In the course of the derivation of the
  Reggeon-exchange amplitude we have made a few technical assumptions
  on the path-integral representation of the propagators, which need to
  be investigated in detail. In particular, the identification of the
  nature of the parameter along the path requires a detailed study of
  the integration over momenta in the path integral in Minkowski
  space, which would yield an explicit expression for the Minkowskian 
  spin factor.
\item According to our results, the dependence on energy of the
  Reggeon-exchange amplitude is affected by the behaviour of the
  mesonic wave functions near the value $0$ for the longitudinal
  momentum fractions of the fermions which are exchanged in the
  process. In order to reconcile this result with the experimentally
  observed universality of the subleading contributions to total cross
  sections, we are led to assume that such a behaviour is a universal
  feature of the nonperturbative wave functions describing the mesons
  in terms of colourless dipoles. On the other hand, it would be
  interesting to understand to what extent the universality of the
  subleading contribution is established experimentally.  
\item A study of the analyticity domain of the relevant quantities is
  necessary, in order to properly justify the analytic continuation
  relations. The features of this analyticity domain are expected to
  be related with the convergence properties of the path integral for
  complex values of the relevant variables, similarly to what has been
  discussed in Ref.~\cite{EMduality} in the Pomeron-exchange case.
\item As regards the gauge/gravity duality approach employed in
Refs.~\cite{Jani,reggeon_duality}, further work is needed
in order to obtain a precise formulation of the saddle-point
approximation of the relevant path integral. Such a formulation would
allow to write down the saddle-point equation for the whole range of
paths, and not only for the class of paths which gives a
predetermined, helicoidal geometry for the Euclidean minimal surface. 
It would also clarify how the fluctuations of the
trajectories around the saddle-point solution discussed in
Refs.~\cite{Jani,reggeon_duality} have to be properly taken into
account. 
\end{itemize}
In conclusion, we hope that further work in these directions could
help in a better understanding of {soft} high-energy scattering,
and the related issue of a first-principle explanation of Regge
phenomenology. 

\nonsection{Acknowledgements}

I am indebted to R.~Peschanski, who has brought my 
attention to this problem, for countless remarks and suggestions.
I thank O.~Nachtmann for reading the manuscript and for
useful suggestions. I also benefitted from useful discussions with
G.~Korchemsky, E.~Meggiolaro, and F.~Mercati. I am grateful for the
hospitality to the IPhT/CEA-Saclay, where part of this work has been
carried out. This work has been partly funded by a grant of the
``Fondazione Angelo Della Riccia'' (Firenze, Italy).  
MG is supported by MICINN under the CPAN project CSD2007-00042 from the
Consolider-Ingenio2010 program, as well as under the grant FPA2009-09638.


\appendix

\nsection{Eikonal approximation for straight-line trajectories}
\label{app:straight}

In this Appendix we rederive the eikonal approximation
Eq.~\eqref{eq:eikonal_3} for the truncated-connected propagator of a
fermion in an external field, using the path-integral formalism. 

Using the trick described in~\cite{Fabbr}, appropriately generalised
to the case of fermions, the truncated-connected propagator for a
quark $Q$ of ``physical'' mass $\m_Q$ can be written in the
path-integral representation as
\begin{equation}
  \label{eq:sand_8}
Z_Q\hS{Q} = \Lim \f{1}{\nu_f-\nu_i} 
\bar{u}(p_Q')\left(\tilde{F}(\nu_f,\nu_i) -
  \tilde{F}(\nu_f,\nu_f)\right) u(p_Q) \, , 
\end{equation}
where
\begin{equation}
  \label{eq:conn}
  \begin{aligned}
  \tilde{F}(\nu_f,\nu_i) &= \int \DX \int \DP\,
  e^{i[p_Q'\cdot X(\nu_f) - p_Q\cdot 
     X(\nu_i)]} e^{-i(\nu_f-\nu_i)(m-i\epsilon)} \\ & \phantom{\int
     \DX \int \DP \,
}\times {\cal  
     M}_{\nu_i,\nu_f} [X,\Pi]{ W}_{\nu_i,\nu_f}[X]  \,,\\
 \tilde{F}(0,0) &= \int d^4x\, 
  e^{i(p_Q'-p_Q)\cdot x}\,,   
  \end{aligned}
\end{equation}
bispinors are normalised as in Eq.~\eqref{eq:bisp_norm}, and we recall that
\begin{equation}
  \label{eq:defin_4}
  \begin{aligned}
  {\cal M}_{\eta,\nu}[X,\Pi] &= \Texp\left[i\int_{\eta}^{\nu} d\tau
   \left(\sla{\Pi}(\tau) - \Pi(\tau)\cdot\dot{X}(\tau)\right)\right]\,
 ,\\
 { W}_{\eta,\nu}[X] &= \Texp\left[-ig \int_{\eta}^{\nu} d\tau A(X(\tau))\cdot
   \dot{X}(\tau) \right]\, .
  \end{aligned}
\end{equation}
Assuming the dominance of the classical trajectory of the
quark, 
\begin{equation}
  \label{eq:eikonal_1}
  X(\tau)= \f{p_Q}{\m_Q}(\tau-\nu_i) + X(\nu_i)\,, 
\end{equation}
where $p_Q$ has been defined in Eq.~\eqref{eq:dipoles}, the path integral
for $\tilde{F}(\nu_f,\nu_i)$ reduces to an integration over the
initial point $x_i\equiv X(\nu_i)$, the final point being determined
by the relation $x_f\equiv X(\nu_f) = \f{p_Q}{\m_Q}(\nu_f-\nu_i) +
x_i$. In the description of mesons in terms of colourless
$q\barq$ dipoles, the ``physical'' quark 
mass $\m_Q$ is identified with the fraction of meson mass
carried by the quark in the initial state, $\m_Q=\zeta m$, which in
principle can be different from the fraction $\m_Q'=\zeta' m$ carried
in the final state. However, we have seen that for {soft}
high-energy scattering one finds $\zeta=\zeta'$, when the eikonal
propagator is inserted in the expressions for the scattering
amplitude. Here we proceed by keeping them distinct: this does not affect 
the trajectory, since $\f{p_Q}{\m_Q} =\f{p}{m}=u$ and
$\f{p_Q'}{\m_Q'}=\f{p'}{m}=u$,  
where $p\simeq p'$ and $m$ are the initial and final momentum and the mass of
the meson. As for the integration over $\Pi$, the saddle point is
given by $\Pi(\tau)=p_Q$, since in that case, given the fact that we
are ``sandwiching'' between bispinors,
$\sla{\Pi}(\tau)-\Pi(\tau)\cdot\dot{X}(\tau)=\slap_Q-\m_Q\to 0$;
choosing $\Pi(\tau)=p_Q'$ yields the same result.\footnote{
Here we are using the Minkowskian spin factor. A
more careful treatment, starting with the Euclidean spin
factor and imposing an appropriate regularisation, yields the
same result when performing the analytic continuation back to
Minkowski space-time.} 
The integration along the direction parallel to $u$ 
is trivial, since in the limit of infinite length we have
translational invariance along $u$. In practice, writing
$x_{i,f}=b+\nu_{i,f}u$, so that $b = (\nu_f x_i -\nu_i
x_f)/(\nu_f-\nu_i)$, we have
\begin{equation}
  p_Q'\cdot x_f-p_Q\cdot x_i = (p'-p)\cdot b + (\m_Q'\nu_f - \m_Q\nu_i)\,,
\end{equation}
and thus, replacing the spin factor with unity, we find that
\begin{equation}
  \label{eq:eikonal_2}
  \begin{aligned}
\tilde{F}(\nu_f,\nu_i)  \simeq & \int d^4x_i\, e^{i(p_Q'\cdot x_f-p_Q\cdot
  x_i)} e^{-i(\nu_f-\nu_i)(m_Q-i\epsilon)}\,
{W}_{\nu_f,\nu_i}\left[\textstyle\f{p_Q}{\m_Q}(\tau-\nu_i) + x_i\right]= \\
&  e^{i[(\m_Q'-m_Q)\nu_f-(\m_Q-m_Q)\nu_i]}\int d^4b \,e^{i (p'-p)\cdot b}\, W_u(b)= \\
&  e^{i[(\m_Q'-m_Q)\nu_f-(\m_Q-m_Q)\nu_i]}\int_{\nu_i}^{\nu_f} d\nu
\int d^3b \, e^{-i (\vec{p}^{\,\prime}-\vec{p})\cdot \vec{b}}\, W_u(b) =\\ & 
 (\nu_f-\nu_i)  e^{i[(\m_Q'-m_Q)\nu_f-(\m_Q-m_Q)\nu_i]}\int d^3b\, e^{i(p'-p)\cdot b}\,
W_u(b)  \, , 
  \end{aligned}
\end{equation}
where through a Lorentz transformation we have set $b^0$ to be the
coordinate parallel to $u$ and $\vec{b}$ the spatial coordinates in
the rest frame of the particle, and we have used the notation $W_u(b)$
for a straight-line Wilson line parallel to $u$ and centered at $b$
(the value of $b^0$ is of course arbitrary in the 
limit of infinite length). As for the disconnected term, we have in
the $b$-coordinates  
\begin{equation}
  \label{eq:delta_term}
  \begin{aligned}
\tilde{F}(0,0) & = \int d^4b e^{i(p_Q'-p_Q)\cdot b} = 
\int_{\nu_i}^{\nu_f} d\nu e^{i(p_Q^{\prime 0}-p_Q^0) \nu}
 \int d^3b
  e^{-i(\vec{p}_Q^{\,\prime}-\vec{p}_Q)\cdot \vec{b}} \\ &=
  (\nu_f-\nu_i)(2\pi)^3 \delta^{(3)}(\vec{p}_{Q\,{\rm
      r.f.}}^{\,\prime}-\vec{p}_{Q\,{\rm r.f.}}) 
  {\cal K}(\nu_i,\nu_f;\m_Q,\m_Q')\,,
  \end{aligned}
\end{equation}
where the subscript ``r.f.'' stands for ``rest frame'',\footnote{Of
  course $\vec{p}_{Q\,{\rm r.f.}}=0$ in the rest frame; we prefer
  however to keep the notation in Eq.~\eqref{eq:delta_term} as
  covariant as possible.} and where  
\begin{equation}
  \begin{aligned}
     {\cal K}(\nu_i,\nu_f;\m_Q,\m_Q') &=
   e^{i\f{\m_Q'-\m_Q}{2}(\nu_f+\nu_i)}\f{\sin\left(\f{\m_Q'-\m_Q}{2}
       (\nu_f-\nu_i)\right)}{\f{\m_Q'-\m_Q}{2}(\nu_f-\nu_i)}\,, \\ 
{\cal K}(\nu_i,\nu_f;\m_Q,\m_Q')& =1\,,\quad {\rm if}\,\,\m_Q'=\m_Q\,,
\\ \lim_{\nu_f\to\infty,\,\nu_i\to
  -\infty} {\cal
  K}(\nu_i,\nu_f;\m_Q,\m_Q')&= 0\,, \quad {\rm 
       if}\,\,\m_Q'\neq\m_Q\,.
   \end{aligned}
\end{equation}
Plugging the results above in Eq.~\eqref{eq:sand_8} we finally obtain
\begin{equation}
  \label{eq:sand_9}
  \begin{aligned}
   Z_Q \hS{Q} &= 
    \delta_{s_Q's_Q}\, 2\sqrt{\m_Q\m_Q'}
    \Big[e^{i[(\m_Q'-m_Q)\nu_f-(\m_Q-m_Q)\nu_i]}\int d^3b
      e^{i(p'-p)\cdot b} 
    W_u(b) \\ &\phantom{2\sqrt{\m_Q\m_Q'}
    \Big[e^{i[(\m_Q'-m_Q)\nu_f-(\m_Q-m_Q)\nu_i]}} - (2\pi)^3 \delta^{(3)}(\vec{p}_{\rm
      r.f.}^{\,\prime}-\vec{p}_{\rm r.f.})\Delta(\m_Q'-\m_Q)\Big] \\
  &=   \delta_{s_Q's_Q}\,2\sqrt{\m_Q\m_Q'}
  e^{i((\m_Q'-m_Q)\nu_f-(\m_Q-m_Q)\nu_i)}\int d^3b 
  e^{i(p'-p)\cdot b} W_u(b) - \delta_Q\, ,
\end{aligned}
\end{equation}
where the function $\Delta(x)$ is defined as $\Delta(x)=1$ if
$x=0$, and 0 otherwise. In the last passage we have recognised that
the disconnected term is the expression in the rest frame of the
invariant delta-function Eq.~\eqref{eq:rel_norm}, in which we have included
also $\Delta(\m_Q'-\m_Q)$, which is essentially a superselection rule on
the particle species. The prefactor comes from the contraction of the
Dirac bispinors, which in the high-energy limit gives
($\vec{p}_Q \para x^1$)
\begin{equation}
  \begin{aligned}
  \bar{u}^{s_Q'}(p_Q')u^{s_Q}(p_Q) &=
  \delta_{s_Q's_Q}\,\sqrt{(E_Q'+\m_Q')(E_Q+\m_Q)} 
\left(1 - \f{|\vec p_Q^{\,\prime}||\vec p_Q|}{(E_Q'+\m_Q')(E_Q+\m_Q)}\right)  \\
&= \delta_{s_Q's_Q}\,2m\sqrt{\zeta\zeta'} = \delta_{s_Q's_Q}\,2\sqrt{\m_Q\m_Q'}\,.
  \end{aligned}
\end{equation}
The expression for the eikonal propagator is only apparently different
from the one already known in the literature (up to the phase factor).
Indeed, if we replace the integration over $b^1$, which is the
coordinate in the longitudinal plane orthogonal to $u_1$, with the
integration over the longitudinal coordinate parallel to $u_2$, and
rescale it so that it becomes a light-cone coordinate in the
high-energy limit $\chi\to\infty$, we recover the well-known
result~\cite{Nacht,Fabbr,Meggiolaro96}. Explicitly, setting $q=p'-p$,
we have 
\begin{equation}
  \begin{aligned}
  (q\cdot u_{1}^\perp) b^1 &= \left[q\cdot\left(\coth\chi u_1 -
    \f{1}{\sinh\chi} u_2\right)\right] b^1 = 
 - (q\cdot u_2)  \f{ b^1}{\sinh\chi} \\ &
= - q\cdot
 \f{u_2}{\cosh\f{\chi}{2}} \f{ b^1}{2\sinh\f{\chi}{2}}   \,.
  \end{aligned}
\end{equation}
We introduce now lightcone coordinates as follows, $b=(b_+ u_+ + b_- u_-)/2 +
b_\perp$, where  $b_\pm = b^0\pm b^1$ and
$b_\perp=(0,0,\vec{b}_\perp)$, and the lightcone vectors are defined
as $u_\pm=(1,\pm 1,\vec{0}_\perp)$. The Minkowskian scalar product is
rewritten as $q\cdot b = (q_+b_-+q_-b_+)/2 +
q_\perp\cdot b_\perp $. Since in the large-$\chi$ limit
$\f{u_2}{\cosh\f{\chi}{2}}\to u_-$ , and moreover $q_-\simeq 0$, we
have that $b_-=-\f{b^1}{\sinh\f{\chi}{2}}$, and thus for
$\chi\to\infty$ we have  
\begin{equation}
  \label{eq:eikonal_3_bis}
  \begin{aligned}
    Z_Q\hS{Q} + \delta_Q = 
  2 \sqrt{\zeta\zeta'} E e^{i(\m_Q'+\m_Q -2m_Q)T}
\int [d^3b]\, e^{iq\cdot b}\, {W}_{u_+}(b) \, , 
\end{aligned}
\end{equation}
where now ${W}_{u_+}(b)$ is a Wilson line running along the $+$ lightcone
direction, $[d^3b]$ includes trasverse coordinates and the $-$
lightcone coordinate, and we have set $\nu_f=-\nu_i=T$ for
simplicity.

\end{document}